\documentclass[hyper,vecarrow]{andp2012}

\usepackage[english]{babel}
\usepackage{amsmath,microtype}
\usepackage{mydefs,journals}
\usepackage{xparse}
\usepackage{diagramLines}


\title{Cosmic Structure Formation with Kinetic Field Theory}

\author[M. Bartelmann]{Matthias Bartelmann\inst{1,}\footnote{corresponding author}}
\author[E. Kozlikin]{Elena Kozlikin\inst{1}}
\author[R. Lilow]{Robert Lilow\inst{1,2}}
\author[C. Littek]{Carsten Littek\inst{1}}
\author[F. Fabis]{Felix Fabis\inst{1}}
\author[I. Kostyuk]{Ivan Kostyuk\inst{1}}
\author[C. Viermann]{Celia Viermann\inst{1,3}}
\author[L. Heisenberg]{Lavinia Heisenberg\inst{4}}
\author[S. Konrad]{Sara Konrad\inst{1}}
\author[D. Geiss]{Daniel Geiss\inst{1,5}}
\address[1]{Universit\"at Heidelberg, Zentrum f\"ur Astronomie, Institut f\"ur Theoretische Astrophysik, Philosophenweg 12, 69120 Heidelberg, Germany}
\address[2]{Department of Physics, Technion, Haifa 3200003, Israel}
\address[3]{Universit\"at Heidelberg, Kirchhoff-Institut f\"ur Physik, Im Neuenheimer Feld 227, 69120 Heidelberg, Germany}
\address[4]{Institut f\"ur Theoretische Physik, ETH Z\"urich, Wolfgang-Pauli-Strasse 27, 8093 Z\"urich, Switzerland}
\address[5]{Institut f\"ur Theoretische Physik, Universit\"at Leipzig, Br\"uderstr.\ 16, 04103 Leipzig, Germany}

\shortauthors{M. Bartelmann et al.}

\begin{abstract}
  Kinetic Field Theory (KFT) is a statistical field theory for an ensemble of point-like classical particles in or out of equilibrium. We review its application to cosmological structure formation. Beginning with the construction of the generating functional of the theory, we describe in detail how the theory needs to be adapted to reflect the expanding spatial background and the homogeneous and isotropic, correlated initial conditions for cosmic structures. Based on the generating functional, we develop three main approaches to non-linear, late-time cosmic structures, which rest either on the Taylor expansion of an interaction operator, suitable averaging procedures for the interaction term, or a resummation of perturbation terms. We show how an analytic, parameter-free equation for the non-linear cosmic power spectrum can be derived.

  We explain how the theory can be used to derive the density profile of gravitationally bound structures and use it to derive power spectra of cosmic velocity densities. We further clarify how KFT relates to the BBGKY hierarchy. We then proceed to apply kinetic field theory to fluids, introduce a reformulation of KFT in terms of macroscopic quantities which leads to a resummation scheme, and use this to describe mixtures of gas and dark matter. We discuss how KFT can be applied to study cosmic structure formation with modified theories of gravity. As an example for an application to a non-cosmological particle ensemble, we show results on the spatial correlation function of cold Rydberg atoms derived from KFT.
\end{abstract}

\shortabstract

\begin{document}

\maketitle

\section{Introduction}
\label{sec:1}

In our cosmic neighbourhood, we see ourselves surrounded by rich and pronounced, large-scale structures. Considering individual galaxies as the smallest constituents of these structures, we see large, almost empty regions, so-called voids, surrounded by filaments which intersect in knots. The typical size of the voids is of order $10\,h^{-1}\,\mathrm{Mpc}$\footnote{Astronomical units of length: $1\,\mathrm{Mpc} = 3.1\cdot10^{24}\,\mathrm{cm}$; $h$ is the Hubble constant in units of $100\,\mathrm{km\,s^{-1}\,Mpc^{-1}}$}. The knots form the so-called galaxy clusters. In addition to this final, present-day state of cosmic structures, we can also observe seed structures in the very early universe. According to the well-established standard model of cosmology (see \cite{2010RvMP...82..331B} for a review), the universe originated in a hot, dense state called the Big Bang. The electromagnetic heat radiation left over from this event was predicted in \cite{1949PhRv...75.1089A}, observed in \cite{1965ApJ...142..419P} and confirmed to have a near-perfect black-body spectrum with a temperature of $2.7\,\mathrm{K}$ in \cite{1990ApJ...354L..37M}. It forms the so-called cosmic microwave background (CMB), which was released when the cosmic plasma recombined, approximately 400,000 years after the Big Bang.

\begin{figure}[ht]
  \includegraphics[width=\hsize]{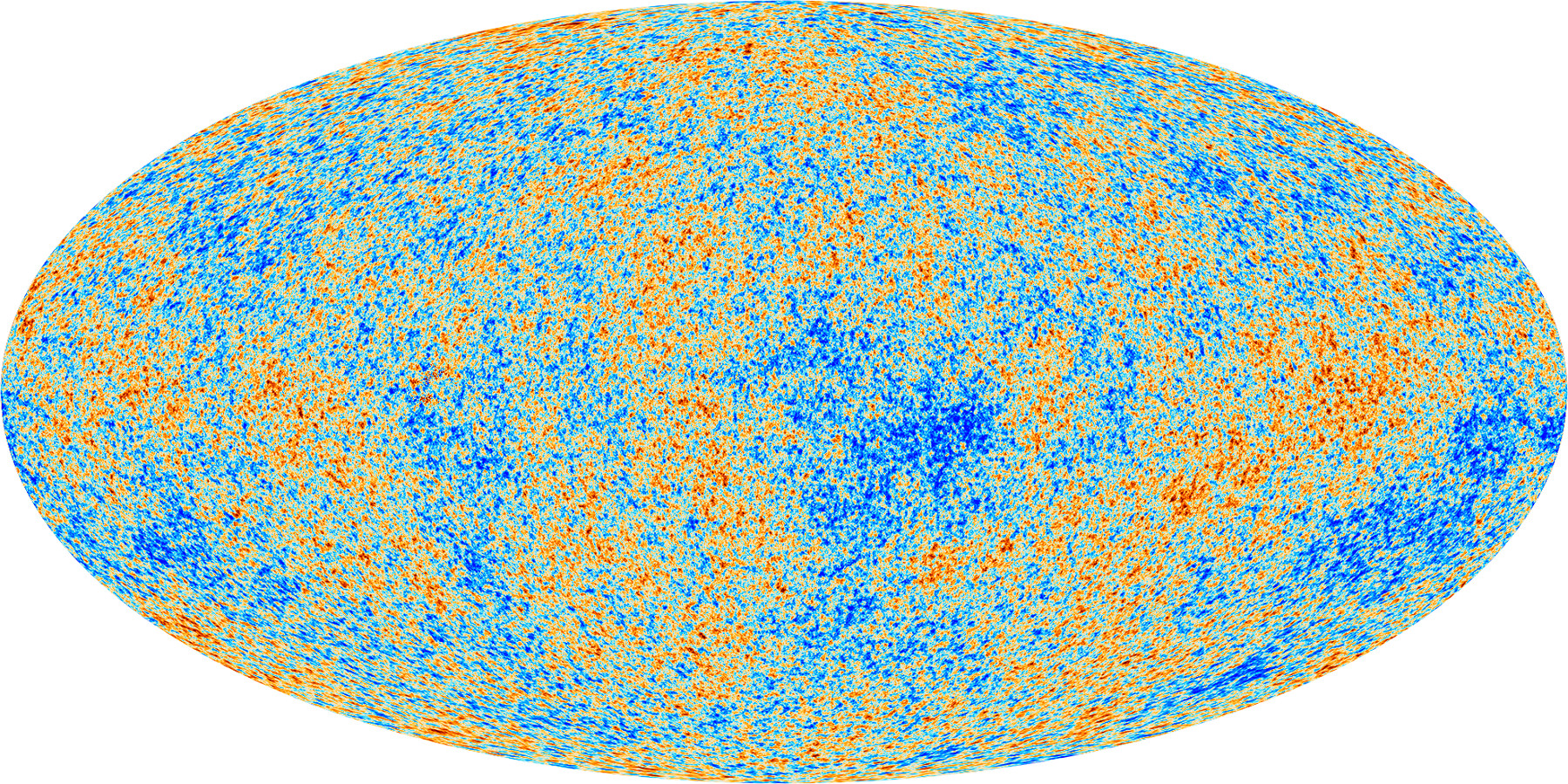}
  \includegraphics[width=\hsize]{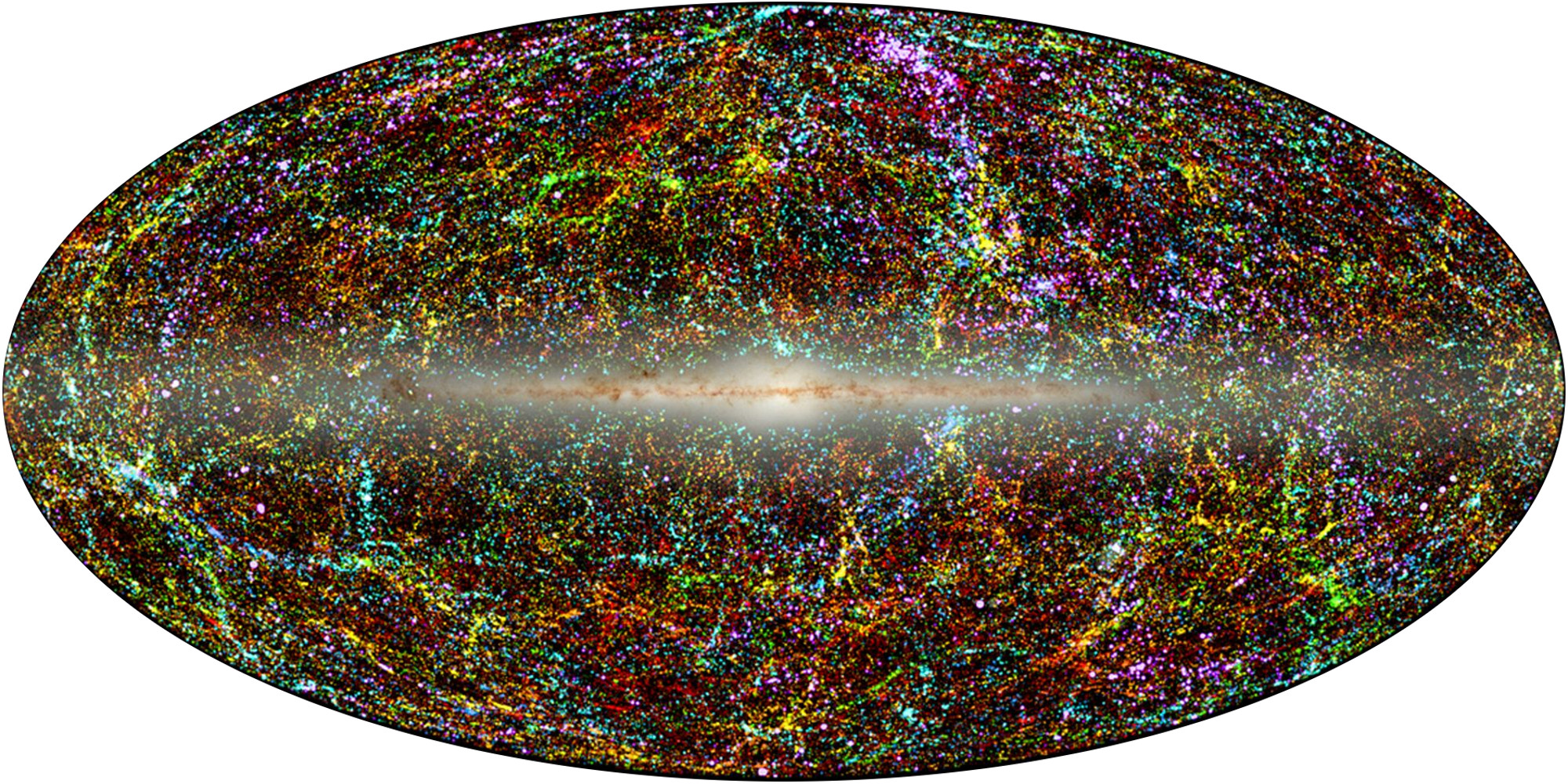}
\caption{These two full-sky maps illustrate the problem of cosmic structure formation. The top panel shows the temperature fluctuations in the Cosmic Microwave Background as observed by the Planck satellite \cite{2016A&A...594A...9P}. They reflect the initial conditions for cosmic structures at approximately 400,000 years after the Big Bang. The bottom panel shows the distribution of galaxies in our cosmic neighbourhood as observed by the 2-MASS survey \cite{2006AJ....131.1163S}.}
\label{fig:1}
\end{figure}

The CMB is almost perfectly isotropic. At closer inspection, it reveals temperature fluctuations with a relative amplitude of $\approx10^{-5}$. We have good reasons to believe that these temperature fluctuations trace the cosmic structures present very shortly after the Big Bang. Notwithstanding the important question as to how these structures originated, we consider them as reflecting the initial state of the evolved cosmic structures we see today. Figure \ref{fig:1} shows both, the initial state as reflected by the CMB temperature fluctuations as observed by the Planck satellite\cite{2016A&A...594A...9P}, and the final state as shown by a galaxy survey conducted at $2\,\mu\mathrm{m}$ wavelength (the 2MASS survey, \cite{2006AJ....131.1163S}).

It is for several reasons most important for cosmology to understand the evolution, the amplitude and the morphology of these structures. In the cosmological standard model, reconciling the amplitudes of the initial and the final state is possible only if we assume that by far most of the matter in the universe does not partake in the electromagnetic interaction, and is therefore called dark matter \cite{1982ApJ...263L...1P}. If this was not the case, the amplitude of the CMB temperature fluctuations would have to be two orders of magnitude larger. Within the limits of even the most precise measurements, the initial state can be considered as a Gaussian random field \cite{2016A&A...594A..17P}. The formation of filamentary structures from a Gaussian random field has been shown to be a natural consequence of gravitational collapse of overdense regions against the cosmic expansion \cite{1970Ap......6..320D, 1986ApJ...304...15B}.

It is quite straightforward to analyze the early stages of cosmic structure formation analytically as long as the relative density fluctuations are small, more precisely as long as the so-called density contrast $\delta$ (cf.\ Eq.\ \ref{eq:45}) is less than unity. The standard procedure begins with the equations of ideal hydrodynamics and evaluates them on an expanding background to show that the density contrast grows in place proportional to the so-called linear growth factor $D_+$,
\begin{equation}
  \delta(a) = \delta_0\,D_+(a)\;,
\label{eq:1}
\end{equation}
with $a$ being the cosmic scale factor describing the expansion of the universe (see Sect.\ \ref{sec:3} below), and $\delta_0$ the density contrast at an arbitrary reference time when $D_+$ is set to unity. As long as the growth remains linear, structures do not change their size against the expanding background. A Fourier analysis in wave numbers co-moving with the mean cosmic expansion shows that the Fourier modes of the density-contrast field neither couple to each other nor change their wave number with time. Large-scale density-fluctuation modes are still linear today.

Despite formidable efforts and successes, it has turned out to be quite hard to follow cosmic structure formation into the non-linear regime with analytic means. See \cite{2002PhR...367....1B} for an excellent review on standard perturbation theory, \cite{1995ApJ...455....7M, 2001A&A...379....8V, 2004ApJ...612...28M, 2004A&A...421...23V, 2006PhRvD..73f3519C, 2006PhRvD..73f3520C, 2008JCAP...10..036P, 2011JCAP...06..015A, 2012JCAP...01..019P, 2012JCAP...12..013A} as examples for several innovative approaches, \cite{1992MNRAS.254..729B, 1993MNRAS.264..375B, 1994MNRAS.267..811B, 1995A&A...296..575B, 1997GReGr..29..733E, 2008PhRvD..77f3530M, 2008PhRvD..78h3503B, 2012JCAP...06..021R, 2013PhRvD..87h3522V} as examples for Lagrangian perturbation theory and \cite{2014JCAP...05..022P, 2014JCAP...07..057C, 2014PhRvD..89d3521H} for recent developments towards an effective field theory of cosmic structure formation. One of the main reasons for this is that the fluid description of the predominantly dark matter must ultimately fail where and when the matter flow converges. When streams meet in a fluid, a discontinuity or a shock forms, keeping the velocity field of the flow unique. This is a consequence of the central assumption of (ideal) hydrodynamics that the mean-free path of the fluid particles is negligibly small. When dark-matter streams meet, however, multiple streams will form that simply cross each other, leading to a multi-valued velocity field. This notorious shell-crossing problem of cosmic structure formation is most likely the most important reason hampering progress in analytic theories of cosmic structure formation.

Of course, one can resort to fully numerical simulations. Corresponding algorithms have reached an impressive level of sophistication and have delivered overwhelmingly detailed results that, by and large, agree very well with observations. Yet, good reasons remain for trying to understand the formation of late-time, non-linear cosmic structures on an analytical basis. Conceptually the most important of these reasons is that numerical simulations strictly speaking do not \emph{explain} the properties of cosmic structures, even though they succeed in \emph{reproducing} them in astounding detail. Of particular importance are universal features of cosmic structures, such as the density profiles of gravitationally-bound objects. Analytic theories, on the other hand, could be expected to trace physical properties of non-linear structures back to their foundations in fundamental physics.

A second convincing reason in favour of an analytic theory is its flexibility against changing assumptions. Running sufficiently detailed and well-resolved numerical simulations is computationally expensive, and scanning wider parameter ranges is thus often forbiddingly time consuming. A third good reason is that shot noise is inevitable in numerical simulations, caused by the finite number of particles in any simulation volume. An analytic theory could be expected to allow taking the thermodynamic limit. Higher-order measures for the statistics of the cosmic structures, such as three- or four-point correlation functions which are indispensable to quantify non-linear and in particular non-Gaussian structures, could then be calculated without uncertainties due to shot noise and the finite number of samples that can be drawn from any numerical simulation.

Kinetic field theory is based on field-theoretical descriptions of classical particles \cite{1973PhRvA...8..423M, 1977PhRvA..16..732F} and was developed mainly for studying glasses, fluctuation-dissipation theorems and the ergodic-non-ergodic transition \cite{2010PhRvE..81f1102M, 2011PhRvE..83d1125M, 2012JSP...149..643D, 2013JSP...152..159D}. We have adapted it to cosmology and applied it to various aspects of cosmological structure formation \cite{2015PhRvD..91h3524B, 2015PhRvE..91f2120V, 2016NJPh...18d3020B, 2017NJPh...19h3001B, 2018JSMTE..04.3214F}, but also to completely different kinds of physical systems whose elementary constituents can be described as classical degrees of freedom. For cosmology, the KFT approach has several decisive advantages, the most important of which is that the motion of the classical particles is described in phase space by Hamilton's equations. Since trajectories in phase space do not cross, this approach avoids the shell-crossing problem by construction. KFT incorporates the dynamics of the particles into a generating functional. Structurally, this generating functional resembles the generating functionals of statistical quantum field theories. One important simplification of the application to classical particles is due to the symplectic nature of Hamilton's equations, which gives rise to Liouville's theorem. Thus, by construction, KFT is built on a diffeomorphic map with unit functional determinant of an initial phase-space configuration to any later time.

We begin this review in Sect.\ \ref{sec:2} with the foundations of the theory and the construction of its central object, i.e.\ its generating functional. We specialize it to cosmology in Sect.\ \ref{sec:3} and summarize several further developments in Sect.\ \ref{sec:4}, mainly offering different ways of taking particle interactions into account in approximate or averaged ways. Here, we derive a closed, analytic, non-perturbative, parameter-free equation for the non-linear density-fluctuation power spectrum which agrees very well with the results from numerical simulations up to wave numbers of $k\approx10\,h^{-1}\,\mathrm{Mpc}$. In Sect.\ \ref{sec:5}, we review how KFT can be used to explain the density profiles of dark-matter haloes. Section \ref{sec:6} presents the derivation of velocity power spectra. In Sect.\ \ref{sec:7}, we establish the connection between KFT and the BBGKY hierarchy of kinetic theory, offering a closure condition for the hierarchy. We discuss in Sect.\ \ref{sec:8} how fluids can be incorporated into KFT. In Sect.\ \ref{sec:9}, we give an exact re-formulation of KFT purely in terms of macroscopic fields and show how this leads to an efficient resummation of perturbation terms. The fluid description and the macroscopic re-formulation are then combined in Sect.\ \ref{sec:10} to describe mixtures of dark matter and gas with KFT. We outline in Sect.\ \ref{sec:11} how KFT can be extended to be applied within modified theories of gravity, and illustrate this outline with results for one specific example. Finally, in Sect.\ \ref{sec:12}, we give an outlook on how to apply KFT to ensembles of cold Rydberg atoms as one example for a completely different physical system. We summarize this review in Sect.\ \ref{sec:13}.

\section{Kinetic Field Theory}
\label{sec:2}

\subsection{Generating Functional}
\label{sec:2.1}

Kinetic field theory is a statistical field theory for classical particle ensembles in or out of equilibrium. Its central mathematical object is a generating functional $Z$. In close analogy to the partition sum of equilibrium thermodynamics, this generating functional integrates the probability distribution $P(\varphi)$ for system states $\varphi$ over the state space,
\begin{equation}
  Z = \int\mathcal{D}\varphi\,P(\varphi)\;.
\label{eq:2}
\end{equation}
It is generally a path integral if the states space is a function space of ﬁeld conﬁgurations.

For classical (canonical) ensembles consisting of $N$ point particles, we can immediately specify the generating functional further. The state space is then the phase space $\Gamma$ of the $N$ particles whose trajectories $(q_j, p_j) =: x_j$ with $1\le j\le N$ are tuples of position $q_j$ and momentum $p_j$. To allow a compact notation, we bundle the phase-space trajectories $x_j$ into a tensorial object
\begin{equation}
  \vc x = x_j\otimes e_j\;,
\label{eq:3}
\end{equation}
where summation over $j$ is implied and the vector $e_j$ has components $(e_j)_i = \delta_{ij}$, $1\le i\le N$. For such tensors, we define the scalar product
\begin{equation}
  \langle\vc x, \vc y\rangle =
  \left(x_i\cdot y_j\right)\left(e_i\cdot e_j\right) = x_j\cdot y_j\;.
\label{eq:4}
\end{equation}
Point particles are described by Dirac delta distributions instead of smooth fields. The path integral in (\ref{eq:2}) then turns into an ordinary integral over the $N$-particle phase space,
\begin{equation}
  Z = \int\D\vc x\,P(\vc x)\;.
\label{eq:5}
\end{equation}
We introduce a generator field $\vc J$ conjugate to the phase-space trajectories $\vc x(t)$ into the generating functional $Z$,
\begin{equation}
  Z \to Z[\vc J] = \int\D\vc x\,P(\vc x)\,\exp\left\{
    \I\int_0^\infty\D t'\left\langle\vc J(t'), \vc x(t')\right\rangle
  \right\}\;,
\label{eq:6}
\end{equation}
such that the functional derivative of $Z[\vc J]$ with respect to the component $J_j(t)$ of the generator field returns the average position $\langle x_j(t)\rangle$ of particle $j$ at time $t$,
\begin{equation}
  \langle x_j(t)\rangle =
  -\I\left.\frac{\delta}{\delta J_j(t)}Z[\vc J]\right|_{\vc J = 0}\;.
\label{eq:7}
\end{equation}
Like $x$, the generator field has components conjugate to position $q$ and momentum $p$,
\begin{equation}
  \vc J = J_j\otimes e_j =
  \begin{pmatrix} J_{q_j} \\ J_{p_j} \end{pmatrix} \otimes e_j\;.
\label{eq:8}
\end{equation}

\begin{figure}[ht]
  \includegraphics[width=\hsize]{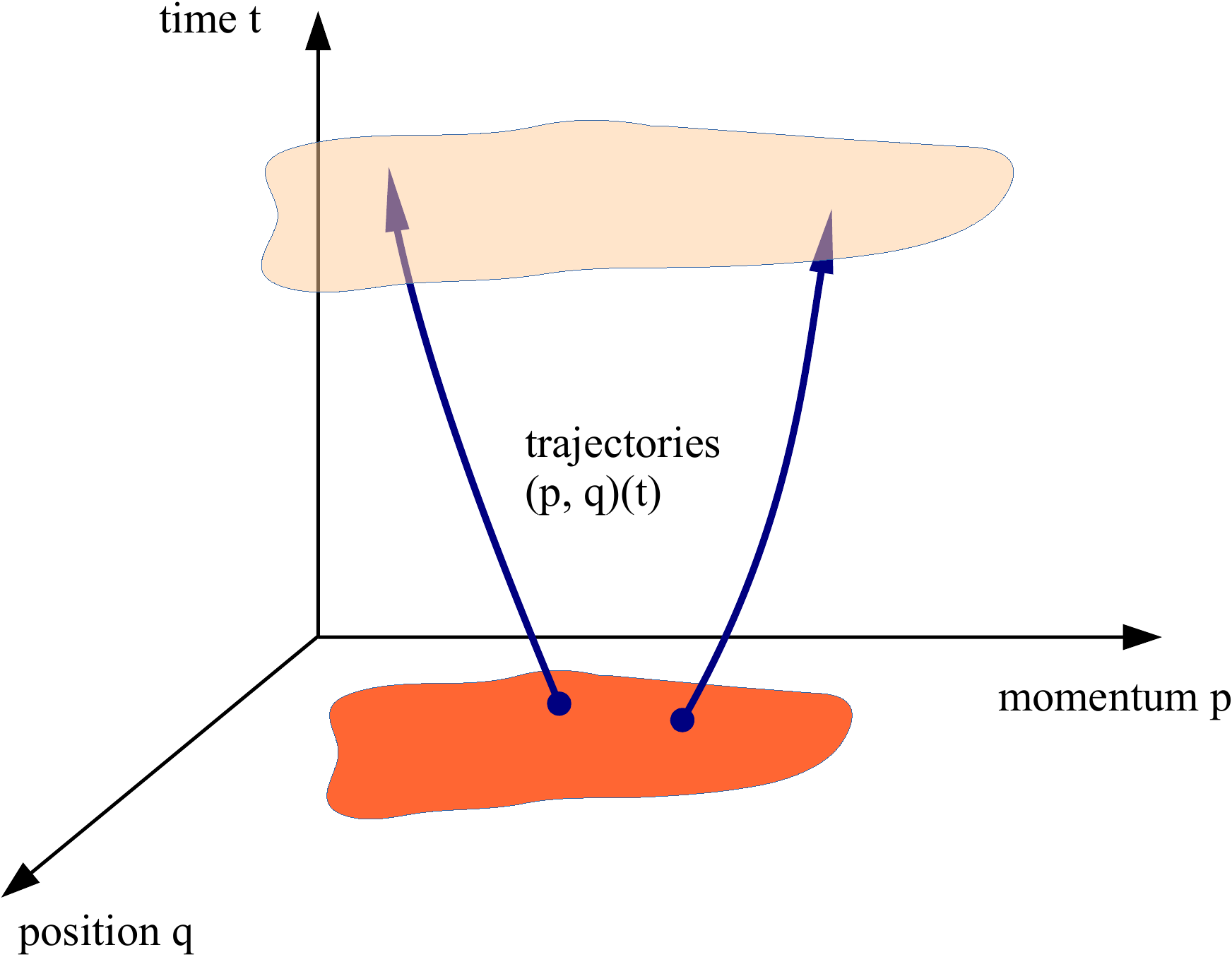}
\caption{Illustration of the main idea of KFT. An initial probability distribution on the phase space is mapped to any later time by means of the Hamiltonian flow. Since phase-space trajectories do not cross, this approach avoids the notorious shell-crossing problem by construction.}
\label{fig:2}
\end{figure}

We further split the probability $P(\vc x)$ for the state $\vc x$ to be occupied into a probability $P(\vc x^\mathrm{(i)})$ for the particle ensemble to occupy an initial state $\vc x^\mathrm{(i)}$ at time $t = 0$, and the conditional probability $P(\vc x|\vc x^\mathrm{(i)})$ for the ensemble to move from there to the time-evolved state $\vc x$,
\begin{equation}
  P(\vc x) = \int\D\vc x^\mathrm{(i)}\,
    P\left(\vc x|\vc x^\mathrm{(i)}\right)P\left(\vc x^\mathrm{(i)}\right)\;.
\label{eq:9}
\end{equation}
For particles on classical trajectories, the transition probability $P(\vc x|\vc x^\mathrm{(i)})$ must be a functional delta distribution of the classical equation of motion,
\begin{equation}
  P\left(\vc x|\vc x^\mathrm{(i)}\right) =
    \delta_\mathrm{D}\left[
      \vc x-\Phi_\mathrm{cl}\left(\vc x^\mathrm{(i)}\right)
    \right]\;,
\label{eq:10}
\end{equation}
where $\Phi_\mathrm{cl}(\vc x^\mathrm{(i)})$ denotes the classical (Hamiltonian) flow on the phase space, beginning at the initial phase-space points $\vc x^\mathrm{(i)}$ of the particle ensemble. Writing the equation of motion in the form $E(\vc x) = 0$, the Hamiltonian flow consists of all solutions of this equation for initial points within a certain domain of phase space, as illustrated in Figure \ref{fig:2}.

Denoting the solution of the classical equations of motion beginning at $\vc x^\mathrm{(i)}$ as
\begin{equation}
  E\left(\vc x, \vc x^\mathrm{(i)}\right) = 0\;,
\label{eq:11}
\end{equation}
we can write the transition probability as
\begin{equation}
  P\left(\vc x|\vc x^\mathrm{(i)}\right) =
    \delta_\mathrm{D}\left[E\left(\vc x, \vc x^\mathrm{(i)}\right)\right]
\label{eq:12}
\end{equation}
with a functional delta distribution singling out the deterministic trajectories solving the equations of motion.

For classical point particles, the equation of motion for a single particle is the Hamiltonian equation,
\begin{equation}
  \dot x-\mathcal{J}\nabla_x H = 0\;,
\label{eq:13}
\end{equation}
where $H$ is the Hamiltonian function on the phase space and
\begin{equation}
  \mathcal{J} := \begin{pmatrix} 0 & \mathcal{I}_3 \\ -\mathcal{I}_3 & 0 \\ \end{pmatrix}
\label{eq:14}
\end{equation}
is the usual symplectic matrix, with $\mathcal{I}_n$ denoting the unit matrix in $n$ dimensions. Let us first assume the particles to be non-interacting, in which case the Hamiltonian equations are linear and admit the construction of a Green's function $G(t,t')$. To prepare for the inclusion of interactions, we augment this free equation of motion by an inhomogeneity or source term $K$,
\begin{equation}
  \dot x-\mathcal{J}\nabla_x H = K\;,
\label{eq:15}
\end{equation}
and write its solution $\bar x(t)$ beginning at $x^\mathrm{(i)}$ in terms of the Green's function as
\begin{equation}
  \bar x(t) = G(t,0)x^\mathrm{(i)}+\int_0^t\D t'G(t,t')K(t')\;.
\label{eq:16}
\end{equation}
We shall henceforth assume that the inhomogeneity is caused by an interaction with the potential $V$. Then, we can write
\begin{equation}
  K(t') = \begin{pmatrix} 0 \\ -\nabla V(t') \end{pmatrix}\;.
\label{eq:17}
\end{equation}
The Green's function $G$ itself is a $6\times6$ matrix which we write as
\begin{equation}
  G(t,t') := 
  \begin{pmatrix}
    g_{qq}(t,t')\,\mathcal{I}_3 & g_{qp}(t,t')\,\mathcal{I}_3 \\
    0 & g_{pp}(t,t')\,\mathcal{I}_3 \\
  \end{pmatrix}\;.
\label{eq:18}
\end{equation}
Its component functions $g_{qq}(t,t')$, $g_{qp}(t,t')$, and $g_{pp}(t,t')$ will be specified later in a cosmological context. For simplicity, we abbreviate $g_{qp}(t,0) =: g_{qp}(t)$.

For the entire particle ensemble, we introduce the $N$-particle Green's function $\vc G(t,t') := G(t,t')\otimes\mathcal{I}_N$ and the $N$-component source field $\vc K := K_j\otimes e_j$. Then, the solutions of the equations of motion for the particle ensemble are
\begin{equation}
  E\left(\vc x, \vc x^\mathrm{(i)}\right) = \vc x(t)-\bar{\vc x}(t) = 0\;.
\label{eq:19}
\end{equation}
Inserting this expression into (\ref{eq:12}), the result into (\ref{eq:9}) and the probability distribution $P(\vc x^\mathrm{(i)})$ into the generating functional $Z[\vc J]$ from (\ref{eq:6}) gives
\begin{equation}
  Z[\vc J, \vc K] = \int\D\Gamma\,
  \exp\left\{\I\int_0^\infty\D t'\,\left\langle
    \vc J(t'), \bar{\vc x}(t')
  \right\rangle\right\}\;,
\label{eq:20}
\end{equation}
where we have introduced the initial phase-space measure
\begin{equation}
  \D\Gamma := \D\vc x^\mathrm{(i)}P\left(\vc x^\mathrm{(i)}\right)\;.
\label{eq:21}
\end{equation}

\subsection{Density Operator}
\label{sec:2.2}

The standard application of KFT in this review will concern the calculation of density power spectra. We shall thus proceed to define a density operator and study its action on the generating functional (\ref{eq:20}). The number density of the particles at time $t_1$ is
\begin{equation}
  \rho(q,t_1) = \sum_{j=1}^N\delta_\mathrm{D}\left(q-q_j(t_1)\right)\;.
\label{eq:22}
\end{equation}
In a Fourier representation, defined by the convention
\begin{equation}
  \tilde f(k) = \int\D^3q\,f(q)\,\E^{-\I kq}\;,\quad
  f(q) = \int\frac{\D^3k}{(2\pi)^3}\,\tilde f(k)\,\E^{\I kq}\;,
\label{eq:23}
\end{equation}
the density mode with the wave number $k_1$ is
\begin{equation}
  \tilde\rho(k_1,t_1) =: \tilde\rho(1) = \sum_{j=1}^N\,\E^{-\I k_1q_j(t_1)}\;,
\label{eq:24}
\end{equation}
where we have introduced the conventional short-hand notation $(k_j,t_j) =: (j)$. We shall abbreviate the integral expressions in (\ref{eq:23}) as
\begin{equation}
  \int_q := \int\D^3q\;,\quad\int_k := \int\frac{\D^3k}{(2\pi)^3}\;.
\label{eq:25}
\end{equation}
Replacing the particle position $q_j(t_1)$ by a functional derivative with respect to the generator-field component $J_{q_j}(t_1)$, we find the density operator
\begin{equation}
  \hat\rho(1) = \sum_{j=1}^N\hat\rho_j(1)\;,
\label{eq:26}
\end{equation}
composed of the one-particle density operators
\begin{equation}
  \hat\rho_j(1) := 
  \exp\left(-k_1\frac{\delta}{\delta J_{q_j}(t_1)}\right)\;.
\label{eq:27}
\end{equation}
For indistinguishable particles, which we shall henceforth assume, we can set the particle index $j$ to an arbitrary value without loss of generality. Since the operator (\ref{eq:27}) is an exponential of a derivative with respect to the generator field, it corresponds to a finite translation of the generator field. After applying $r\ge1$ of these operators, the generator field is translated by
\begin{equation}
  \vc J \to\vc J-\sum_{j=1}^r\delta_\mathrm{D}\left(t'-t_j\right)\,
  k_j\cdot\begin{pmatrix} 1 \\ 0 \end{pmatrix} \otimes e_j\;.
\label{eq:28}
\end{equation}
Setting the generator field $\vc J$ to zero after application of the density operator turns the generating functional into
\begin{equation}
  Z[\vc L] = \int\D\Gamma\,\E^
    {\I\left\langle\vc L_q,\vc q\right\rangle+
     \I\left\langle\vc L_p,\vc p\right\rangle+\I S_\mathrm{I}}
\label{eq:29}
\end{equation}
with the components
\begin{align}
  \vc L_q &= -\sum_{j=1}^rk_j\otimes e_j\;,\nonumber\\
  \vc L_p &= -\sum_{j=1}^rk_jg_{qp}(t_j)\otimes e_j\;,
\label{eq:30}
\end{align}
of the translations and the interaction term
\begin{equation}
  S_\mathrm{I} = \sum_{j=1}^rk_j\cdot \int_0^{t_j}\D t'\,g_{qp}(t_j,t')\,\nabla_jV(t')\;.
\label{eq:31}
\end{equation}
In (\ref{eq:29}), $\vc q$ and $\vc p$ in the exponent are the \emph{initial} particle positions and momenta,
\begin{equation}
  \begin{pmatrix} \vc q\\ \vc p\end{pmatrix} = \vc x^\mathrm{(i)}\;.
\label{eq:32}
\end{equation}

The generating functional (\ref{eq:20}) and the result (\ref{eq:29}) after applying $r$ one-particle density operators and setting the source field $\vc J$ to zero are fully general in the sense that neither the initial particle distribution in phase space, nor the Green's function, nor the interaction potential have been specified yet. So far, we have just assumed that a Green's function and an interaction potential exist.

\subsection{Interaction Operator}
\label{sec:2.3}

We suppose that the potential $V$ is a linear superposition of contributions by individual particles,
\begin{align}
  V(q,t) &= \sum_{j=1}^Nv\bigl(q-q_j(t)\bigr) =
  \sum_{j=1}^N\int_yv(q-y)\delta_\mathrm{D}\bigl(y-q_j(t)\bigr) \nonumber\\
  &= \int_y\,v(q-y)\rho(y,t)\;,
\label{eq:33}
\end{align}
where we have identified the density $\rho$ from (\ref{eq:22}). The potential gradient at position $q_i$ is
\begin{align}
\label{eq:34}
  \nabla_iV(q,t) &=
  \int_q\delta_\mathrm{D}\left(q-q_i(t)\right)\,\nabla V(q,t) \\ &=
  \int_q\,\rho_i(q,t)\,\nabla V(q,t) =: -\int_q\,B_i(q,t)\,V(q,t)\;, \nonumber
\end{align}
where we have introduced the \emph{response field}
\begin{equation}
  B_i(q,t) := -\nabla\delta_D\left(q-q_i(t)\right)
\label{eq:35}
\end{equation}
after partial integration. With (\ref{eq:33}), we can write this result as
\begin{equation}
  \nabla_iV(q,t) = \int_q\int_y\,B_i(q,t)\,v(q-y)\,\rho(y,t)\;.
\label{eq:36}
\end{equation}
Applying the convolution theorem, this expression turns into
\begin{equation}
  \nabla_iV(q,t'_1) = \int_{k_1'}\,
  \tilde B_i(-1')\,\tilde v(1')\,\tilde\rho(1')
\label{eq:37}
\end{equation}
where $(-1') =: (-k_1',t_1')$. The Fourier transform of the density $\rho$ is expressed by the operator $\hat\rho$ from (\ref{eq:26}), while we can assign the operator
\begin{equation}
  \hat B_i(-1') = \I k_1'\,\E^{\I k_1'q_i(t_1')} = \I k_1'\,\hat\rho_i(-1')
\label{eq:38}
\end{equation}
to the response field.

\begin{figure}[ht]
  \includegraphics[width=\hsize]{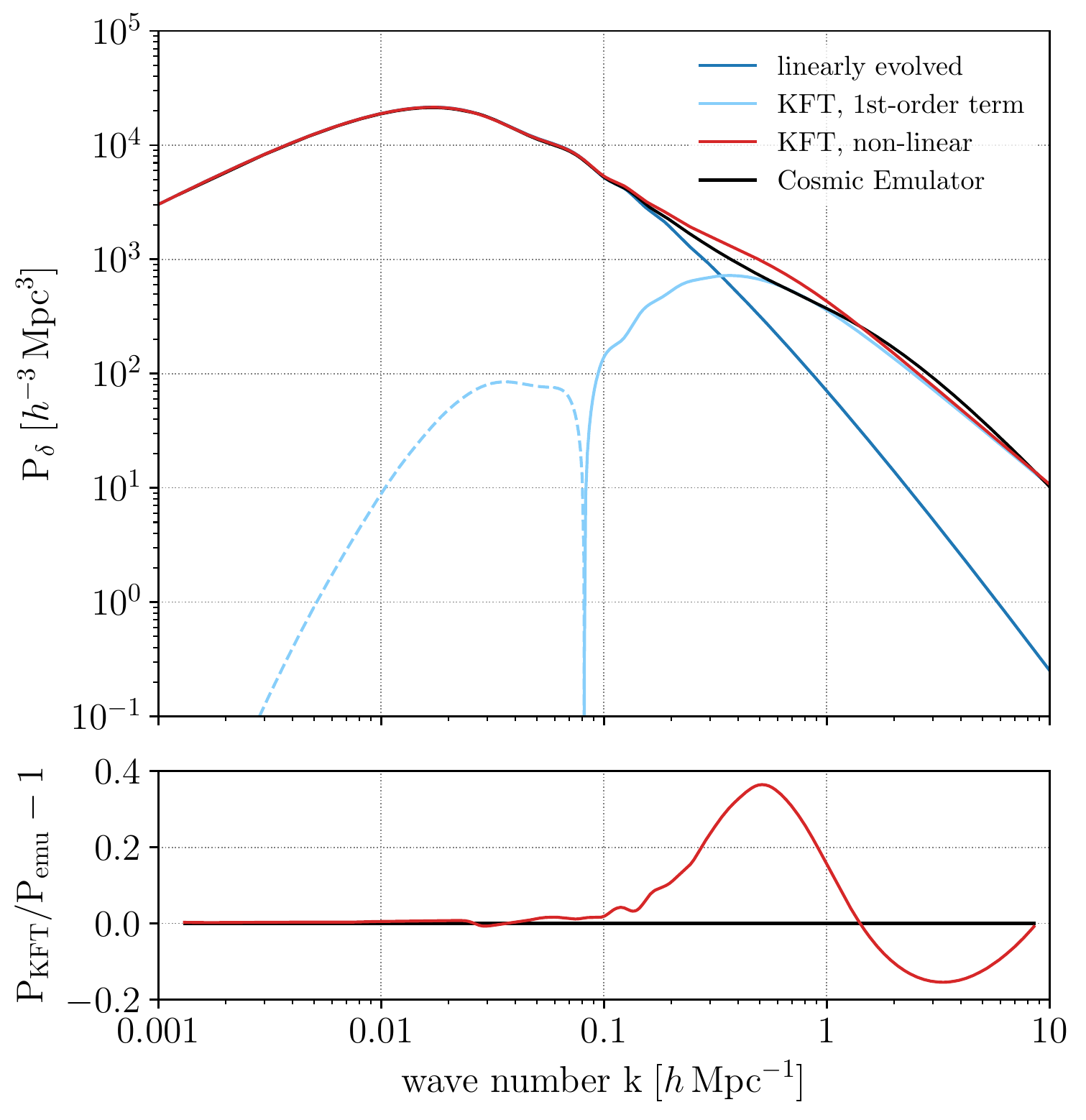}
\caption{Example of a perturbative result in a cosmological application. (Upper panel) The dark-blue curve shows the initial CAMB density-fluctuation power spectrum \cite{2000ApJ...538..473L} ($z_\mathrm{i} = 1100$), linearly evolved to the present time. The light-blue curve is the first-order perturbation term with improved Zel'dovich trajectories, with dashed parts illustrating negative values. The curve thus illustrates how power is removed from intermediate and transported to smaller scales. The red curve is the sum of the linearly evolved power spectrum and the first-order perturbation term. The black curve finally shows the non-linearly evolved power spectrum obtained from numerical simulations. Already at first order, the analytic result comes close to the fully numerical result. (Lower panel) The relative deviation of the KFT result as compared to the numerical simulation result of Cosmic Emulator \cite{2014ApJ...780..111H, 2010ApJ...715..104H, 2010ApJ...713.1322L, 2009ApJ...705..156H} is shown for the $k$-range provided by Cosmic Emulator.}
\label{fig:3}
\end{figure}

With the definition $\D1' := \D t_1'\D^3k_1'/(2\pi)^3$, this allows us to write the interaction term $S_\mathrm{I}$ as the \emph{interaction operator}
\begin{equation}
  \hat S_\mathrm{I} = \sum_{j=1}^rk_j\cdot\int_0^{t_j}\D1'\,g_{qp}(t_j,t'_1)\,
  \hat B_j(-1')\tilde v(1')\hat\rho(1')
\label{eq:39}
\end{equation}
acting on the free generating functional,
\begin{equation}
  Z[\vc L] = \E^{\I\hat S_\mathrm{I}}Z_0[\vc L]\;,
\label{eq:40}
\end{equation}
with
\begin{equation}
  Z_0[\vc L] = \int\D\Gamma\,\E^
  {\I\left\langle\vc L_q,\vc q\right\rangle+
   \I\left\langle\vc L_p,\vc p\right\rangle}\;.
\label{eq:41}
\end{equation}
One approach to perturbation theory can now begin with a Taylor expansion of exponential factor in terms of the interaction operator. Anticipating the specialisation of KFT to cosmology detailed in Sect.\ \ref{sec:3}, Fig.\ \ref{fig:3} shows cosmological results obtained in first-order perturbation theory \cite{2016NJPh...18d3020B}. A summary of this calculation is presented in Appendix A.

\section{Specialisation to Cosmology}
\label{sec:3}

We shall now specify the theory to its cosmological application. Where not stated otherwise, we use the cosmological parameters $\Omega_\mathrm{m} = 0.3$, $\Omega_{\Lambda} = 0.7$, $\Omega_\mathrm{b} = 0.04$ for the density parameters of matter, the cosmological constant, and baryonic matter, respectively; further $h = 0.7$ for the dimension-less Hubble constant and $\sigma_8 = 0.8$ for the normalisation of the power spectrum. Moreover, we choose the form of the cold-dark matter transfer function provided by \cite{1986ApJ...304...15B} to specify the initial power spectrum.

\subsection{Equation of Motion for Point Particles}
\label{sec:3.1}

The cosmological standard model is built upon the theory of general relativity and two symmetry assumptions. They assert that there exists a mean flow in the Universe with respect to which all sufficiently averaged observable properties appear spatially isotropic and homogeneous. This mean flow can then only either expand or shrink isotropically and must thus be characterised by a scale factor $a(t)$, depending only on cosmic time $t$. The dynamics of $a(t)$ is determined by Einstein's field equations which, under the symmetry assumptions made, simplify to Friedmann's equations. The relative cosmic expansion rate is the Hubble function $H(a) = \dot a/a$. Usually, $a$ is normalised to unity today, but it is more convenient for our purposes to normalise $a$ to unity at the initial time $t=0$ such that $a(0) =: a_\mathrm{i} = 1$.

The existence of the mean flow suggests to introduce coordinates $q$ comoving with the flow, defined in terms of the physical coordinates $r$ via $r = aq$. The Lagrange function of point particles of mass $m$, expressed in terms of the comoving coordinates, is
\begin{equation}
  L(q,\dot q, t) = \frac{m}{2}a^2\dot q^2-m\Phi
\label{eq:42}
\end{equation}
(see, e.g.\ \cite{1980lssu.book.....P}), where $\Phi$ is the gravitational potential sourced by the fluctuations of the matter density $\rho$ around its time-dependent mean $\bar\rho$ via the Poisson equation
\begin{equation}
  \nabla^2\Phi = 4\pi Ga^2\left(\rho-\bar\rho\right)\;.
\label{eq:43}
\end{equation}
The mean cosmic density after the radiation-dominated epoch is
\begin{equation}
  \bar\rho(a) = \frac{3H_\mathrm{i}^2}{8\pi G}a^{-3}\Omega_\mathrm{mi}\;,
\label{eq:44}
\end{equation}
where $H_\mathrm{i}$ and $\Omega_\mathrm{mi}$ are the Hubble function and the cosmic matter-density parameter at the initial time. If we set this time early in the matter-dominated epoch, as we intend to do throughout, we can safely set $\Omega_\mathrm{mi} = 1$ and thus bring the Poisson equation (\ref{eq:43}) into the form
\begin{equation}
  \nabla^2\Phi = \frac{3}{2}H_\mathrm{i}^2\frac{\delta}{a}\;,\quad
  \delta := \frac{\rho-\bar\rho}{\bar\rho}\;,
\label{eq:45}
\end{equation}
with the density fluctuations now being expressed by the dimension-less density contrast $\delta$.

Linear perturbation theory shows that the density fluctuations $\delta$ grow proportional to the so-called linear growth factor $D_+$. It is convenient to replace the cosmic time $t$ by the growth factor,
\begin{equation}
  t\to D_+(t)-D_+(0)\;,
\label{eq:46}
\end{equation}
where the time $t=0$ is now set to some initial time which is late enough for matter to dominate over radiation, but early enough for density fluctuations to be very small. A suitable choice is the time when the cosmic microwave background was released. In this new time coordinate, the Lagrange function reads
\begin{equation}
  \bar{L}(q,\dot q, t) = \frac{g}{2}\dot q^2-\bar\Phi(q)
\label{eq:47}
\end{equation}
(see e.g.\ \cite{2015PhRvD..91h3524B}), with the potential $\bar\Phi$ satisfying the Poisson equation
\begin{equation}
  \nabla^2\bar\Phi = \frac{3}{2}\frac{a}{g}\delta\;.
\label{eq:48}
\end{equation}
Here, $g(t)$ is defined to be the function
\begin{equation}
  g(t) := a^2D_+fH\;,\quad f := \frac{\D\ln D_+}{\D\ln a}\;,
\label{eq:49}
\end{equation}
normalised to unity at $t = 0$ or $a = a_\mathrm{i}$. Notice that the potential $\bar\Phi$ defined in (\ref{eq:48}) now has the dimension of a squared length. The equation of motion reads
\begin{equation}
  \ddot q(t)+\frac{\dot g}{g}\dot q(t)+\nabla\varphi = 0\;,\quad
  \varphi := \frac{\bar\Phi}{g}\;,
\label{eq:50}
\end{equation}
with the potential $\varphi$ now obeying the Poisson equation
\begin{equation}
  \nabla^2\varphi = \frac{3}{2}\frac{a}{g^2}\delta\;.
\label{eq:51}
\end{equation}

The usual Legendre transform of the Lagrange function (\ref{eq:47}) leads to the Hamiltonian
\begin{equation}
  \bar{H}(q,p) = \frac{p^2}{2g}+\bar\Phi
\label{eq:52}
\end{equation}
and thus to the Newtonian equation of motion
\begin{equation}
  \dot x = \left(\begin{array}{c} p/g \\ -\nabla\bar\Phi \end{array}\right)\;,
\label{eq:53}
\end{equation}
which is solved by the Green's function
\begin{equation}
  \bar G(t,t') =
  \begin{pmatrix}
    \mathcal{I}_3 & \bar g_{qp}(t,t') \\ 0 & \mathcal{I}_3
  \end{pmatrix}\;,\quad
  \bar g_{qp}(t,t') = \int_{t'}^t\frac{\D \bar{t}}{g(\bar{t})}\;.
\label{eq:54}
\end{equation}
It is important to note that $\bar g_{qp}(t,t')$ is limited from above because of the cosmic expansion. Consequently, the inertial trajectories described by this Green's function deviate strongly from the true, fully interacting trajectories. It will thus be advantageous to find a replacement for $\bar g_{qp}$ that already captures part of the gravitational interaction. An example for such a replacement is given by the Zel'dovich approximation \cite{1970A&A.....5...84Z}, but we prefer a slightly more general approach here.

\subsection{Particle Trajectories and Effective Force}
\label{sec:3.2}

We wish to solve the equation of motion (\ref{eq:50}) with an expression of the form
\begin{equation}
  q(t) = q_0+g_{qp}(t)\dot q_0+\int_0^t\D t'\,g_{qp}(t,t')\,f(t')
\label{eq:55}
\end{equation}
such that the propagator $g_{qp}(t,t')$ can play the role of the $q$-$p$ component of the Green's function $G(t,t')$. Taking the first two time derivatives of (\ref{eq:55}) gives
\begin{equation}
  \ddot q(t) = \ddot g_{qp}(t)\dot q_0+\dot g_{qp}(t,t)f(t)+
  \int_0^t\D t'\,\ddot g_{qp}(t,t')f(t')\;.
\label{eq:56}
\end{equation}
For (\ref{eq:56}) to agree with the equation of motion (\ref{eq:50}), the effective force $f(t)$ has to be appropriately adapted once $g_{qp}$ has been chosen. A frequent and convenient choice in cosmology is the Zel'dovich approximation \cite{1970A&A.....5...84Z}, which describes particle trajectories as inertial in the time coordinate $t = D_+$,
\begin{equation}
  g_{qp}(t,t') = t-t'\;.
\label{eq:57}
\end{equation}
Inserting this particular choice into (\ref{eq:56}) and comparing to (\ref{eq:50}) immediately results in
\begin{equation}
  f(t) = -\frac{\dot g}{g}\dot q-\nabla\varphi\;.
\label{eq:58}
\end{equation}
This is thus the effective force term in the Zel'dovich approximation. It is chosen such that the \emph{total} initial force, i.e.\ the sum of the potential gradient and the velocity-dependent contribution, vanishes initially, corresponding to the initial inertial motion of the particles.

Another choice for the propagator $g_{qp}$ begins with defining the effective force term such that only its velocity-dependent contribution is initially absent,
\begin{equation}
  f(t) = \frac{\dot g}{g}h\dot q-\nabla\varphi\;,\quad
  h := \frac{1}{g}-1\;.
\label{eq:60}
\end{equation}
The remaining initial force then causes the particles to lag behind the inertial motion. This choice has been suggested to avoid part of the re-expansion in the Zel'dovich approximation of cosmic structures after their formation \cite{2015PhRvD..91h3524B}. Relative to this effective force, the equation of motion (\ref{eq:50}) reads
\begin{equation}
  \ddot q(t)-\dot h(t)\dot q(t) = f(t)\;,
\label{eq:60a}
\end{equation}
which implies the homogeneous solution $\dot q(t) = \dot q(0)\E^h$ and thus the propagator
\begin{equation}
  g_{qp}(t,t') = \int_{t'}^t\D\bar t\,\E^{h(\bar t)-h(t')}\;.
\label{eq:59}
\end{equation}
At late times, $h\to-1$, and the propagator (\ref{eq:59}) approaches $\E^{-1}$ times the Zel'dovich propagator.

Since the peculiar velocity $\dot q$ itself depends on the time-integrated force $f(t)$, (\ref{eq:58}) and (\ref{eq:60}) are integral equations for the force. After transforming them to differential equations, they can be solved by variation of constants. The solution of (\ref{eq:60}), which we will use here together with the propagator (\ref{eq:59}), turns out to be
\begin{equation}
  f(t) = -\nabla\varphi-\frac{\dot gh}{g^2}\int_0^tg\nabla\varphi\D t'\;.
\label{eq:62}
\end{equation}
We shall use this expression later as a starting point for an averaging and an approximation scheme.

\subsection{Initial Conditions}
\label{sec:3.3}

Supported by observations of the temperature fluctuations in the cosmic microwave background, we can assume that the density fluctuations in the early universe can be characterised as a Gaussian random field (for recent empirical support of this assumption, see \cite{2016A&A...594A..17P}). By the Helmholtz theorem, the peculiar velocity field can be decomposed into a curl and a gradient. By the cosmic expansion and angular-momentum conservation, the curl component will quickly decay, so we can model the initial peculiar velocity as the gradient of a velocity potential $\Psi$. Then, by the continuity equation, the density contrast $\delta$ has to satisfy the Poisson equation $\delta = -\nabla^2\Psi$. Thus, given the velocity potential, both density contrast and peculiar velocity will be determined. The velocity potential also has to be a Gaussian random field. If the density contrast is statistically characterised by its power spectrum $P_\delta(k)$, the velocity potential has the power spectrum
\begin{equation}
  P_\Psi(k) = k^{-4}P_\delta(k)
\label{eq:66}
\end{equation}
according to the Poisson equation. With the peculiar velocity and the density contrast both being Gaussian random fields derived from the velocity potential, either of the power spectra $P_\Psi(k)$ or $P_\delta(k)$ completely determines their statistical properties. Note that we assume irrotationality of the velocity field and thus absence of vorticity at the initial time \emph{only}. During the subsequent nonlinear evolution of the particles vorticity can and will nevertheless be created.

To specify the generating functional of KFT, we need the probability distribution $P(\vc x^\mathrm{(i)})$ for the initial phase-space coordinates of the point particles of our ensemble.

Drawing particle positions randomly by a Poisson process with a probability proportional to the density contrast $\delta = -\nabla^2\Psi$, and assigning momenta to these particles proportional to $\nabla\Psi$, the probability distribution $P(\vc x^\mathrm{(i)}) = P(\vc q, \vc p)$ turns out to be
\begin{equation}
  P(\vc q, \vc p) = \frac{V^{-N}}{\sqrt{(2\pi)^{3N}\det C_{pp}}}\,
  \mathcal{C}(\vc p)\,
  \exp\left(-\frac{1}{2}\vc p^\top C_{pp}^{-1}\vc p\right)\;,
\label{eq:67}
\end{equation}
where $C_{pp}$ is the momentum-correlation matrix depending on the particle positions \cite{2016NJPh...18d3020B}. For late-time cosmological applications using the unbound Zel'dovich propagator (\ref{eq:57}) or its improvement (\ref{eq:59}), the factor $\mathcal{C}(\vc p)$, which is a polynomial in the momenta, can safely be set to unity. We shall see in Sect.\ \ref{sec:9} how $\mathcal{C}(\vc p)$ can be fully taken into account.

The momentum-correlation matrix is given by
\begin{equation}
  C_{pp} = \frac{\sigma_1^2}{3}\mathcal{I}_3\otimes\mathcal{I}_N+
  C_{p_jp_k}\otimes E_{jk}\;,
\label{eq:68}
\end{equation}
where $j\ne k$ and $E_{jk} = e_j\otimes e_k$. The variance $\sigma_1^2$ is defined by
\begin{equation}
  \sigma_1^2 = \int_k\,k^2\,P_\Psi(k) =
  \int_k\,\frac{P_\delta(k)}{k^2}\;.
\label{eq:69}
\end{equation}
The first term on the right-hand side of (\ref{eq:68}) is the momentum dispersion caused by the momenta being drawn from a Gaussian random field. The second term on the right-hand side is the correlation between the momenta of different particles, separated by $q_{jk} := |q_j-q_k|$. Since we assume that the initial density fluctuations are a statistically isotropic and homogeneous random field, the momentum correlations can only depend on the absolute value of the relative particle separation.

\section{Further Developments}
\label{sec:4}

After these preparations, we return to some further developments simplifying the evaluation of the free generating functional (\ref{eq:41}) and the interaction term (\ref{eq:31}).

\subsection{Factorisation of the Free Generating Functional}
\label{sec:4.1}

With the Gaussian initial distribution (\ref{eq:67}) of particle positions and momenta, the momenta can immediately be integrated out in the free generating functional (\ref{eq:41}), leading to
\begin{equation}
  Z_0[\vc L] = V^{-N}\int\D\vc q\,
  \exp\left(-\frac{1}{2}\vc L_p^\top C_{pp}\vc L_p\right)\,
  \E^{\I\langle\vc L_q,\vc q\rangle}\;.
\label{eq:70}
\end{equation}
With the momentum-correlation matrix depending only on the relative particle separations, this remaining expression can be fully factorised. In the simplest case of a two-point function, leaving out an irrelevant delta distribution, the result is
\begin{equation}
  Z_0[\vc L] = (2\pi)^3\delta_\mathrm{D}\left(k_1+k_2\right)\,V^{-2}\,
  \E^{-Q_0}\,\mathcal{P}(k_1,t)\;,
\label{eq:71}
\end{equation}
where the non-linearly evolved power spectrum $\mathcal{P}(k_1,t)$ is given by
\begin{equation}
  \mathcal{P}(k_1,t) := \int_q\,\left(\E^{Q}-1\right)\,\E^{\I k_1\cdot q}
\label{eq:72}
\end{equation}
with the expressions
\begin{equation}
  Q_0 := \frac{\sigma_1^2}{3}\,g_{qp}^2(t)\,k_1^2\;,\quad
  Q := -g_{qp}^2(t)k_1^2a_\parallel(q)\;.
\label{eq:73}
\end{equation}
appearing in the exponentials. The quadratic form $Q_0$ independent of $q$ appearing in the exponential damping term derives from the momentum dispersion.

The function $a_\parallel(q)$ appearing in this expression is the correlation function of those momentum components parallel to the line connecting the positions of the two momenta being correlated. In terms of the power spectrum $P_\delta(k)$ of the density fluctuations, it is given by
\begin{equation}
  a_\parallel(q) = a_1(q)+\mu^2a_2(q)
\label{eq:74}
\end{equation}
with
\begin{align}
  a_1(q) &= -\frac{1}{2\pi^2}\int_0^\infty\D k\,
  P_\delta(k)\,\frac{j_1(kq)}{kq}\;, \nonumber\\
  a_2(q) &= \frac{1}{2\pi^2}\int_0^\infty\D k\,P_\delta(k)\,j_2(kq)\;.
\label{eq:75}
\end{align}
Here, the spherical Bessel functions $j_{1,2}(kq)$ appear, and $\mu$ is the cosine of the angle enclosed by $k_1$ and $q$. Notice that, since $j_1(x)\approx x/3$ and $j_2(x)\approx x^2/5$ for $x\ll1$, the functions $a_{1,2}(q)$ and $a_\parallel(q)$ have the limits
\begin{equation}
  a_1(q)\to-\frac{\sigma_1^2}{3}\;,\quad a_2(q)\to0\;,\quad
  a_\parallel(q)\to-\frac{\sigma_1^2}{3}
\label{eq:76}
\end{equation}
for $q\to0$, and thus $Q\to Q_0$ in the same limit.

If the exponential in (\ref{eq:72}) can be approximated by its Taylor expansion to first order, the power spectrum $\mathcal{P}$ evolves linearly,
\begin{equation}
  \mathcal{P}(k_1,t) \approx g_{qp}^2(t)\,P_\delta(k_1)\;.
\label{eq:77}
\end{equation}

\subsection{Averaged Interaction Term}
\label{sec:4.2}

We now return to the interaction term (\ref{eq:31}) and evaluate it for a two-point function at equal times, $t_1=t_2=t$, taking the constraint $k_1+k_2 = 0$ into account which is enforced by the delta distribution in (\ref{eq:71}). We then have
\begin{equation}
  S_\mathrm{I}(t) = -k_1\cdot\int_0^t\D t'\,g_{qp}(t,t')\,\left(
    f_1(t')-f_2(t')
  \right)\;,
\label{eq:78}
\end{equation}
where $f_i$ is the force (\ref{eq:62}) acting on particle $i$. Dropping the time argument for brevity, we bring the force terms into the form
\begin{equation}
  f_1-f_2 = 2f_{12}+\sum_{j=3}^N\left(f_{1j}-f_{2j}\right)\;,
\label{eq:79}
\end{equation}
where $f_{ij}$ is the force on particle $i$ due to particle $j$, and we have used $f_{12}-f_{21} = 2f_{12}$ by Newton's third law. If we can neglect three-point correlations for now, the second term on the right-hand side of (\ref{eq:79}) can be neglected in an isotropic random field since the forces exerted by particles $j$ with $j\ge3$ on particles $1$ and $2$ will vanish on average because there is no preferred direction they could point to. We can then simplify the interaction term to
\begin{equation}
  S_\mathrm{I}(t) = -2k_1\cdot\int_0^t\D t'\,g_{qp}(t,t')\,f_{12}(t')
\label{eq:80}
\end{equation}
containing the projection of the force $f_{12}$ between particles $1$ and $2$ on the wave vector $k_1$.

The Fourier transform of the potential $v$ of a unit point mass satisfying the Poisson equation (\ref{eq:51}) is
\begin{equation}
  \tilde v(t) = -\frac{A(t)}{\bar{\rho} \, k^2}\;,\quad
  A(t) = \frac{3a}{2g^2}\;,
\label{eq:82}
\end{equation}
thus the gradient of the potential of particle $2$ at the position of particle $1$ is
\begin{equation}
  \nabla_1\varphi_2(t) = -\I \frac{A(t)}{\bar{\rho}}\int_k
  \frac{k}{k^2}\,\E^{\I k(q_1(t)-q_2(t))}\;.
\label{eq:83}
\end{equation}

A seemingly radical approximation of the interaction term (\ref{eq:80}) consists in replacing the potential gradient $\nabla_1\varphi_2$ between the particles $1$ and $2$ by a suitable average. Simply averaging $\nabla_1\varphi_2$ over all particle pairs would return zero because of the statistical isotropy of the particle distribution. It is thus important to realise that the interaction term (\ref{eq:80}) contains the projection of the potential gradient on the wave vector $k_1$ of the density mode considered. We thus need to calculate the projection $k_1\cdot\nabla_1\varphi_2$ in a suitable average which does not vanish for particles correlated with the density mode $k_1$.

The probability $P_{12}$ for finding particles $1$ and $2$ at a separation $q_{12} := |q_1-q_2|$ has a spatially independent mean contribution $\bar\rho$ and a conditional contribution due to the particle correlations, expressed by the spatial correlation function $\xi(q_{12})$ of the particles,
\begin{equation}
  P_{12} = \bar\rho\left(1+\xi_{12}(q_{12})\right)\;.
\label{eq:84}
\end{equation}
For uncorrelated particles in a homogeneous random field, the direction of $\nabla_1\varphi_2$ is random with respect to $k_1$, thus its contribution to the average must vanish, while the correlated part remains. We thus weigh the potential gradient $\nabla_1\varphi_2$ with the conditional probability $\xi(q_{12})$ and take the Fourier transform of the result to obtain the average Fourier component of the potential gradient at wave number $k_1$,
\begin{equation}
  \langle\nabla_1\varphi_2\rangle(k_1) =
  \int_{q_{12}}\,\xi(q_{12})\,\nabla_1\varphi_2\,
  \E^{-\I k_1\cdot\left(q_1-q_2\right)}\;.
\label{eq:85}
\end{equation}
By means of the Fourier convolution theorem, we can then write the Fourier transform of the \emph{averaged} potential gradient as a convolution of the Fourier transforms of the potential gradient itself with that of the correlation function, i.e.\ with the power spectrum. For the latter, we take the linearly time-evolved initial density-fluctuation power spectrum $P_\delta(k)$, damped on the free-streaming scale. Thus, we insert
\begin{equation}
  D_+^2\,\bar P_\delta(k) := D_+^2\,\E^{-\sigma_1^2g_{qp}^2k^2/3}P_\delta(k)
\label{eq:88}
\end{equation}
for the Fourier transform of the correlation function and obtain
\begin{equation}
  \left\langle\nabla_1\varphi_2\right\rangle(k_1,t) =
  -\I A(t)D_+^2
  \int_\kappa\frac{k_1-\kappa}{(k_1-\kappa)^2}\,
  \bar P_\delta(\kappa)\;.
\label{eq:87}
\end{equation}

The angular integral remaining in (\ref{eq:87}) can be carried out and results in
\begin{equation}
  \int_\kappa\frac{k_1-\kappa}{(k_1-\kappa)^2}\,
  \bar P_\delta(\kappa) = k_1^2\hat k_1\int_0^1\frac{y^2\D y}{(2\pi)^2}\,
  \bar P_\delta(k_1y)\,J(y)\;,
\label{eq:89}
\end{equation}
where $\hat k_1$ is the unit vector in direction of $k_1$, and $y := \kappa/k_1$. We truncate the integration at $y = 1$ ($\kappa = k_1$) to suppress modes on scales smaller than the density fluctuation considered. The function $J(y)$ is
\begin{equation}
  J(y) := \int_{-1}^1\D\mu\frac{1-\mu y}{1+y^2+2\mu y} =
  1+\frac{1-y^2}{2y}\ln\frac{1+y}{|1-y|}\;.
\label{eq:90}
\end{equation}
Inserting the averaged potential gradient (\ref{eq:87}) into the force term (\ref{eq:62}), we obtain a scale-dependent, mean force term $\langle f_{12}\rangle(k_1,t)$ weighed by the correlation function between correlated particle pairs, which defines the mean interaction term via (\ref{eq:80}),
\begin{equation}
  \left\langle S_\mathrm{I}(t)\right\rangle =
  -2k_1\cdot\int_0^t\D t'\,g_{qp}(t,t')\,
  \left\langle f_{12}\right\rangle(k_1,t')\;,
\label{eq:91}
\end{equation}
which by construction does not depend on the particle positions or momenta any more.

\begin{figure}[ht]
  \includegraphics[width=\hsize]{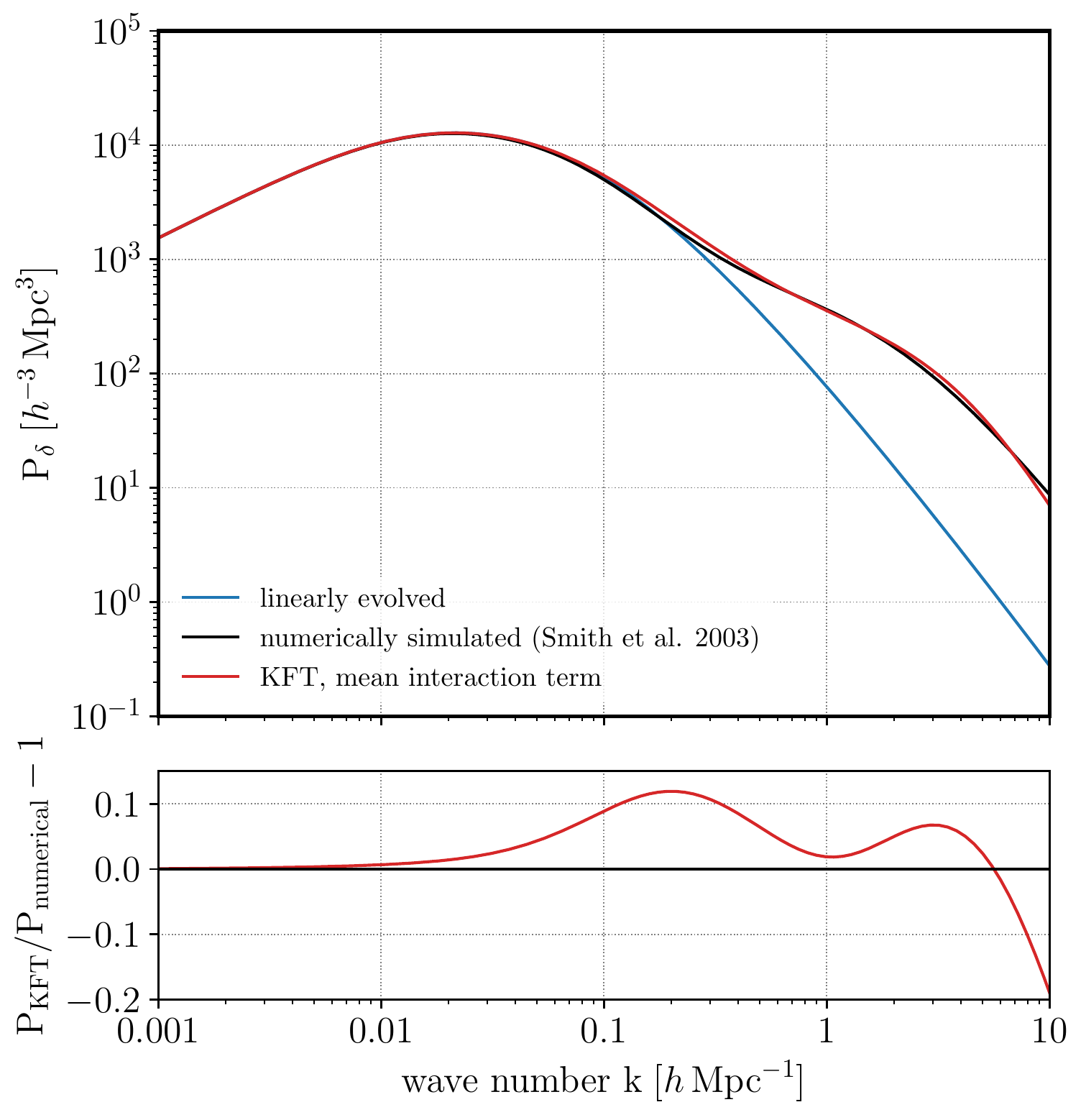}
\caption{Upper panel: The analytic result (\ref{eq:92}), assuming an initial power spectrum as provided by \cite{1986ApJ...304...15B} at $z_\mathrm{i} = 1100$, is shown here for the present epoch (red curve) and compared to the linearly evolved power spectrum (blue curve) and the non-linear power spectrum obtained from numerical simulations\cite{2003MNRAS.341.1311S} (black curve). Lower panel: The relative deviation of the KFT result from the numerical simulation result of \cite{2003MNRAS.341.1311S} is shown. The agreement up to wave number $k\approx10\,h\,\mathrm{Mpc}^{-1}$ is very good.}
\label{fig:4}
\end{figure}

Replacing $S_\mathrm{I}$ by this average $\langle S_\mathrm{I}\rangle$, we can thus pull the interaction term in front of the integral (\ref{eq:29}) and use the form (\ref{eq:71}) of the free generating functional to obtain the closed expression
\begin{equation}
  \bar{\mathcal{P}}(k,t) = \E^{-Q_0+\I\langle S_\mathrm{I}\rangle}\mathcal{P}(k,t)
\label{eq:92}
\end{equation}
for the non-linearly evolved, density-fluctuation power spectrum. Note that this expression is non-perturbative and parameter-free. Comparing the perturbative first-order result shown in Figure \ref{fig:3} with the power spectrum obtained from (\ref{eq:92}) shown in Figure \ref{fig:4} clearly demonstrates the significant improvement.

The non-linear power spectrum in the mean-field approximation shown in this Figure falls below the numerically simulated spectrum for $k\gtrsim10\,h\,\mathrm{Mpc}^{-1}$. This does not indicate a conceptual breakdown of KFT, but rather a limitation of the mean-field approximation of the force term. The scale of $\approx10\,h\,\mathrm{Mpc}^{-1}$ at $z = 0$ is thus not of fundamental importance, but depends on the normalisation of the power spectrum.

\subsection{Interaction in the Born Approximation}
\label{sec:4.3}

Let us now return to the force term (\ref{eq:58}) with the potential gradient expressed by its Fourier transform (\ref{eq:83}). If we wish to avoid averaging over particle positions, as we did before, the essential difficulty is that the actual particle positions $q_{1,2}(t)$ appear in the Fourier phase. This difficulty can be by-passed by approximating the particle trajectories by their unperturbed or inertial trajectories,
\begin{equation}
  q_j(t) \to q_j+g_{qp}(t)p_j\;,\quad j = 1,2\;,
\label{eq:93}
\end{equation}
where $q_j$ and $p_j$ without a time argument are meant to indicate initial particle positions and momenta. We abbreviate $q_{ij} = q_i-q_j$, $p_{ij}=p_i-p_j$ and introduce $K = g_{qp}(t)k$. Exchanging the order of integration over $k$ and $t$, we find
\begin{align}
  f_{ij}(t) &= \frac{\I}{\bar\rho}\int_k\frac{k}{k^2}\E^{\I k\cdot q_{ij}}
  \left[
    A\E^{\I K\cdot p_{ij}}+\frac{\dot gh}{g^2}\int_0^t\D t'\,
    gA\E^{\I K\cdot p_{ij}}
  \right]
\label{eq:94}
\end{align}

\begin{figure*}[ht]
  \includegraphics[width=0.48\hsize]{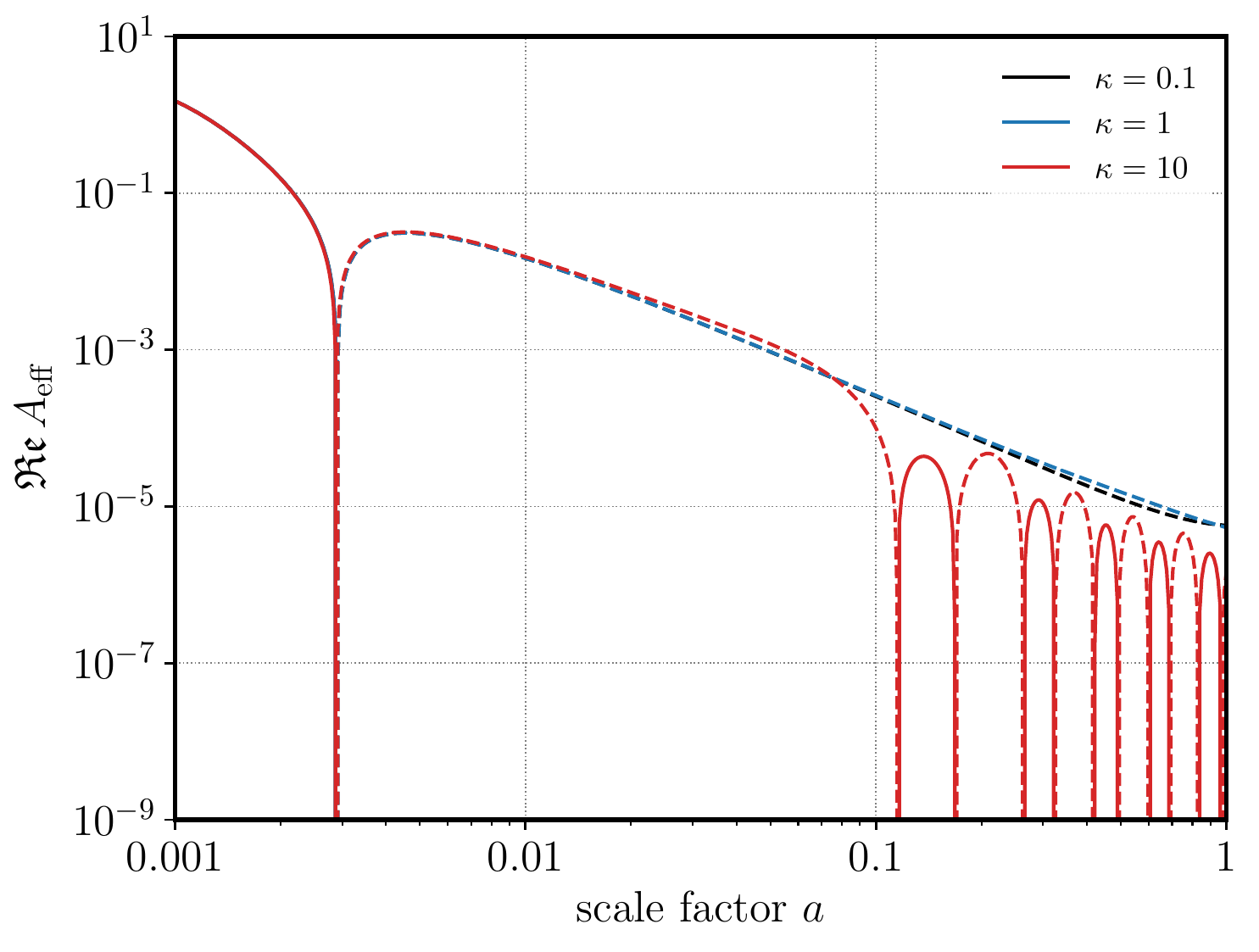}\hfill
  \includegraphics[width=0.48\hsize]{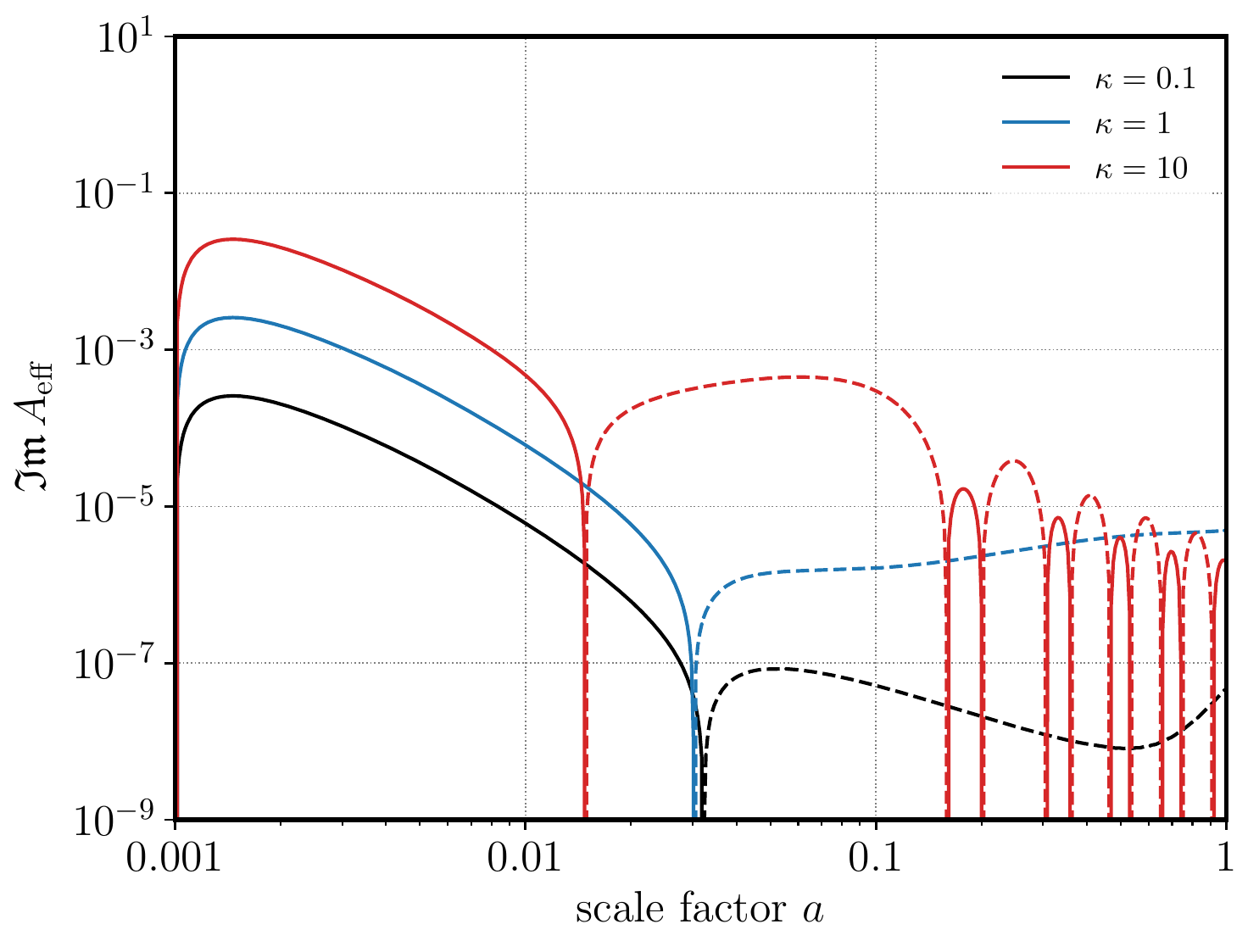}
\caption{Real (left) and imaginary (right) parts of the effective amplitude $A_\mathrm{eff}(\kappa, t)$ of the interaction potential defined in (\ref{eq:94b}). The curves are shown in dependence of the scale factor $a$ and are parameterised by the scalar product $\kappa := k\cdot p_{ij}$ between wave vector $k$ and initial momentum difference $p_{ij}$. The momentum difference is scaled by $\sigma_1$ here. Dashed parts of the curves indicate negative values.}
\label{fig:5}
\end{figure*}

\begin{figure}[ht]
  \includegraphics[width=\hsize]{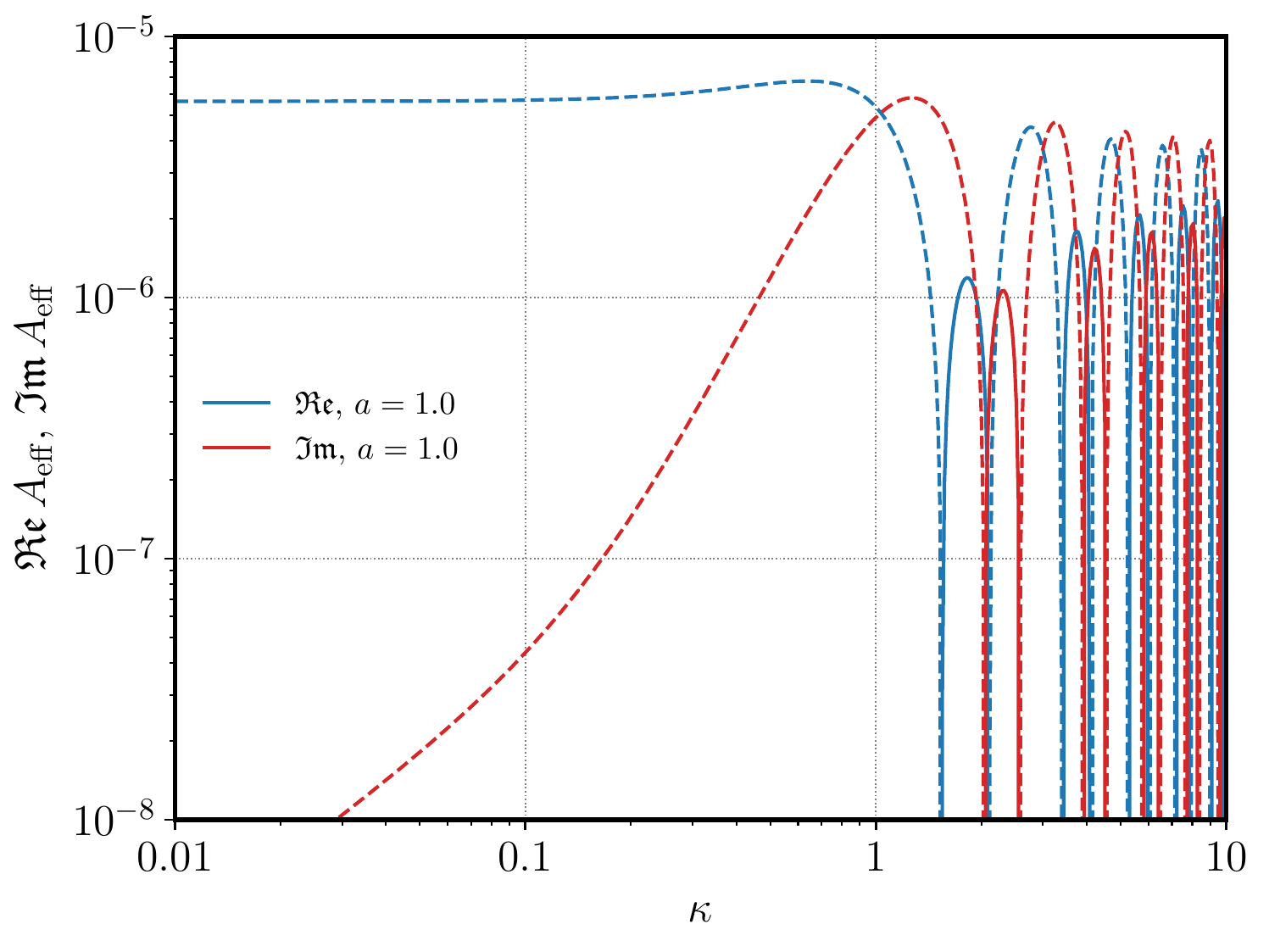}
\caption{Real and imaginary parts of the effective potential $V_\mathrm{eff}(\kappa, t)$ for $a = 1$ in dependence of $\kappa$. As in Fig.~\ref{fig:5}, dashed sections indicate negative values.}
\label{fig:6}
\end{figure}

The second term in brackets in (\ref{eq:94}),
\begin{equation}
  B(\kappa, t) := \frac{\dot gh}{g^2}\int_0^t\D t'\,gA\E^{\I K\cdot p_{ij}}
\label{eq:94a}
\end{equation}
with $\kappa := k\cdot p_{ij}$, is easily evaluated by numerically integrating its real and imaginary parts. The potential of the force term $f_{ij}$ from (\ref{eq:94}) thus has the effective, $k$-dependent amplitude
\begin{equation}
  A_\mathrm{eff}(\kappa, t) := A\E^{\I K\cdot p_{ij}}+B(\kappa, t)\;,
\label{eq:94b}
\end{equation}
whose real and imaginary parts we show in Figs.~\ref{fig:5} and \ref{fig:6}. Figure \ref{fig:5} shows the dependence on the scale factor for three different choices of $\kappa$, while Fig.~\ref{fig:6} displays the dependence on $\kappa$ for $a = 1$.

These Figures show several interesting properties of the effective potential amplitude. First, for large values of $\kappa$, i.e.\ for small scales or large initial momenta, the real part of $V_\mathrm{eff}$ begins oscillating at late times. This indicates that the displacement of the particles becomes comparable or larger than the structures they belong to, thus reducing the effective potential amplitude at late times: structures on such scales are then not built up any more, but particles oscillate within them. Alternatively, these oscillations can be interpreted as the effect of an oscillating effective propagator $g_{qp}$, which mimicks the motion of particles in gravitationally-bound structures.

Second, the imaginary part of the potential amplitude is typically small at early times and for small values of $\kappa$, but becomes large at late times for large $\kappa$. This indicates damping by the transport of particles and the associated mixing of phases of particle trajectories. This damping justifies the exponential cut-off in the power spectrum (\ref{eq:88}) which we have used for calculating the mean force term before.

It remains to be seen in comparison with numerical simulations how useful this variant of the Born approximation will be for cosmological structure formation. The agreement between numerical simulations and the non-linear power spectrum (\ref{eq:92}) including the averaged interaction, however, suggests that the Born-approximated interaction term may return similarly or even more accurate results.

\section{Halo profiles in KFT}
\label{sec:5}

The formation of large-scale, dark-matter structures in our KFT formulation as well as in numerical simulations is governed by a few physical properties only: the interaction potential between dark matter particles, the equations of motion governing particle trajectories, the underlying cosmological background model, i.e.\ the expansion history of the model universe, and the initial density and momentum correlations set, as is most commonly believed, by quantum fluctuations during inflation.

It would appear that in numerically simulated as well as in observed, gravitationally-bound dark-matter structures, these physical properties or their interplay lead to a particular shape of the radial density profile \cite{1996ApJ...462..563N, 2012ApJ...757...22C, 2013SSRv..177....3B} that is valid for halos on scales ranging from dwarf galaxies to massive galaxy clusters. Since the KFT formalism allows to easily change all the ingredients for large-scale structure formation, it is a natural playground for investigating the origin of the dark-matter, halo-density profile.

We found in \cite{2017NJPh...19h3001B} that taking into account initial momentum correlations leads to a characteristic deformation of the non-linear power spectrum on scales of the order of $k\approx0.3\,h\,\mathrm{Mpc}^{-1}$ but not at wave numbers large enough for dark-matter halos to appear. Furthermore, we have shown that the full initial density correlations play a role for the shape of the density-fluctuation power spectrum at early times, but are washed out at late times when $a\to1$, and therefore are unlikely responsible for the shape of density profiles of highly non-linear objects \cite{2018JSMTE..04.3214F}.

In fact, it is the Newtonian gravitational potential that is often charged with being responsible for the shape and universality property of dark matter halos. We have therefore used the KFT formalism to investigate whether the shape of the interaction potential is responsible for the shape of the non-linear power spectrum on small scales, $k\ge1\,h\,\mathrm{Mpc}^{-1}$, where contributions to the non-linear power spectrum from inner structures of dark-matter halos begin to dominate.

According to the widely used halo model, the non-linear power spectrum is a convolution of Fourier-trans\-formed Navarro-Frenk-White (NFW) profiles \cite{1996ApJ...462..563N}, weigh\-ed with the mass function (for an extensive review see \cite{2002PhR...372....1C}). Since the non-linear power spectrum is reproduced with KFT at least up to $k\approx10\,h\,\mathrm{Mpc}^{-1}$, un-weighing with the mass function leads to the density profile of dark-matter halos.

The non-linear contributions to the power spectrum obtained from KFT approximately correspond to the one-halo term of the halo model. Thus, leaving the relative abundances of haloes with different mass unchanged, we can study with KFT how different choices for the gravitational potential affect the density-profile shape.

We point out that, while the halo model must be provided with the explicit form of the halo density profile, the KFT approach does not need any input of this kind. The non-linear density power spectrum and the halo density profile that is produced in the KFT approach are thus the result of particle dynamics including particle interactions and correlations.

To account for the fact that a general power-law potential leads to a linear growth factor that depends on the wave number $k$, the Poisson equation has to be modified accordingly. This is achieved through an algebraic, ad-hoc modification of Poisson's equation in Fourier space,
\begin{equation}
  -f(k)\,\tilde\Phi = 4\pi G\bar\rho a^2\tilde\delta\;,
\label{eq:112}
\end{equation}
with $\tilde\Phi$ the gravitational potential, $G$ the gravitational constant and $\bar\rho$ the mean matter density. The modifying function $f(k)$ is the inverse of the Fourier-transformed particle interaction potential and is, in that sense, fixed. For the Newtonian gravitational interaction, it is simply $f(k) = k^2$.

\begin{figure}[t]
  \includegraphics[width=\hsize]{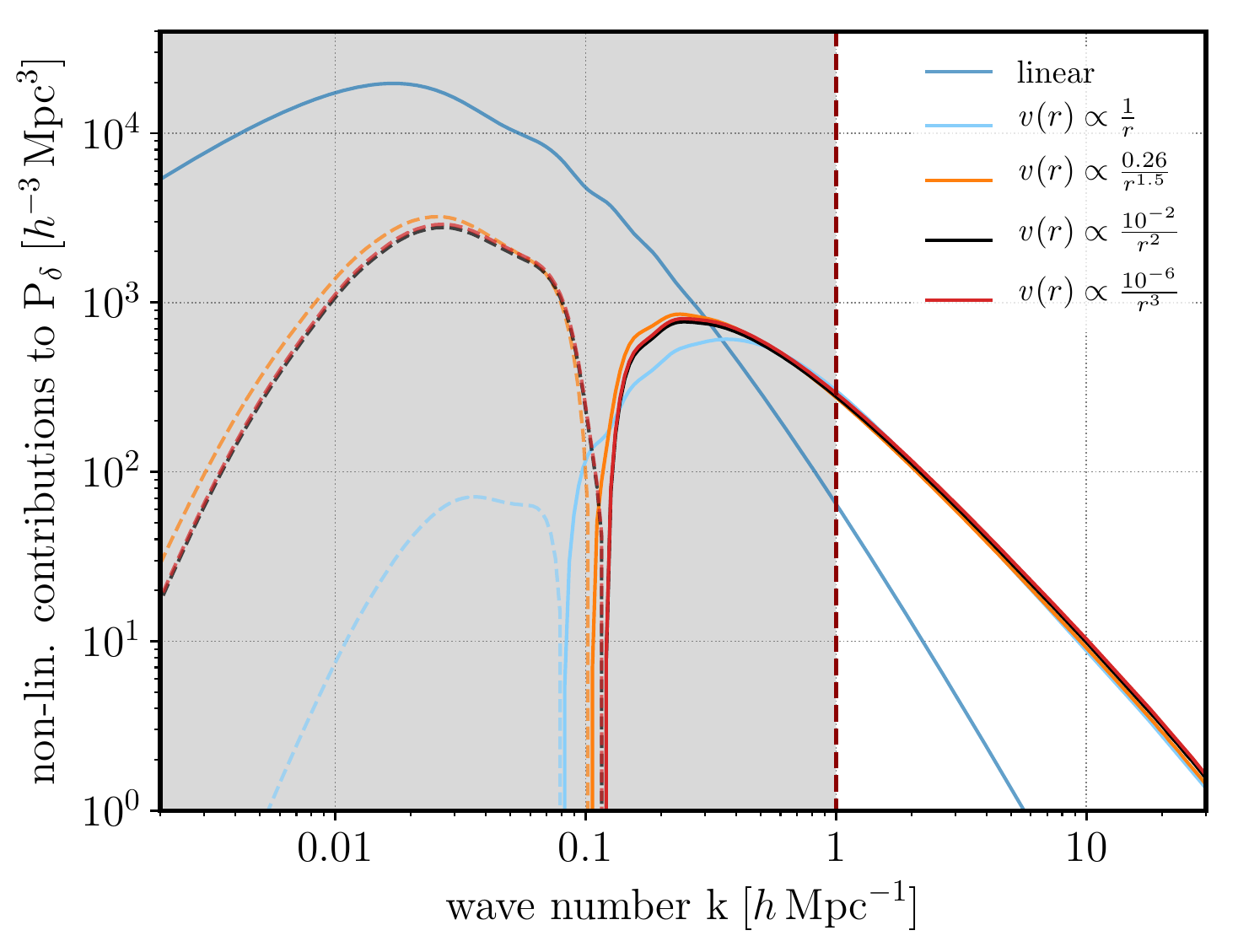}
\caption{The non-linear contributions to the power spectrum today ($z=0$) are shown together with the linear power spectrum for Newtonian gravity which serves here as a reference only, using an initial power spectrum as obtained with CAMB \cite{2000ApJ...538..473L}. The power spectrum in the gray-shaded area, to a large degree, does not depend on the inner correlations on dark matter halos and is therefore neglected in our analysis. Starting from $k=1\,h\,\mathrm{Mpc^{-1}}$, it is assumed that inner halo structures begin to affect the shape of the non-linear power spectrum. On scales $k>1\,h\,\mathrm{Mpc^{-1}}$, there is very good agreement of the non-linear corrections independent of the shape of the interaction potential.}
\label{fig:7}
\end{figure}

A first analysis of the small-scale behaviour of the non-linear contributions to the power spectrum under different gravitational laws using the perturbative approach described in Sect.\ \ref{sec:2.3} yields the results shown in Fig.\ \ref{fig:7} for interaction potentials of the form
\begin{equation}
  v(r) = A\left(r^2 + \varepsilon^2\right)^{-n/2}\;,
\label{eq:113}
\end{equation}
where the parameter $\varepsilon$ is introduced for technical reasons in order to perform the Fourier transforms of the interaction potential for all $n$. It can be interpreted as a smoothing scale. We set $\varepsilon = 10^{-4}$, which corresponds to scales much smaller than the scales we are currently interested in. We have verified that a small enough $\varepsilon$ does not influence the power spectrum on the much larger scales that we are considering.

Particle interactions are included up to first order in the interaction operator (\ref{eq:39}) in the same way as for the results shown in Fig.\ \ref{fig:3} since the agreement with non-linear power spectra from Cosmic Emulator \cite{2014ApJ...780..111H, 2010ApJ...715..104H, 2010ApJ...713.1322L, 2009ApJ...705..156H} that uses state-of-the-art $N$-body simulations is already very good.

The initial correlations appear in a quadratic form in an exponential in the KFT formalism (cf.\ \ref{eq:67}) and must therefore be Taylor-expanded before the integration over initial particle positions and momenta can be performed in (\ref{eq:40}). Initial correlations are included up to second order here. Including the full hierarchy of initial momentum correlations, as shown in Sect.\ \ref{sec:4.1}, proves to be much more cumbersome in the perturbative approach compared to the non-perturbative schemes of Sects.\ \ref{sec:3.2} and \ref{sec:3.3}. Including the full hierarchy of initial momentum correlations in this context would not alter the results shown here, as they only affect the shape of the power spectrum on intermediate scales ($k\approx0.3\,h\,\mathrm{Mpc^{-1}}$) as shown in \cite{2017NJPh...19h3001B}.

The amplitudes of the interaction potential (\ref{eq:113}) and of the initial density-fluctuation power spectrum are chosen such that the amplitudes of the non-linear contributions to the power spectrum for any potential match that of the Newtonian case, and that the amplitude of linear growth on large scales, i.e.\ $k\ll 1\,h\,\mathrm{Mpc}^{-1}$, also matches the one observed today.

The linear power spectrum can be obtained, as usual, by multiplying the initial power spectrum $P^\mathrm{(i)}_\delta$ with the appropriate linear growth factor corresponding to the interaction potential being used,
\begin{equation}
  P_\delta (k) = D^2_+(a, k) P^\mathrm{(i)}_\delta(k)\;.
\label{eq:114}
\end{equation}
It is clear from this expression that the linear growth for non-Newtonian potentials will be different from the Newtonian case due to the wave-number dependence of the growth factor. Therefore, the linear growth together with the non-linear growth for non-Newtonian gravity will produce power spectra that look quite different from the Newtonian case. However, this is irrelevant for this analysis, because only those contributions to the density-fluctuation power spectrum which are due to the inner structure, i.e.\ density profile, of dark matter halos are of interest here.

Switching to the picture of the halo model, the argument becomes clearer: While the linear power spectrum, corresponding to the two-halo term, describes the correlations between two different dark matter halos, the one-halo term describes the contributions from the inner structures of halos, which are highly non-linear. Therefore, also in KFT the information on the density profile of dark matter halos will be contained exclusively in the non-linear contributions.

Figure \ref{fig:7} clearly shows that, for smaller scales, i.e.\ $k>1\,h\,\mathrm{Mpc}^{-1}$, the non-linear contributions are almost insensitive to the slope of the interaction potential. In the gray-shaded area in Fig.\ \ref{fig:7}, there is some sensitivity to the potential slope, however these scales are too large to be relevant for the density profiles of even the largest dark matter halos.

Since the slopes of the non-linear contributions remain the same even for strongly varying interactions laws, it appears that different choices for the gravitational potential would not affect the density-profile shape of dark matter halos.

This result is surprising and should still be perceived with some caution. Although the non-linear power spectrum for Newtonian gravity obtained with the first-order perturbation term in KFT agrees well with predictions from numerical simulations as shown in Fig.\ \ref{fig:3}, it is still uncertain whether the perturbation series converges regardless of the interaction potential. If the perturbation series does converge, for which there are good reasons, then higher-order terms should yield ever smaller corrections to the non-linear power spectrum.

However, for short-ranged interaction potentials, most of the structure is expected to be accumulated on very small scales, and higher-order interaction terms should become more dominant. Hence, the perturbative ansatz is expected to perform worse and may eventually break down. To test whether this will indeed be the case, higher-order terms and a check of the convergence of the perturbation series are needed.

We emphasise that the above results were interpreted assuming that the relative abundances of haloes with different mass remains unchanged. This assumption was made because we also assumed that only non-linear contributions corresponding to the one-halo term were considered. In actuality, this does not need to be the case. We could and should as well have contributions from the two-halo term in the non-linear part of the power spectrum. This poses a considerable problem for the analysis using the perturbative approach as well as for the same analysis employing the non-perturbative approaches presented in Sects.\ \ref{sec:3.2} and \ref{sec:3.3}. To consider the fully non-linear density-fluctuation power spectrum which corresponds to the one-halo and two-halo terms combined, the mass function in the halo model would have had to be adjusted accordingly, i.e.\ more low-mass halos and less high-mass halos for short ranged interaction potentials. In short, the halo mass function has to be known for each particle interaction potential that is considered. It is only then possible to make conclusive statements about the halo density profiles.

\section{Momentum-Density Correlations}
\label{sec:6}

\subsection{Operator Expression}
\label{sec:6.1}

Another important and conceptually straightforward application of KFT is the calculation of momentum power spectra \cite{2018cldd.book.....L}. These are hardly accessible with the standard analytic approaches to cosmic structure formation. The momentum field could naively be constructed as
\begin{equation}
  p(t_1) = \sum_{j=1}^Np_j(t_1) = \frac{1}{m}\sum_{j=1}^Nu_j(t_1)\;,
\label{eq:115}
\end{equation}
with the velocities $u_j(t_1)$ of the particles $j$, assuming all particle masses to be equal. This momentum field lacks any spatial dependence, since the microscopic degrees of freedom are the phase-space coordinates of all particles. In order to retain any spatial information, we have to impose that each particle can contribute to the momentum at a position $q$ if and only if it is at this position,
\begin{align}
  \Pi(q,t_1) &= \sum_{j=1}^Np_j(t_1)\delta_\mathrm{D}\left(q-q_j(t_1)\right)
  \nonumber\\ &=
  \sum_{j=1}^Np_j(t_1)\rho_j(q,t)\;.
\label{eq:116}
\end{align}
In the last step, we have identified the Dirac delta distribution with the density of the $j$-th particle at position $q$ and time $t_1$. Thus, the new field $\Pi$ is a momentum density.

Fourier-transforming $\Pi(q,t_1)$ and replacing the phase-space coordinates of particle $j$ by functional derivatives with respect to the corresponding source fields, we find the momentum-density operator
\begin{equation}
  \hat\Pi(1) = \sum_{j=1}^N\hat\Pi_j(1)\;,
\label{eq:117}
\end{equation}
composed of the one-particle operators
\begin{equation}
  \hat\Pi_j(1) := \frac{\delta}{\mathrm{i}\delta J_{p_j}(t_1)}\hat\rho(1)
\label{eq:118}
\end{equation}
with $\hat\rho(1)$ from (\ref{eq:27}). At this point, we note that components of the momentum-density field are available by specifying the operator to
\begin{equation}
  \hat\Pi_j^\alpha(1) := 
  \frac{\delta}{\mathrm{i}\delta J_{p_j}^\alpha(t_1)}\hat\rho(1)\;,
\label{eq:119}
\end{equation}
where $\alpha=(1,2,3)$ enumerates the Cartesian vector components.

The application of $r\ge1$ of the operators (\ref{eq:118}) to the generating functional translates the generator field, as discussed in Sect.\ \ref{sec:2.2}, and pulls down the momentum trajectories from the phase factor in (\ref{eq:6}). Having applied these operators and setting the generator field $\mathbf{J}$ to zero afterwards, we arrive at the expression
\begin{align}
  Z'[\vc L] &=
  \int\D\Gamma\prod_{j=1}^r\left[
    g_{pp}(t_j)p_j-\int_0^{t_j}\D t'\,g_{pp}(t_j,t')\nabla_jV(t')
  \right]\nonumber\\ &\cdot \E^
  {\I\left\langle\vc L_q,\vc q\right\rangle+
   \I\left\langle\vc L_p,\vc p\right\rangle+\I S_\mathrm{I}}\;,
\label{eq:120}
\end{align}
where we have used the definitions (\ref{eq:30}) and (\ref{eq:31}).

In order to facilitate results obtained for density power spectra from Sect.\ \ref{sec:4.1}, we employ the Fourier transform of the potential gradient (\ref{eq:37}) and replace the initial momentum of particle $j$ by a partial derivative with respect to the shift vector,
\begin{equation}
  p_j = -\I\frac{\partial}{\partial L_{p_j}}\;.
\label{eq:121}
\end{equation}
In this way, we can write the momentum operator as
\begin{align}
\label{eq:122}
  \hat p_j(t_j) &= -\mathrm{i}g_{pp}(t_j)
  \frac{\partial}{\partial L_{p_j}} \\ &-
  \int_0^{t_j}\D t'\,g_{pp}(t_j,t')\int_{k_1'}
  \hat B_j(-1')\tilde v(1')\hat\rho_j(1')\;, \nonumber
\end{align}
such that the previous expression may be written as
\begin{equation}
  Z'[\vc L] = \prod_{j=1}^r\left[\hat p_j(t_j)\right]Z[\vc L]\;,
\label{eq:123}
\end{equation}
with $Z[\vc L]$ given by (\ref{eq:40}). Thus, the calculation of $r$-point correlations of the momentum-density differs from the calculation of density correlations only by the application of the product of $\hat p_j$-operators. In this way, the factorisation of the free generating functional can be applied in this context as well.

\subsection{Perturbation Theory and Diagrams}
\label{sec:6.2}

In order to include particle interactions, a perturbative approach is necessary. One such approach begins with the Taylor expansion of the exponential factor in $Z[\vc L]$ in terms of the interaction operator $\hat S_\mathrm{I}$. In this Section, we formulate rules allowing the systematic calculation of all terms in the perturbation series by employing diagrams representing these terms. These rules have been stated for pure density correlations in \cite{2017NJPh...19h3001B}. We extend them here to also include momentum-density correlations. We note that this is effectively a perturbative expansion in the interaction potential, and thus corrections to the momentum trajectories need to be accounted for at the appropriate order. Thus, the linear correction to a momentum-density correlation is given by the acceleration of one particle in (\ref{eq:122}), which is of zero-th order in the Taylor expansion of the exponential, plus the linear term of the perturbative expansion. At the end of this Section, we will give a simple example of the linear correction to a $\left\langle\rho\Pi\right\rangle$ power spectrum.

Formulating the rules below, we assume that an $n$-point correlation function is to be calculated at $m$-th perturbation order. The diagrammatic representation of the terms correcting for forces between the particles can then be constructed following these rules:
\begin{itemize}
  \item[(i)]
  \begin{itemize}
    \item[a.] Attach $s=n+2m$ wave vectors pointing outward (represented by arrows) to a circle marking the free generating functional $Z_0$, where $s$ is the total number of density, momentum-density, and response field operators.
    \item[b.] Of these, distinguish $r$ by dashed arrows, representing the momentum-density operators $\hat\Pi_{j_l}$, from the $(s-r)$ solid arrows representing density fields associated with either $\rho_{j_l}$ or $B_{j_l}$.
  \end{itemize}
  \item[(ii)] The time ordering is counter-clockwise along the circle, with the latest time being at the top. Interactions are represented by two lines attached to $Z_0$ at the same point in time. If the correlator is simultaneous, the external wave vectors are also attached to the same point.
  \item[(iii)] The interaction potential is represented by a circled $v$ connected to a pair of internal wave vectors, which are marked with a prime. If the potential is translation-invariant, the connected internal wave vectors point in opposite directions and have the same magnitude.
  \item[(iv)]
  \begin{itemize}
    \item[a.] Each response field is marked by a circle segment starting at a negative internal wave vector and connecting two different wave vectors. The circle segments always end at a later time.
    \item[b.] Distinguish dashed circle segments, representing a deviation of the actual from the freely evolved particle position and given by the application of $\hat S_\mathrm{I}$, from solid circle segments corresponding to a change of the particle momentum by the second term in (\ref{eq:122}). These lines can only connect a response field (solid arrow) with a momentum-density (dashed arrow).
    \item[c.] At a dashed wave vector, either no or one solid circle segment must end, but arbitrarily many dashed circle segments may end at the same wave vector, internal or external.
  \end{itemize}
  \item[(v)]
  Each diagram is assigned a multiplicity counting the equivalent diagrams.
\end{itemize}

\begin{figure}[ht]
  \includegraphics[width=\hsize]{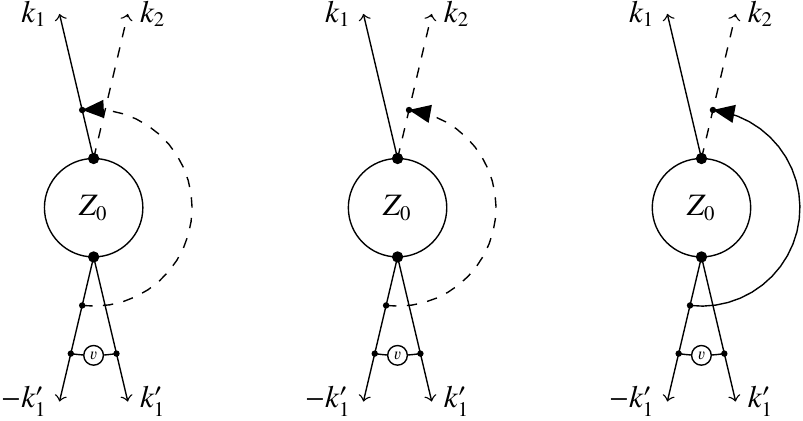}
\caption{At linear order, three terms contribute to the correction of the 2-point cumulant of a density and momentum-density field. The first two come from the Taylor expansion representing a deviation from the final freely-evolved position due to interactions and the last diagram accounts for the deviation from the free momentum trajectory.}
\label{fig:8}
\end{figure}

To illustrate the construction of the diagrams and the physical meaning of the terms, we consider the linear-order correction to an equal-time 2-point cross-correlation function of a density $\rho$ and a momentum-density field $\Pi$. The respective diagrams are shown in Fig.\ \ref{fig:8}.

By rules (i) and (ii), we attach one solid density line and one dashed momentum-density line at the same point to the circle representing the free generating functional for an equal-time correlation. Rule (iii) connects the internal wave vectors by the interaction potential. Following rule (iv), we construct all possible particle identifications. The first two diagrams represent the linear term of the Taylor expansion
\begin{align}
\label{eq:124}
  \mathcal{D}_{S_\mathrm{I}}^\mathrm{(lin)} = \sum_{j=1}^2
  \delta_{j,4}\,k_j\cdot &\int_0^t\D 1'\,
  \tilde v(1')\,g_{qp}(t,t') \\ &\cdot
  k_1'g_{pp}(t) \frac{\partial}{\partial L_{p_2}}\,Z_0[\mathbf{L}]\;.
  \nonumber
\end{align}
Without loss of generality, we here enumerate the particles in clockwise order starting at the top left. This term accounts for the displacement of a particle from its position according to inertial motion at the time of evaluation. The third diagram represents the deviation of the particle's momentum due to the interaction with other particles. The term is
\begin{equation}
  \mathcal{D}_p^\mathrm{(lin)} = \delta_{2,4}\int_0^t\D 1'\,
  \tilde v(1')\,k_1'\,g_{pp}(t,t')\,Z_0[\mathbf{L}]\;.
\label{eq:125}
\end{equation}
The translation tensor is given by (\ref{eq:30}), and the Kronecker symbols identify an internal with an external particle.

At quadratic order in the interaction potential, considerably more diagrams contribute because two interactions of particles occur in the evolution of the system. Three scenarios can be distinguished: (1) one external particle undergoes two interactions, (2) each external particle undergoes one interaction, (3) one external particle scatters from a previously scattered internal particle.

\subsection{Momentum-Density Power Spectra}
\label{sec:6.3}

In a similar fashion as for the density power spectra (\ref{eq:72}) and (\ref{eq:92}), we can write down a closed expression for the momentum-density power spectrum. We focus here on one scalar function that can be constructed from the $\left\langle\Pi\times\Pi\right\rangle$ correlation tensor. More specifically, we consider the projection of $\Pi$ perpendicular to the mode $k$. This power spectrum is needed in order to calculate the amplitude of secondary temperature fluctuations in the cosmic microwave background caused by Thomson scattering off free electrons moving with the bulk of structures. This is called the kinetic Sunyaev-Zel'dovich effect \cite{1969Ap&SS...4..301Z}.

The result is
\begin{equation}
  \left\langle\Pi_\bot\Pi_\bot\right\rangle(k,t) =
  -2g^2_{pp}(t)\,\E^{-Q_0+i\left\langle S_\mathrm{I}(t)\right\rangle}
  \int_qa_\bot(q)\E^{Q-\I k\cdot q}\;,
\label{eq:126}
\end{equation}
with the definitions for $Q_0$, $Q$ and $\left\langle S_I(t)\right\rangle$ of equations (\ref{eq:73}) and (\ref{eq:91}).

The function $a_\bot(q)$ is the correlation function of the initial momentum components perpendicular to the line connecting the positions of the two momenta being correlated. It can also be expressed by the initial power spectrum $P_\delta(k)$ of density fluctuations and is given by
\begin{equation}
  a_\bot(q) = a_1(q)+\frac{1-\mu^2}{2}a_2(q)
\label{eq:127}
\end{equation}
using the definitions for $a_1(q)$ and $a_2(q)$ in (\ref{eq:75}).

Taking the limit of expression (\ref{eq:126}) for large scales or small wave numbers, $k\rightarrow0$, we recover the standard result of \cite{1987ApJ...322..597V} for the slope of the power spectrum,
\begin{equation}
  \lim_{k\to 0}\left\langle\Pi_\bot\Pi_\bot\right\rangle(k,t)\propto k^2.
\end{equation}

\begin{figure}[ht]
  \includegraphics[width=\hsize]{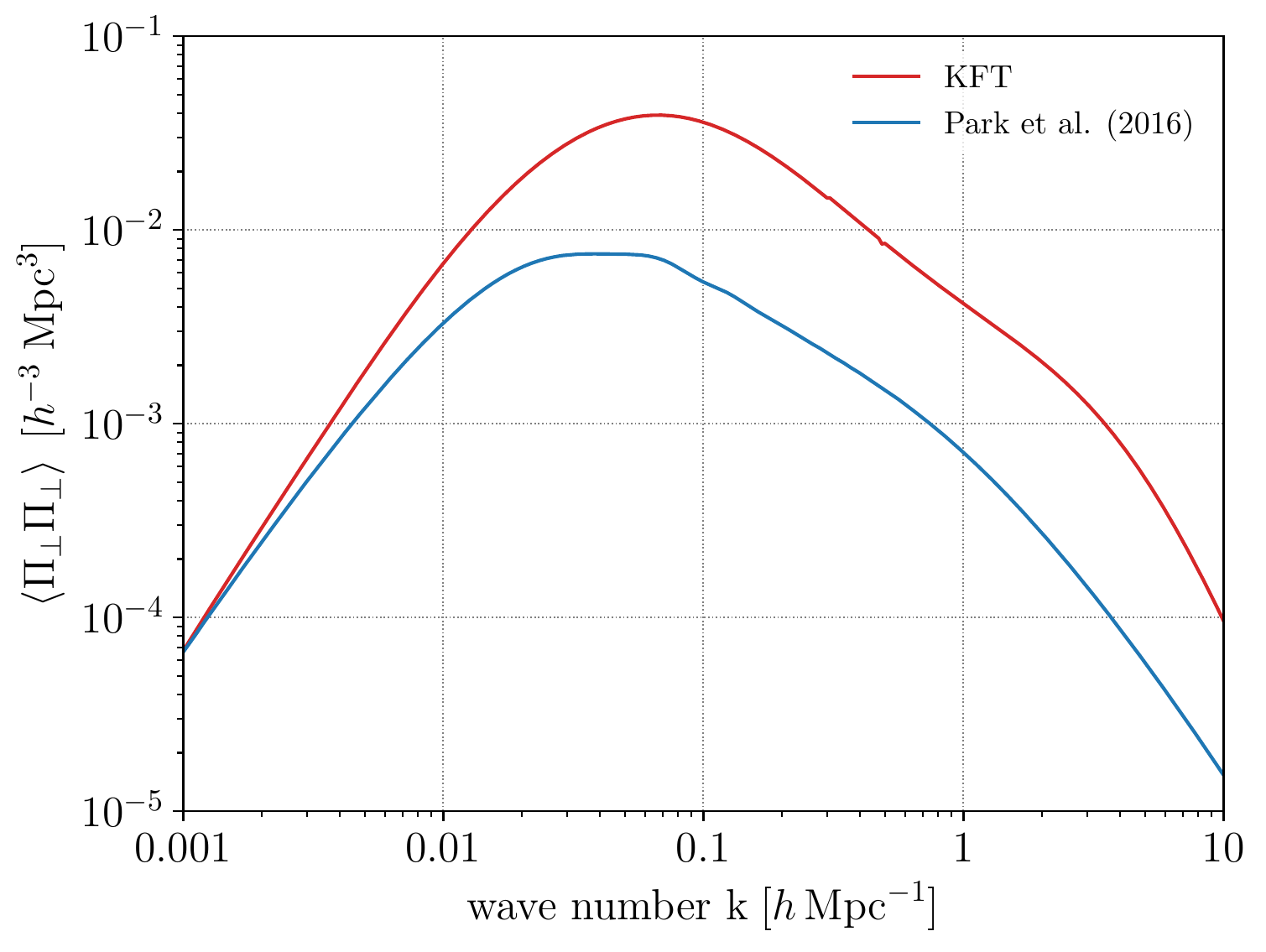}
\caption{Power spectrum for the projection of the momentum-density $\Pi$ perpendicular to the mode $k$ at present time. The red curve shows the KFT result (\ref{eq:126}), the blue curve the results obtained by \cite{2016ApJ...818...37P} using Eulerian standard perturbation theory at one-loop order. The cosmological parameters are chosen to be $\Omega_\mathrm{m} = 0.279, \Omega_\Lambda = 0.721, h = 0.701, \Omega_\mathrm{b} = 0.0462, \sigma_8 = 0.8$, and we use an initial BBKS spectrum \cite{1986ApJ...304...15B} in our calculations.}
\label{fig:9}
\end{figure}

In Fig.\ \ref{fig:9}, we compare our results with those obtained by \cite{2016ApJ...818...37P} who used Eulerian standard perturbation theory. The authors of \cite{2016ApJ...818...37P} calculate the power spectrum of the momentum-density by separating it into unconnected and connected terms using Wick's theorem and evaluating these at one-loop order. This captures the interactions only partially, while the KFT approach predicts a higher amplitude when using the averaged interaction term (\ref{eq:91}). In addition, our results suggest that the spectrum increases proportional to $k^2$ at wave numbers $\lesssim10^{-2}\,h/\mathrm{Mpc}$, while the SPT results fall below this slope approximately at wavenumbers above $10^{-3} \, h/\mathrm{Mpc}$.

\section{The BBGKY hierarchy in KFT}
\label{sec:7}

To relate KFT to conventional kinetic theory, we now want to extract evolution equations for the phase-space density and its higher-order correlators. We will first define a phase-space density operator and generalise the generating functional as well as the interaction term from Sect.\ \ref{sec:2} to cover all of phase space.

In analogy to the number density (\ref{eq:22}) and its operator (\ref{eq:27}), we introduce the phase-space density
\begin{equation}
  f(x, t) = \sum_j\delta_\mathrm{D}\left(x-x_j(t)\right)
\label{eq:128}
\end{equation}
and its operator expression in a six-dimensional Fourier space
\begin{equation}
  \hat f(s, t) = \sum_j\exp\left(
    -\I k\cdot\frac{\delta}{\I\delta J_{q_j}(t)}
    -\I\ell\cdot\frac{\delta}{\I\delta J_{p_j}(t)}
  \right)\;,
\label{eq:129}
\end{equation}
where $x=(q,p)$ is again the six-dimensional phase-space position with its Fourier conjugate $s=(k,\ell)$.

In this Section, we will always consider equal-time, $r$-point phase-space density cumulants with a Taylor expansion of the interaction up to $N$-th order,
\begin{equation}
  \left\langle\tilde f(1)\,\tilde f(2)\dots\tilde f(r)\right\rangle^{(N)} =
  \sum_{n=1}^N\frac{\left(\I\bar S_\mathrm{I}(t)\right)^n}{n!}Z_0[\vc L]\;,
\label{eq:130}
\end{equation}
where $\tilde f(1) = \tilde f(t,s_1)$, $\tilde f(2) = \tilde f(t,s_2)$ and so on, and $Z_0[\vc L]$ denotes the free generating functional with $r$ phase-space density operators having been applied and the source fields set to zero. Similar to (\ref{eq:20}), the free generating functional is then
\begin{equation}
  Z_0[\vc L] = \int\D\Gamma\,\E^
   {\I\left\langle\vc L_q, \vc q\right\rangle +
    \I\left\langle\vc{\bar L}_p, \vc p\right\rangle}\;,
\label{eq:131}
\end{equation}
with the shift operators
\begin{equation}
  \vc L_q  = -\sum_{j=1}^r k_j\otimes e_j \;,\quad
  \bar{\vc L}_p = - \sum_{j=1}^r\left(
    k_j\,g_{qp}(t_j)+\ell_j
  \right)\otimes e_j\;.
\label{eq:132}
\end{equation}

The interaction (\ref{eq:39}) can then be extended to cover all of phase space,
\begin{equation}
  \bar S_\mathrm I(t) = \sum_{j=1}^r\int_0^t\D 1'\left(
    k_jg_{qp}(t,t') + \ell_j
  \right)\hat D_j(-1')v'(1')\hat f(1')\;.
\label{eq:133}
\end{equation}
Here, the potential $\tilde v(1')$ from (\ref{eq:39}) is extended to read
\begin{equation}
  v'(1') = \tilde v(k_1')\,(2 \pi)^3\delta_\mathrm{D}\left(\ell_1'\right)\;,
\label{eq:134}
\end{equation}
the integration measure is $\D 1' = \D t_1'\D^6s'_1/(2\pi)^6$, and the phase-space response-field operator is defined by
\begin{equation}
  \hat D_j(-1') = \I k_1'\hat f_j(-1')\;.
\label{eq:135}
\end{equation}

Before computing the evolution equations of phase-space density cumulants, we need to introduce the phase-space density current
\begin{equation}
  f u( x, t)= \sum_j\delta_\mathrm{D}\left( x -  x_j(t)\right)
  \frac{p_j}{m}
\label{eq:137}
\end{equation}
with its Fourier-space operator
\begin{align}
\label{eq:138}
  &\widehat{fu}(s,t) = \\ &\sum_j\exp\left(
    -\I k\cdot\frac{\delta}{\I\delta J_{q_j}(t)}
    -\I\ell\cdot\frac{\delta}{\I\delta J_{p_j}(t)}
  \right)\frac{1}{m}\frac{\delta}{\I\delta J_{p_j}(t)}\;. \nonumber
\end{align}

Applying a phase-space density-current operator $\widehat{fu}$ together with $r-1$ phase-space density operators $f$ to the free generating functional and setting the source fields to zero results in
\begin{equation}
  Z_0[\vc L] = \int\D\Gamma\E^
   {\I\left\langle\vc L_q, \vc q\right\rangle+
    \I\left\langle\bar{\vc L}_p, \vc p\right\rangle}\,\frac{p_r}{m}\;.
\label{eq:139}
\end{equation}
The time evolution of the phase space correlator (\ref{eq:130}) can now be found by direct computation,
\begin{align}
\label{eq:140}
  \partial_t\left\langle
   \tilde  f(1)\dots \tilde f(r)
  \right\rangle^{(N)} &=
  \sum_{n=1}^N\frac{(\I \bar S_\mathrm I(t))^n}{n!}\partial_t Z_0[\vc L] \\ &+
  \sum_{n=1}^N\frac{(\I \bar S_\mathrm I(t))^{n-1}}{(n-1)!}
  \left[\partial_t \bar S_\mathrm I\right] Z_0[\vc L]\nonumber\;.
\end{align}

In the first term, the time derivative acts on the generating functional
\begin{align}
\label{eq:141}
  \partial_tZ_0[\vc L] &= \int\D\Gamma\E^
   {\I\left\langle\vc L_q, \vc q\right\rangle +
    \I\left\langle\bar{\vc L}_p, \vc p\right\rangle}\,
    \I\left\langle\partial_t\bar{\vc L}_p, \vc p\right\rangle \\ &=
  -\sum_{j=1}^r\I k_j\int\D\Gamma\E^
  {\I\left\langle\vc L_q, \vc q\right\rangle +
   \I\left\langle\bar{\vc L}_p, \vc p\right\rangle}
   \frac{p_j}{m} \;, \nonumber
\end{align}
while it acts on the time-integral boundary in the interaction in the second term,
\begin{equation}
  \partial_t\bar S_\mathrm I(t) = \sum_{j=1}^r\int\frac{\D^6s_1'}{(2\pi)^6}\,
  \ell_j\hat D_j(-1')v'(1')\hat f
  (1') \;,
\label{eq:142}
\end{equation}
where we used $g_{qp}(t,t) = 0$ and set $t_1' = t$.

Inserting the derivatives back into (\ref{eq:140}), we can use (\ref{eq:139}) in the first term, and (\ref{eq:135}) in the second term. Identifying applied operators in the form of the free generating functional and collecting them into operators finally results in
\begin{align}
  \partial_t&\left\langle
    \tilde f(1)\dots\tilde f(r)
  \right\rangle^{(N)} = -\sum_{j=1}^r\I k_j\left\langle
    \tilde f(1)\dots\widetilde{fu}(j)\dots\tilde f(r)
  \right\rangle^{(N)} \nonumber\\ &+
  \sum_{j=1}^r\I\ell_j\I k_j\int\frac{\D^6s_1'}{(2\pi)^6}\,v'(1')
  \nonumber\\ &\cdot
  \left\langle
    \tilde f(1)\dots\tilde f(j-1')\dots\tilde f(r)\tilde f(1')
  \right\rangle^{(N-1)} \;,
\label{eq:143}
\end{align}
where $\tilde f(j-1'):= \tilde f(t,s_j-s_1')$.

This shows that the time evolution of the $r$-point phase-space correlator is divided into two effects: The first is the convection of the phase-space density described by the first term. The second term contains all possible interactions with an additional point in phase space. Here, $r+1$-point phase-space correlators appear. Consequently, evolution equations for $r+1$-point phase space density correlators are needed in order to solve the $r$-point evolution equation. These again contain convection terms as well as $r+2$-point phase space density correlators. For the latter, evolution equations are needed that contain $r+3$-point correlators, and so on. An (almost) infinite hierarchy of coupled differential equations unfolds, known as the BBGKY hierarchy. A Fourier transform back into configuration space shows that (\ref{eq:143}) are the exact terms of the conventional BBGKY hierarchy \cite{2015PhRvE..91f2120V}.

A truncation criterion of the BBGKY hierarchy emerges here from the order of the Taylor expansion of the interaction: In the evolution equation for the $r$-point correlator, the $r+1$-point correlators appear only in the $(N-1)$-st order in the interaction, reduced by one. This is repeated in the time evolution of the two $r+1$-point correlators: the $r+2$ point correlator is again reduced by one order in the interaction and so on for each further step up the hierarchy. Once the $0$-th order is reached, only convection terms and no higher-order correlators appear in the evolution equation, and the hierarchy ends.

\section{Fluids in KFT}
\label{sec:8}

\subsection{Introduction}
\label{sec:8.1}

Cosmic structure formation is governed by both dark and baryonic matter. Therefore, it is crucial to extend KFT to capture the physics of both particle types. Unlike dark matter, baryonic interactions are not limited to gravity only. This causes the cosmic matter power spectrum to attain additional features such as baryon acoustic oscillations (BAO) as well as disturbances in the spectrum at small scales due to e.g.\ pressure and baryonic cooling. We describe here one way towards including mixtures of dark and baryonic matter into KFT.

\subsection{Extension of the generating functional}
\label{sec:8.2}

In principle, fluid dynamics should be captured by KFT if microscopic gas-particle interactions were taken into account. However, it is a regime hard to reach in any expansion of the interaction. To acquire access to fluid dynamics nonetheless, we implement a fluid model based on an idea close to the conventional approach to hydrodynamics: We introduce mesoscopic particles similar to conventional fluid elements and demand a hydrodynamical scale hierarchy. Then, on the scale of the mesoscopic particles, local thermal equilibrium can be assumed to have been established. The properties of the mesoscopic particles are then described by state variables such as pressure and energy density. At the same time, we demand that the mesoscopic particles are much smaller than any scale we are interested in and can be modelled as point-like.

For simplicity, we assume an isotropic fluid which is fully described by a spatially dependent velocity field $u(q)$, a pressure field $P(q)$ and an energy-density field $\epsilon(q)$. In addition, we assume that the pressure is only caused by the random velocities of the microscopic particles. Then, both the pressure and the internal energy-density are proportional to the enthalpy-density $h$,
\begin{equation}
  h = \epsilon + P = \frac{5}{3}\epsilon = \frac{5}{2}P\;.
\label{eq:144}
\end{equation}

In order to sample the fluid, the mesoscopic particles need to capture these properties: they need to contain information about their position, the momentum, and the enthalpy at that position. Hence, we endow each particle with these three degrees of freedom,
\begin{equation}
  \varphi_j = \left(q_j, p_j,\mathcal{H}_j\right) \qquad
  \text{for the $j$-th particle} \;,
\label{eq:145}
\end{equation}
and describe their non-interacting dynamics by ballistic motion, conserving momentum and enthalpy. These dynamics are expressed by the Green's function
\begin{equation}
  \tilde G_R(t,t_\mathrm{i}) =
  \begin{pmatrix}
    \mathcal{I}_3 & \frac{t- t_\mathrm{i}}{m} \mathcal{I}_3 & 0 \\
    0 & \mathcal{I}_3 & 0 \\
    0 & 0 & 1 \\
  \end{pmatrix}\,\theta\left(t-t_\mathrm{i}\right)\;.
\label{eq:146}
\end{equation}

From here on, a free generating functional for KFT can be constructed in analogy to (\ref{eq:20})
\begin{equation}
  \tilde Z_0\left[\vc J, \vc K\right] =
  \int\D\Gamma_\mathrm{i}\,\exp\left(
    \I\int_\mathrm{0}^\mathrm{\infty}\D t'\,
    \langle\vc J (t'), \bar{\vc\varphi}(t')\rangle
  \right)\;,
\label{eq:147}
\end{equation}
where the tilde marks the free generating functional for the fluid model, $J_j = (J_{q_j}, J_{p_j}, J_{\mathcal{H}_j})$ and
\begin{align}
  & \bar{\vc\varphi}(t) = \left(
      \tilde G_\mathrm{R}(t,0)\,\varphi^{\,\mathrm{(i)}}_j-
      \int_0^t\D t'\,\tilde{G}_\mathrm{R}(t,t')\,K_j(t')
    \right)\otimes e_j\;, \nonumber \\
  & K(t') = (K_q, K_p, K_\mathcal{H})\;.
\label{eq:148}
\end{align}

\subsection{Interaction operator}
\label{sec:8.3}

In addition, acceleration due to pressure gradients and pressure-volume work must be taken into account. For viscous hydrodynamics, diffusion of energy and momentum must be added. Within KFT, these effects are included via interactions between the mesoscopic particles. The exact form of the interaction operator is found via an approach similar to smooth-particle hydrodynamics (SPH) \cite{2005RPPh...68.1703M}. For the purposes of this review, we only sketch the derivation of the interaction operator for pressure-volume work. A thorough derivation including all terms of ideal and viscous fluid dynamics can be found in \cite{2018arXiv181013324V}.

The pressure field split into particle contributions reads
\begin{equation}
  P(q) = \frac{2}{5}\mathcal{H}\rho(q) =
  \frac{2}{5}\sum_i\mathcal{H}_i\,\delta_\mathrm{D}\left(q-q_i\right)\;,
\label{eq:149}
\end{equation}
where $\mathcal{H}_i$ is the enthalpy and $q_i$ the position of the $i$-th particle.

Following the Euler equation, the time evolution of the momentum field contains the term
\begin{equation}
  \dot p(q_1) = -\frac{\partial_{q_1}P(q_1)}{\rho(q_1)} =
  -\frac{2}{5\rho(q_1)}\sum_i\mathcal{H}_i\,
  \partial_{q_1}\delta_\mathrm{D}\left(q_1- q_i\right)\;,
\label{eq:150}
\end{equation}
where we inserted the discretised pressure field from (\ref{eq:149}).

The momentum change at position $q_1$ is associated to particles according to their (spatial) contribution at that position. To this end, we weigh the momentum change at $q_1$ with a Dirac delta distribution around the $j$-th particle's position and integrate over the entire space,
\begin{align}
\label{eq:151}
  \dot p_j &= \int\D^3q_1\,\dot p(q_1)\,\rho_j(q_1) \\
  &= -\frac{2}{5}\sum_i\int\D^3q_1\,\frac{1}{\rho(q_1)}
  \delta_\mathrm{D}\left(q_1-q_j\right)\,\mathcal{H}_i\,
  \partial_{q_1}\delta_\mathrm{D}\left(q_1-q_i\right) \nonumber\;.
\end{align}

For the interaction operator, this object must be expressed by functional derivatives acting on the free generating functional. However, this is not possible for the inverse density. We handle this complication by approximating the inverse densities in a Taylor series around the mean density $\bar \rho$ of the ensemble. As a first step, we truncate the approximation already at $0$-th order,
\begin{equation}
  \frac{1}{\rho_j(q_1)} \approx \frac{1}{\bar\rho}\;.
\label{eq:152}
\end{equation}

For the pressure-volume work, a similar term for the enthalpy change of the $j$-th particle, $\dot{\mathcal{H}}_j$, can be derived in an analogous calculation. From the momentum and enthalpy changes, the interaction operator for an ideal, isotropic fluid can then be constructed,
\begin{align}
\label{eq:153}
  \hat S_\mathrm{I}^\mathrm{id} &= \frac{2\I}{5\bar\rho}\sum_{(i,j)}
  \int\D 1\,k_1\,\hat\rho_j(-1)\,\hat{\mathcal{H}}_i\,\hat\rho_i(1)\,
  \frac{\delta}{\I\delta K_{p_j}(1)} \\ &+
  \frac{2\I}{3\bar\rho}\,\sum_{(i,j)}
  \int\D 1\,k_1\,\hat\rho_j(-1)\left[
    \hat u_i-\hat u_j
  \right]\hat{\mathcal{H}}_i\,\hat\rho_i(1)\,
  \frac{\delta}{\I\delta K_{\mathcal{H}_j}(1)}\nonumber\;,
\end{align}
with $\D1 = \D t\D^3k_1$ and the one-particle operators
\begin{align}
\label{eq:154}
  &\hat\rho_j(1) = \exp\left(-\I k_1\frac{\delta}{\I\delta J_{q_j}}\right)\;,\\
  &\hat{\mathcal{H}}_j = \frac{\delta}{\I\delta J_{\mathcal{H}_j}}\;,
  \quad \hat u_j = \frac{1}{m}\frac{\delta}{\I\delta J_{p_j}}\;. \nonumber
\end{align}

For an ensemble characterized by the free generating functional (\ref{eq:147}) and the interaction operator (\ref{eq:153}), macroscopic evolution equations for density, momentum density, and energy density can be extracted. As shown in \cite{2018arXiv181013324V}, these are indeed the continuity, Euler, and energy-conservation equations of ideal hydrodynamics. If an interaction operator for diffusive effects is added, the Navier-Stokes and energy conservation of viscous hydrodynamics are recovered.

\section{Macroscopic Reformulation of KFT}
\label{sec:9}

From the perspective of most quantum and statistical field theories, the generating functional (\ref{eq:5}) of KFT is rather unusual in the sense that the path integral is expressed in terms of microscopic degrees of freedom even though we are actually interested in macroscopic fields like the density. As described before, the decisive advantage of this approach is the simplicity of the equations of motion for the microscopic degrees of freedom.

However, to facilitate the application of established perturbative as well as non-perturbative field-theoretical techniques, we would like to reformulate the KFT partition function as a path integral over \emph{macroscopic} fields. It turns out that this will lead to a resummation of an infinite subset of terms appearing in the microscopic perturbative expansion in the interaction operator. This will allow us to treat dark matter particles in terms of their fundamental Newtonian dynamics rather than the improved Zel'dovich dynamics. The presentation here is based on the more detailed and general derivation in \cite{2018arXiv180906942L} which uses the full phase-space density rather than the spatial density to preserve momentum information.

\subsection{Macroscopic action}
\label{sec:9.1}

We begin by using (\ref{eq:9}) and (\ref{eq:12}) to write the partition function as
\begin{align}
  Z &= \int\D\Gamma\int\mathcal{D}\vc x\,
    \delta_\mathrm{D}\bigl[E(\vc x,\vc x^\mathrm{(i)})\bigr] \nonumber\\ &=
  \int\D\Gamma\int\mathcal{D}\vc x\int\mathcal{D}\vc\chi\,
  \E^{\I\vc\chi\cdot E(\vc x,\vc x^\mathrm{(i)})} \;,
\label{eq:155}
\end{align}
where we represented the delta distribution in terms of a functional Fourier integral with respect to an auxiliary field $\vc\chi$ with components $\chi_j = \bigl(\chi_{q_j}, \chi_{p_j}\bigr)$. We further introduce the combined microscopic field $\vc\psi:=(\vc x,\vc\chi)$ and its action
\begin{equation}
  S_\psi:=\vc\chi\cdot E(\vc x,\vc x^\mathrm{(i)})
  \stackrel{!}{=} S_{\psi,0} + S_{\psi,\mathrm{I}} \;,
\label{eq:156}
\end{equation}
which splits into a part $S_{\psi,0}$ describing the free motion of the particles and a part $S_{\psi,\mathrm{I}}$ describing their interactions. Using (\ref{eq:15}) and (\ref{eq:17}), the latter can be expressed as
\begin{equation}
  S_{\psi,\mathrm{I}} = - \int\D t\sum_{j=1}^N\chi_{p_j}(t)\nabla_jV(t) \;.
\label{eq:157}
\end{equation}
Inserting (\ref{eq:37}) and defining the \emph{dressed response field} $\mathcal{F}$ via its Fourier transform,
\begin{equation}
  \tilde{\mathcal{F}}(1) :=
  -\sum_{j=1}^N \chi_{p_j}(t_1) \, \tilde{B}_j(1) \, \tilde{v}(1)\;,
\label{eq:158}
\end{equation}
allows us to write the interaction term as
\begin{equation}
  S_{\psi,\mathrm{I}} = \int\D 1
  \tilde{\mathcal{F}}(-1)\tilde{\rho}(1) \eqqcolon\mathcal{F}\cdot\rho\;,
\label{eq:159}
\end{equation}
where we also introduced the dot product as a short-hand notation for integrating over field arguments.

We now replace the explicitly $\psi$-dependent field $\rho$ with a new formally $\psi$-independent field $\phi_\rho$, using a functional delta distribution,
\begin{equation}
  Z = \int\mathcal{D}\phi_\rho\int\D\Gamma\int\mathcal{D}\psi\,
  \delta_\mathrm{D}\bigl[\phi_\rho - \rho\bigr]
  \E^{\I S_{\psi,0} + \I \mathcal{F} \cdot\phi_\rho} \;.
\label{eq:160}
\end{equation}
This way, the new field $\phi_\rho$ effectively still carries all the information contained in $\rho$. Most importantly, $\phi_\rho$- and $\rho$-correlation functions are identical. However, to emphasise their different origin we will deliberately call $\phi_\rho$ the macroscopic and $\rho$ the collective density field in the following.

Similar to (\ref{eq:112}), we now express the delta distribution as a functional Fourier transform with respect to a new macroscopic auxiliary field $\phi_\beta$,
\begin{equation}
  Z = \int\mathcal{D}\phi_\rho\int\mathcal{D}\phi_\beta
  \int\D\Gamma\int\mathcal{D}\psi\,
  \E^{-\I\phi_\beta\cdot (\phi_\rho-\rho)+
      \I S_{\psi,0}+\I \mathcal{F}\cdot\phi_\rho} \;.
\label{eq:161}
\end{equation}
Pulling all $\psi$-independent parts to the front, we find that the remaining microscopic part of the path integral assumes the form of the free generating functional of $\rho$- and $\mathcal{F}$-correlators,
\begin{equation}
  Z = \int\mathcal{D}\phi_\rho\int\mathcal{D}\phi_\beta\,
  \E^{-\I\phi_\rho\cdot\phi_\beta}
  \underbrace
  {\int\D\Gamma\int\mathcal{D}\psi
   \E^{\I S_{\psi,0}+\I\phi_\beta\cdot\rho+\I\phi_\rho\cdot\mathcal{F}}}_
  {\eqqcolon Z^{\rho,\mathcal{F}}_0\bigl[\phi_\beta,\phi_\rho\bigr]}\;,
\label{eq:162}
\end{equation}
with $\phi_\beta$ and $\phi_\rho$ playing the roles of the source fields for the collective fields $\rho$ and $\mathcal{F}$, respectively.

Finally, defining the combined macroscopic field $\phi:= (\phi_\rho,\phi_\beta)$ and the free generating functional of collective-field cumulants $W^{\rho,\mathcal{F}}_0:=\ln Z^{\rho,\mathcal{F}}_0$, we arrive at the result
\begin{equation}
  Z = \int\mathcal{D}\phi\,\E^{\I S_{\phi}}
\label{eq:163}
\end{equation}
with the macroscopic action
\begin{equation}
  S_\phi:= - \phi_\rho\cdot\phi_\beta -
  \I W^{\rho,\mathcal{F}}_0\bigl[\phi_\beta,\phi_\rho\bigr] \;.
\label{eq:164}
\end{equation}

We emphasise that this reformulation is \emph{exact} and hence (\ref{eq:163}) still contains the complete information on the microscopic dynamics, even though $S_\phi$ does not depend on $\psi$ any more. The microscopic information is now encoded in the free generating functional $W^{\rho,\mathcal{F}}_0\bigl[\phi_\beta,\phi_\rho\bigr]$ and thus, by means of a functional Taylor expansion, in the free collective-field cumulants
\begin{align}
\label{eq:165}
  &G^{(0)}_{\rho\dotsm\rho\mathcal{F}\dotsm\mathcal{F}}
  (1,\dotsc,n_\rho,1',\dotsc,n'_\mathcal{F}) \\
  &\quad:=\bigl\langle
    \tilde{\rho}(1)\dotsm\tilde{\rho}(n_\rho)
    \tilde{\mathcal{F}}(1')\dotsm\tilde{\mathcal{F}}(n'_\mathcal{F})
  \bigr\rangle_{0,\mathrm{c}} \nonumber \\
  &\quad = \left.\prod_{u=1}^{n_\rho}
    \left(\frac{\delta}{\I\delta\tilde\phi_\beta(u)}\right)
  \prod_{r=1}^{n_\mathcal{F}}
    \left(\frac{\delta}{\I\delta\tilde\phi_\rho(r')}\right)
  W^{\rho,\mathcal{F}}_0\bigl[\phi_\beta,\phi_\rho\bigr]
  \right|_{\mathrlap{\phi=0}} \;. \nonumber
\end{align}

\subsection{Including density-density and density-momentum correlations}
\label{sec:9.2}

Since we will use this macroscopic reformulation of KFT in Sect.\ \ref{sec:9.4} to treat Newtonian dynamics, whose Green's function (\ref{eq:54}) is limited from above, we have to take the complete expression for the initial probability distribution (\ref{eq:67}) into account, which includes all correlations between the initial phase-space coordinates of particles. This means that the lengthy polynomial $\mathcal{C}(\vc p)$ introduced in (\ref{eq:67}) must be included. This makes the explicit calculation of the above cumulants anything but straightforward. By adopting the essential ideas of the so-called Mayer cluster expansion \cite{1941JChPh...9....2M}, we have however been able to condense this process into a small set of rules for Feynman-like diagrams. Here, we will restrict ourselves to reviewing the recipe for cumulants involving the particle number density $\rho$, while the complete technical derivation for cumulants involving the phase-space density $f$ can be found in \cite{2018JSMTE..04.3214F}.

The general form of the diagrams in this approach is very simple: particles are represented by dots, and correlations between them by
different types of connecting lines. Since we are only interested in connected contributions to correlations, we need to consider
connected diagrams only, i.e.\ such through which we can trace a continuous path. A general free cumulant as in (\ref{eq:165}) is then ordered in terms of the number of particles $\ell$ being correlated in such a connected way,
\begin{align}
  & G^{(0)}_{\rho\cdots\rho\mathcal{F}\cdots\mathcal{F}}
  (1,\ldots,n_{\rho},1',\ldots,n'_{\mathcal{F}}) = \nonumber\\
  &\quad \sum_{\ell=1}^{n_\rho}
    G^{(0,\ell)}_{\rho\cdots\rho\mathcal{F}\cdots\mathcal{F}}
    (1,\ldots,n_{\rho},1',\ldots,n'_{\mathcal{F}}) \;.
\label{eq:166}
\end{align}
Note that, as a consequence of statistical homogeneity, the sum over particle numbers truncates at the number of density fields in the cumulant. Combined with the diagram rules, this ensures that all cumulants have a finite number of terms, i.e.\ KFT can exactly describe the highly non-linear effects of initial correlations on the free-streaming evolution with a finite number of explicit expressions.

We begin with the calculation of pure density cumulants since mixed cumulants involving the dressed response field $\mathcal{F}$ are derived from the former by the insertion of simple response factors. We first need the crucial concept of a \emph{field-label grouping}. For any fixed number of particles $\ell$, we group the field labels $(1,\ldots,n_\rho,1',\ldots,n_\mathcal{F}')$ of the cumulant into $j=1,\ldots,\ell$ non-empty sets $\mathrm{I}_j$. Any such collection of sets is called a field-label grouping. We define a variant of the phase-translation vectors (\ref{eq:30}) in terms of these groupings as
\begin{align}
\label{eq:167}
   L_{q,\mathrm{I}_j}(t) &= \sum_{r\in\mathrm{I}_j}{L}_{q,r}(t) :=
  \sum_{r\in\mathrm{I}_j} k_rg_{qq}(t_r,t) \;, \\
   L_{p,\mathrm{I}_j}(t) &= \sum_{r\in\mathrm{I}_j}{L}_{p,r}(t) :=
  \sum_{r\in\mathrm{I}_j} k_rg_{qp}(t_r,t) \;.
\label{eq:168}
\end{align}
A general $\ell$-particle density cumulant is then given as a sum over all possible field-label groupings,
\begin{align}
  & G^{(0,\ell)}_{\rho\cdots\rho}(1,\ldots,n_\rho) =
  \bar\rho^\ell (2\pi)^3 \delta_\mathrm{D}\left(
    \sum_{r=1}^{n_\rho}{L}_{q,r}(t_\mathrm{i})
  \right) \nonumber\\ & \quad
  \sum_{\{\mathrm{I}_1,\ldots,\mathrm{I}_\ell\}}
  \E^{-Q(\mathrm{I}_1,\ldots,\mathrm{I}_\ell)}
  \tilde{\Sigma}_{\mathrm{CDiag}}^{(\ell)}
  (\mathrm{I}_1,\ldots,\mathrm{I}_\ell) \;,
  \label{eq:169}
\end{align}
where we introduced the generalisation of the damping amplitude $Q_0$ from (\ref{eq:73}) as
\begin{equation}
  Q(\mathrm{I}_1,\ldots,\mathrm{I}_\ell) :=
  -\frac{\sigma_1^2}{6}
  \sum_{j=1}^\ell L_{p,\mathrm{I}_j}^{\,2}(t_\mathrm{i})\;.
\label{eq:170}
\end{equation}

The term $\tilde\Sigma_\mathrm{CDiag}^{(\ell)}(\mathrm{I}_1,\ldots,\mathrm{I}_\ell)$ represents the sum over all possible connected $\ell$-particle correlation diagrams. The possible diagram line types are
\begin{align}
\label{eq:171}
  \mytikz{\hortwopattern[j][k]\ppline{p1}{p2}} &=
  \mathcal{P}_{p_jp_k}\bigl({\mathcal{K}}^{\,(\mathrm{i})}_{jk}\bigr) :=
  \int_{q^{(\mathrm{i})}_{jk}}\E^{-\I\theta_q}
  \left(\E^{-\theta_p}-1\right)\;, \\
  \mytikz{\hortwopattern[j][k]\ddline{p1}{p2}} &=
  \mathcal{P}_{\delta_j\delta_k}\bigl({\mathcal{K}}^{\,(\mathrm{i})}_{jk}\bigr)
  :=
  \int_{q^{(\mathrm{i})}_{jk}}C_{\delta_j\delta_k}\E^{-\theta} \;,
  \nonumber \\
  \mytikz{\hortwopattern[j][k]\dpline{p1}{p2}} &=
  \mathcal{P}_{\delta_jp_k}\bigl({\mathcal{K}}^{\,(\mathrm{i})}_{jk}\bigr)
  :=
  \int_{q^{(\mathrm{i})}_{jk}}
  \left(-\I{C}_{\delta_jp_k}\cdot L_{p,\mathrm{I}_k}\right)
  \E^{-\theta} \;,
  \nonumber \\
  \mytikz{\hortwopattern[j][k]\ddlline{p1}{p2}} &=
  \mathcal{P}_{(\delta p)^2_{jk}}\bigl({\mathcal{K}}^{\,(\mathrm{i})}_{jk}\bigr)
  \nonumber\\ &:=
  \int_{q^{(\mathrm{i})}_{jk}}
  \left(-\I{C}_{\delta_jp_k}\cdot L_{p,\mathrm{I}_k}\right)
  \left(-\I{C}_{\delta_kp_j}\cdot L_{p,\mathrm{I}_j}\right)
  \E^{-\theta} \;, \nonumber
\end{align}
with the phase $\theta := \I\theta_q+\theta_p$ and
\begin{equation}
  \theta_q := 
    {\mathcal{K}}^{\,(\mathrm{i})}_{jk}\cdot
    {q}^{\,(\mathrm{i})}_{jk} \;,\quad
  \theta_p :=  L_{p,\mathrm{I}_j}^\top C_{p_ip_j} L_{p,\mathrm{I}_k}\;.
\label{eq:172}
\end{equation}
In addition to the momentum correlations introduced in Sect.\ (\ref{sec:3.3}), $C_{\delta_j\delta_k}$ represents density autocorrelations, and $C_{\delta_jp_k}$ density-momentum correlations between the initial positions of particles $j$ and $k$. Due to statistical homogeneity, the above Fourier integrals only need to be performed over the relative initial coordinates $q^{\,(\mathrm{i})}_{jk} = q^{\,(\mathrm{i})}_j-q^{\,(\mathrm{i})}_k$ of particles, where $j<k$ in all cases by convention. All translation vectors are evaluated at the initial time. One may draw these lines between particle dots according to the following rules:

\begin{itemize}
  \item No self-correlations in the form of subdiagrams \tikz[baseline=-0.1ex,radius=0.85cm]{ \coordinate[pdot] (p) at (0,0); \coordinate (center) at (0,0.2\gs); \draw[gline] circle[radius=0.2\gs, at=(center)];} may occur for any kind of correlation line.
  \item Any pair of particles can be connected directly by at most one line, i.e.\ no subdiagrams $\mytikz{\hortwopattern \ggline{p1}{p2}[0.3cm] \ggline{p1}{p2}[-0.3cm]}$ may occur for any combination of correlation lines.
  \item No particle may have more than one solid $\delta$-line attached to it, i.e.\ no subdiagrams my occur which include $\mytikz[0.5ex]{\coordinate[pdot] (p) at (0,0); \ddline{p}{0.75\gs,0} \ddline{p}{0.5\gs,0.5\gs} }$.
\end{itemize}

By convention, there is a flow of a general \emph{Fourier momentum} $\mathcal{K}^{\,(\mathrm{i})}_{jk}$ along each line from smaller to larger labels. The algorithm of \emph{Feynman rules} for evaluating $\tilde\Sigma_\mathrm{CDiag}^{(\ell)}(\mathrm{I}_1,\ldots,\mathrm{I}_\ell)$ for a given grouping of field labels is then as follows.

\begin{enumerate}[(1)]
  \item Choose a fixed graphical arrangement of $\ell$ dots labeled $j=1,\ldots,\ell$, representing the particles carrying the label sets $\mathrm{I}_j$. These are the vertices of the diagrams.
  \item Given this fixed set of vertices, draw all possible diagrams using the lines in (\ref{eq:171}), subject to the above rules. Repeat the next three steps for all these diagrams.
  \item For any closed loop in a diagram, assign one of the lines in the loop connecting vertices labeled $j<k$ a loop momentum $ {k}^{\,(\mathrm{i})}_{jk}$ and introduce an integral $\int_{k^{(\mathrm{i})}_{jk}}$.
  \item Pick a vertex $j$ which has only one line left with undetermined momentum and use the conservation law
  \begin{equation}
      L_{q,\mathrm{I}_j} +
    \sum_{i=1}^{j-1}\mathcal{K}^{\,(\mathrm{i})}_{ij} -
    \sum_{i=j+1}^\ell\mathcal{K}^{\,(\mathrm{i})}_{ji} = 0
  \label{eq:173}
  \end{equation}
  to fix the momentum of this line. Incoming momenta from vertices with smaller labels are counted positive while momenta outgoing to vertices with larger labels are counted negative. Momenta associated with vertices not connected to vertex $j$ are zero.
  \item Consecutively go through all other vertices to fix the remaining undetermined momenta by repeatedly applying the previous step.
\end{enumerate}

Once a pure density $n$-point cumulant is known any mixed $(n = n_\rho+n'_\mathcal{F})$-point cumulant is simply obtained by first inserting $n'_\mathcal{F}$ response factors and interaction potentials as
\begin{align}
  & G^{(0,\ell)}_{\rho\cdots\rho\mathcal{F}\cdots\mathcal{F}}
  (1,\ldots,n_{\rho},1',\ldots,n'_{\mathcal{F}}) = \nonumber\\ &\quad
  \bar\rho^\ell (2\pi)^3 \delta_\mathrm{D}\left(
    \sum_{r=1}^n L_{q,r}(t_\mathrm{i})
  \right) \tilde{v}(1')\ldots\tilde{v}(n'_\mathcal{F}) \nonumber \\ &\quad
  \sum_{\{\mathrm{I}_1,\ldots,\mathrm{I}_\ell\}}
  \E^{-Q(\mathrm{I}_1,\ldots,\mathrm{I}_\ell)}\left(
    b_{\mathrm{I}(1')}\ldots b_{\mathrm{I}(n'_B)}
  \right)
  \tilde\Sigma_\mathrm{CDiag}^{(\ell)}
  (\mathrm{I}_1,\ldots,\mathrm{I}_\ell) \;.
\label{eq:174}
\end{align}
The response factors are defined simply as
\begin{equation}
  b_{\mathrm{I}(r)} = \I L_{\mathrm{I}(r)}(t_r) \cdot k_r\;,
\label{eq:175}
\end{equation}
where $\mathrm{I}(r)$ is the set of field labels containing $r$. The retarded nature of the particle propagators contained in these response factors also leads to the general property that only such field-label groupings $\{\mathrm{I}_1,\ldots,\mathrm{I}_\ell\}$ give non-vanishing contributions to mixed cumulants (\ref{eq:174}) which have the following property: For all $\mathrm{I}_j, j=1,\ldots,\ell$, there must be at least one $\rho$-label $r\in\mathrm{I}_j$ such that $t_{r} \geq t_{u'}$ for all $\mathcal{F}$-labels $u'\in\mathrm{I}_j$. This can be understood as the causal flow of responses having to terminate at some density field label $r$ at later time since there is no instant response of the spatial density to forces acting on particles.

\subsection{Macroscopic perturbation theory}
\label{sec:9.3}

The reformulated path integral (\ref{eq:163}) allows us to set up a new perturbative approach to KFT following the standard procedure familiar from quantum and statistical field theory, i.e.\ in terms of propagators and vertices. For this purpose, we first split the action (\ref{eq:164}) into a propagator part $S_\Delta$, collecting all terms quadratic in $\phi$, and a vertex part $S_\mathcal{V}$, containing the remaining terms,
\begin{equation}
  S_\phi\stackrel{!}{=} S_\Delta + S_\mathcal{V}
\label{eq:176}
\end{equation}
with the definitions
\begin{align}
  \I S_\Delta &:= -\frac{1}{2}\int\D 1\int\D 2\,
  \tilde\phi(-1)\Delta^{-1}(1,2)\tilde\phi(-2) \;,
\label{eq:177} \\
  \I S_\mathcal{V} &:=
  \sum_
    {\mathclap{\substack{n_\beta, n_\rho=0 \\ n_\beta+n_\rho\neq 2}}}^\infty
  \quad\frac{1}{n_\beta! n_\rho!}
  \prod_{u=1}^{n_\beta}\left(\int\D u\,\tilde\phi_\beta(-u)\right)
  \prod_{r=1}^{n_\rho}\left(\int\D r'\tilde\phi_\rho(-r')\right)
  \notag\\[-0.5\baselineskip]
  &\qquad\qquad\times\mathcal{V}_{\beta\dotsm\beta\rho\dotsm\rho}
  (1,\dotsc,n_\beta,1',\dotsc,n'_\rho) \;,
\label{eq:178}
\end{align}
introducing the inverse macroscopic propagator $\Delta^{-1}$ and the macroscopic $(n_\beta+n_\rho)$-point vertices $\mathcal{V}_{\beta\dotsm\beta\rho\dotsm\rho}$.

We further define the macroscopic generating functional $Z^\phi$ by introducing a source field ${M  = \bigl(M_\rho, M_\beta\bigr)}$ for the combined macroscopic field $\phi$ into the partition function,
\begin{equation}
  Z^\phi[M] := \int\mathcal{D}\phi\,\E^{\I S_\phi + \I M \cdot\phi} \;.
\label{eq:179}
\end{equation}
Then, the vertex part of the action can be pulled in front of the path integral by replacing its $\phi$-dependence with functional derivatives with respect to $M$, $\hat{S}_\mathcal{V}:= S_\mathcal{V}\bigr|_{\phi\rightarrow\frac{\delta}{\I\delta M}}$, acting on the remaining path integral,
\begin{align}
  Z^\phi[M] &= \E^{\I\hat{S}_\mathcal{V}}
  \int\mathcal{D}\phi\,
  \E^{-\frac{1}{2}\phi\cdot\Delta^{-1}\cdot\phi+\I M\cdot\phi} \nonumber \\
  &= \E^{\I\hat{S}_\mathcal{V}}
  \E^{\frac{1}{2}(\I M)\cdot\Delta\cdot(\I M)} \;.
\label{eq:180}
\end{align}
Expanding the first exponential in (\ref{eq:180}) in powers of the vertices now gives rise to a new perturbative approach that we will refer to as the \emph{macroscopic perturbation theory}.

Within this approach, the interacting macroscopic-field cumulants are obtained by taking appropriate functional derivatives of the macroscopic cumulant-generat\-ing functional $W^\phi[M] :=\ln{Z^\phi[M]}$. In particular, the lowest perturbative order of the 2-point density cumulant $G_{\rho\rho}$ is given by the $\rho\rho$-component of the macroscopic propagator,
\begin{align}
  G_{\rho\rho}(1,2) &= \left.
    \frac{\delta}{\I\delta M_\rho(1)}\frac{\delta}{\I\delta M_\rho(2)}W^\phi[M]
  \right|_{M=0} \nonumber \\
  &= \Delta_{\rho\rho}(1,2) + \text{terms involving vertices} \;.
\label{eq:181}
\end{align}
Adopting the usual field-theoretical language, we will refer to this as the tree-level expression for $G_{\rho\rho}$. For further detail on the general properties of the macroscopic perturbation theory, we refer the reader to \cite{2018arXiv180906942L}, where we introduce a Feynman-diagram language to compute the higher-order perturbative contributions -- i.e.\ the so-called loop-contributions -- systematically.

To find expressions for the inverse propagator $\Delta^{-1}$ and the vertex term $\mathcal{V}_{\beta\dotsm\beta\rho\dotsm\rho}$, we insert the functional Taylor expansion of $W^{\rho,\mathcal{F}}_0$ into the macroscopic action (\ref{eq:164}) and identify terms with (\ref{eq:177}) and (\ref{eq:178}), respectively,
\begin{align}
  \Delta^{-1}(1,2) &=
  \begin{pmatrix}
    G^{(0)}_{\mathcal{F}\mathcal{F}} & \I\mathcal{I}+G^{(0)}_{\mathcal{F}\rho}\\
    \I\mathcal{I}+G^{(0)}_{\rho\mathcal{F}} & G^{(0)}_{\rho\rho}
  \end{pmatrix}(1,2) \;,
\label{eq:182} \\
  \mathcal{V}_{\beta\dotsm\beta\rho\dotsm\rho}&(1,\dotsc,n_\beta,1',\dotsc,n'_\rho) \notag\\
  &= \I^{n_\beta+n_\rho}G^{(0)}_{\rho\dotsm\rho\mathcal{F}\dotsm\mathcal{F}}
  (1,\dotsc,n_\beta,1',\dotsc,n'_\rho) \;.
\label{eq:183}
\end{align}
Here, $\mathcal{I}$ denotes the identity 2-point function,
\begin{equation}
  \mathcal{I}(1,2) := (2\pi)^3
  \delta_\mathrm{D}\bigl(k_1+k_2\bigr)\delta_\mathrm{D}(t_1-t_2) \;.
\label{eq:184}
\end{equation}

The propagator $\Delta$ is then obtained by a combined matrix and functional inversion of (\ref{eq:182}), defined via the following matrix integral equation,
\begin{equation}
  \int\D 1'\Delta(1,-1')\Delta^{-1}(1',2) \stackrel{!}{=}
  \mathcal{I}(1,2)\,\mathcal{I}_2 \;.
\label{eq:185}
\end{equation}
The matrix part of this inversion can be performed immediately and yields
\begin{equation}
  \Delta(1,2) =
  \begin{pmatrix}
    \Delta_\mathrm{R}\cdot G^{(0)}_{\rho\rho}\cdot\Delta_\mathrm{A} &
    \;\; -\I\Delta_\mathrm{R} \\
    -\I\Delta_\mathrm{A} & \;\; 0
  \end{pmatrix}(1,2) \;,
\label{eq:186}
\end{equation}
where we defined the \emph{retarded} and \emph{advanced} (macroscopic) propagators
\begin{equation}
  \Delta_\mathrm{R}(1,2) = \Delta_\mathrm{A}(2,1) :=
  \Bigl(\mathcal{I} - \I G^{(0)}_{\rho\mathcal{F}}\Bigr)^{-1}(1,2)
\label{eq:187} \;,
\end{equation}
containing the remaining functional inverses. They describe the linear response of the density at time $t_1$ to a perturbation of the interacting system at time $t_2$.

Consequently, $\Delta_{\rho\rho} = \Delta_\mathrm{R}\cdot G^{(0)}_{\rho\rho}\cdot\Delta_\mathrm{A}$ is the 2-point density cumulant resulting from the linear response of the interacting system to the correlated motion of the free-streaming particles. Generally, $\Delta_{\rho\rho}$ will nevertheless be nonlinear in the initial density correlation because the free-streaming of particles in $G^{(0)}_{\rho\rho}$ itself builds up those nonlinearities as long as the initial particle momenta are correlated\cite{2018JSMTE..04.3214F}.

In \cite{2018arXiv180906942L}, we describe how $\Delta_\mathrm{R}$, $\Delta_\mathrm{A}$ and $\Delta_{\rho\rho}$ can be computed explicitly for a given physical system. But even without specifying the system, it is always possible to express (\ref{eq:187}) formally in terms of a Neumann series by expanding the functional inverse in orders of $\I G^{(0)}_{\rho\mathcal{F}}$,
\begin{align}
  \Delta_\mathrm{R}(1,2) &= \Delta_\mathrm{A}(2,1) =
  \sum_{n=0}^\infty\Bigl(\I G^{(0)}_{\rho\mathcal{F}}\Bigr)^n(1,2)
  \nonumber \\
  \begin{split}
    &= \mathcal{I}(1,2) + \I G^{(0)}_{\rho\mathcal{F}}(1,2) \\
    &\hphantom{=}+\int\D 1'\,
      \I G^{(0)}_{\rho\mathcal{F}}(1,-1')\,
      \I G^{(0)}_{\rho\mathcal{F}}(1',2)+
    \dotsb\;.
  \end{split}
\label{eq:188}
\end{align}
From the definition of $\mathcal{F}$ in (\ref{eq:158}), it follows that $G^{(0)}_{\rho\mathcal{F}} \propto v$. Therefore, $\Delta_\mathrm{R}$, $\Delta_\mathrm{A}$ and $\Delta_{\rho\rho}$ contain terms of arbitrarily high order in the interaction potential $v$. This demonstrates that the lowest perturbative order within the macroscopic approach already captures effects which could only be described at infinitely high order within the microscopic perturbation theory.

A closer analysis reveals that, in a statistically homogeneous system, the transition from the micro- to the macroscopic formulation entails a resummation of all contributions which do not lead to any mode-coupling beyond the one already introduced by the free evolution. Due to this property, we will refer to the macroscopic reformulation of KFT as \emph{resummed KFT} (RKFT).

\subsection{Resumming Newtonian dynamics}
\label{sec:9.4}

Using RKFT enables us to treat dark-matter particles in terms of their fundamental Newtonian dynamics rather than the improved Zel'dovich dynamics. Since the inertial motion of particles on Zel'dovich-type trajectories already contains part of the gravitational interaction, even a first-order calculation in the microscopic perturbation theory captures the nonlinear evolution of the power spectrum over a wide range of scales remarkably well, as shown in Fig.\ \ref{fig:3} and discussed in detail in \cite{2016NJPh...18d3020B}.

This is not the case for Newtonian dynamics, and any expansion to finite order in the Newtonian gravitational potential does not even reproduce the linear growth of structures correctly. Working with an averaged interaction term or the Born approximation, as described in Sects.~\ref{sec:4.2} and~\ref{sec:4.3}, does not improve the situation either. This is because both approaches require deviations from the inertial trajectories caused by the (effective) forces acting between them to be small, which is not the case for the truly non-interacting inertial trajectories in Newtonian dynamics.

\begin{figure}[ht]
  \includegraphics[width=\columnwidth]{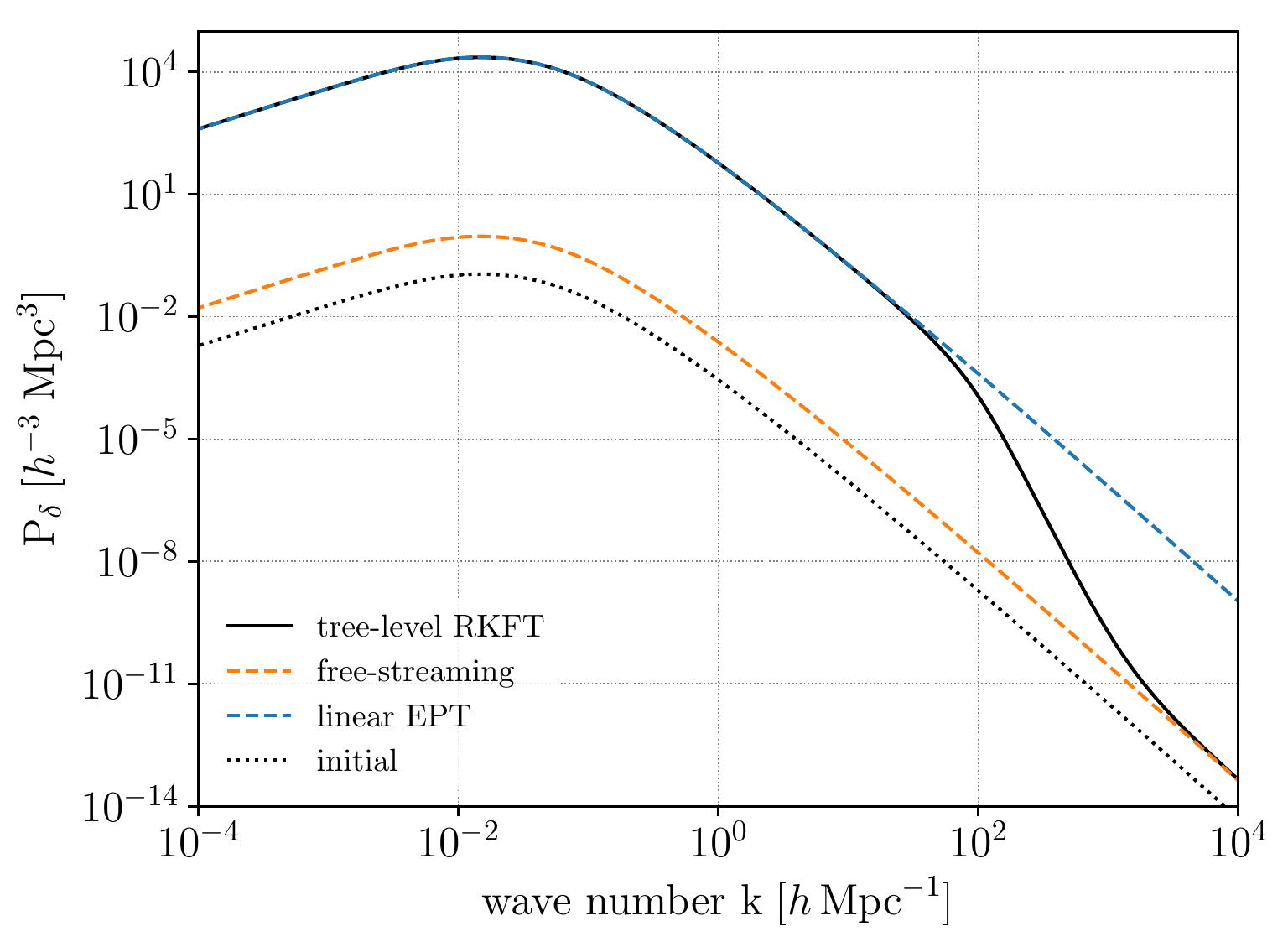}
\caption{Comparison of different density contrast power spectra evolved from the time of CMB decoupling to redshift zero, using cosmological parameters $\Omega_\mathrm{m} = 0.3$, $\Omega_{\Lambda} = 0.7$, $\Omega_\mathrm{b} = 0$, $h = 0.7$, $\sigma_8 = 0.8$: the tree-level Newtonian RKFT spectrum (black solid), the spectrum resulting from Newtonian free-streaming (orange dashed), and the spectrum obtained from linear Eulerian perturbation theory (EPT, blue dashed). For reference, the initial BBKS spectrum \cite{1986ApJ...304...15B} is shown as well (black dotted). The tree-level RKFT result interpolates between linear growth on large scales and free-streaming growth on small scales where the particles' momentum variance becomes relevant.}
\label{fig:10}
\end{figure}

Using RKFT, however, allows to overcome this limitation. In Fig.\ \ref{fig:10}, we show the tree-level RKFT result for the density contrast power spectrum,
\begin{equation}
  P_\delta^{(\Delta)}(1) :=\left.
    \frac{1}{\bar{\rho}^2}\int\frac{\D^3k_2}{(2\pi)^3}\Delta_{\rho\rho}(1,2)
  \right|_{\mathrlap{t_2 = t_1}} \;,
\label{eq:189}
\end{equation}
evolved from the time of CMB decoupling until today assuming Newtonian particle dynamics, and compare it to the power spectrum resulting from Newtonian free-streaming as well as the power spectrum obtained from linear Eulerian perturbation theory (EPT).

At small wavenumbers, $P_\delta^{(\Delta)}$ follows the linear EPT spectrum, before it starts to fall below the linear prediction at wavenumbers $k\gtrsim 20\,h\,\mathrm{Mpc}^{-1}$. When going to even higher wavenumbers it eventually approaches and follows the free-streaming spectrum for $k\gtrsim 10^4\,h\,\mathrm{Mpc}^{-1}$.

First of all, this shows that the partial resummation of Newtonian gravitational interactions captured by the macroscopic propagator allows to precisely recover the linear growth of structures on the largest scales even though the structure growth resulting from non-interacting Newtonian dynamics is orders of magnitude weaker. At the same time, the resummation does not compensate the small-scale damping effect caused by the particles' initial momentum variance $\sigma_1^2$. This behaviour is reminiscent of the small-scale decay of the power spectrum found in Renormalized Perturbation Theory (RPT) \cite{2006PhRvD..73f3519C, 2006PhRvD..73f3520C}.

However, we can also infer that RKFT does not enhance this damping effect either, as $P_\delta^{(\Delta)}$ never drops below the free-streaming spectrum. This differs from the power spectra found in RPT or Zel'dovich dynamics, which overpredict the dissolution of small-scale structures\cite{2006PhRvD..73f3519C, 2006PhRvD..73f3520C, 2002PhR...367....1B, 2017NJPh...19h3001B}.

Altogether, the tree-level RKFT result for the density contrast power spectrum is thus found to capture the linear effects introduced by gravitational interactions in a way that is fully consistent with the underlying non-interacting Newtonian particle dynamics. Higher-order perturbative corrections within RKFT are currently investigated.

\section{Mixtures of gas and dark matter in KFT}
\label{sec:10}

Baryonic matter with its additional non-gravitational interactions has a profoundly different effect on large-scale cosmic structures than dark matter. To gain a complete understanding of structure growth, it is therefore necessary to investigate how a mixture of dark and baryonic matter co-evolves through the cosmic history.

To describe such a mixture, we extend the macroscopic reformulation of KFT to include two different particle species. The long term goal hereby is to use this approach to study the non-linear evolution of structures which contain both baryonic and dark matter. In this formalism, the collisionless character of dark matter is kept while at the same time the fluid properties of baryonic matter needs to be taken into account. 

We demonstrate the functionality of this formalism on structure growth in the linear regime. Our aim hereby is a proof of concept showing that KFT is able to correctly describe the co-evolution of mixtures. The foundations presented here can be used in the future to investigate nonlinear structure growth.

Furthermore, we demonstrate that our formalism can qualitatively reproduce some of the key effects baryons have on linear structure formation. Firstly, the suppression of structures on small scales due to the repulsive effects of baryonic pressure is demonstrated. Secondly, by also taking interactions between photons and baryonic matter into account, essential features of baryon-acoustic oscillations are reproduced.

\subsection{KFT with two particle types}
\label{sec:10.1}

To describe the interactions of two different particle types in the framework of KFT we expand the macroscopically reformulated generating functional (\ref{eq:180}) to a system of two particle species in a straightforward fashion by replacing the macroscopic fields $\phi_\rho$ and $\phi_\beta$ as well as the collective dressed response field $\mathcal{F}$ by two-component vectors containing the respective field contributions for baryons and dark matter,
\begin{equation}
  \phi_\rho \rightarrow \bigl(\phi_\rho^\mathrm{b}, \phi_\rho^\mathrm{d}\bigr) \;,\quad
  \phi_\beta \rightarrow \bigl(\phi_\beta^\mathrm{b}, \phi_\beta^\mathrm{d}\bigr) \;,\quad
  \mathcal{F} \rightarrow \bigl(\mathcal{F}^\mathrm{b}, \mathcal{F}^\mathrm{d}\bigr)\;.
\label{eq:190}
\end{equation}
Proceeding in a completely analogous fashion to Sect.\ \ref{sec:9}, we obtain a macroscopic propagator that takes exactly the same form as in the single-particle species case (\ref{eq:186}). The only difference being that all 2-point functions appearing in the propagator are now replaced by $2 \times 2$ matrices,
\begin{align}
  G^{(0)}_{\rho\rho} &\rightarrow
    \begin{pmatrix}
      G^{(0,\mathrm{bb})}_{\rho\rho} & G^{(0,\mathrm{bd})}_{\rho\rho} \\
      G^{(0,\mathrm{db})}_{\rho\rho} & G^{(0,\mathrm{dd})}_{\rho\rho}
    \end{pmatrix}\;,\quad
  G^{(0)}_{\rho\mathcal{F}} \rightarrow
    \begin{pmatrix}
      G^{(0,\mathrm{bb})}_{\rho\mathcal{F}} & G^{(0,\mathrm{bd})}_{\rho\mathcal{F}} \\
      G^{(0,\mathrm{db})}_{\rho\mathcal{F}} & G^{(0,\mathrm{dd})}_{\rho\mathcal{F}}
    \end{pmatrix}\;,
  \label{eq:191} \\
  \mathcal{I}(1,2) &\rightarrow
    (2\pi)^3\delta_\mathrm{D}(k_1+k_2) \delta_\mathrm{D}(t_1-t_2) \, \mathcal{I}_2\;.
  \label{eq:192}
\end{align}
The individual components of $G^{(0)}_{\rho\rho}$ describe the auto- and cross-correlations between the density fields of baryonic and dark matter, while the components of $G^{(0)}_{\rho\mathcal{F}}$ describe the response of these density fields to the interactions between particles of either the same or different species. These interactions are accordingly collected in a symmetric $2 \times 2$ potential matrix,
\begin{equation}
  v \rightarrow
  \begin{pmatrix}
    v^{\mathrm{bb}} & v^{\mathrm{bd}} \\
    v^{\mathrm{db}} & v^{\mathrm{dd}} \\
  \end{pmatrix}\;.
\label{eq:193}
\end{equation}
To be able to calculate the power spectrum for matter mixtures, we still need to find the appropriate interaction potential $v^{\mathrm{b}\mathrm{b}}$, which is the only entry containing not only gravitational interactions.

\subsection{Baryon interactions}
\label{sec:10.2}

Unlike collisionless dark matter, baryonic matter is well-described by an ideal gas, and hence its dynamics are governed by the ideal fluid dynamics discussed in Sect.\ \ref{sec:8}. Accordingly, we treat baryons in KFT as mesoscopic particles characterised by a position, a momentum and an enthalpy.

However, to facilitate the incorporation of mesoscopic particles into the RKFT framework, we approximate the gas as being isothermal. That is, we approximate the baryon temperature as being spatially homogeneous throughout the Universe, following the evolution of the mean gas temperature $T(t)$. In this case, every mesoscopic particle has the same enthalpy
\begin{equation}
  \mathcal{H}\bigl(T(t)\bigr) = \frac{\gamma}{\gamma-1} m^\mathrm{b} c_\mathrm{s}^2\bigl(T(t)\bigr)\;,
\label{eq:194}
\end{equation}
where $m^\mathrm{b}$ is the mass of a mesoscopic particle, $\gamma$ is the adiabatic index of the gas and $c_\mathrm{s}(T)$ is its sound velocity. This effectively reduces the dynamical degrees of freedom to just a position and a momentum, allowing us to treat baryons like microscopic particles experiencing a gravitational potential
\begin{equation}
  \tilde{v}_\mathrm{g}(k,t) = - \frac{3}{2} \frac{a(t)}{\bar{\rho} g(t)} \frac{1}{k^2}
\label{eq:195}
\end{equation}
as well as an additional effective repulsive interaction potential modelling the effects of pressure
\begin{equation}
  \tilde{v}_\mathrm{p}(k,t) = \frac{c_\mathrm{s}^2\bigl(T(t)\bigr)}{\bar{\rho}^\mathrm{b} H_\I^2}
  \frac{a^2(t)}{g(t)}\;,
\label{eq:196}
\end{equation}
where $\bar{\rho}^\mathrm{b}$ is the mean number density of the mesoscopic particles \cite{2018arXiv181107741G}. The different components of the potential matrix (\ref{eq:192}) are then given by $v^\mathrm{dd}=v^\mathrm{bd}=v^\mathrm{db}=v_\mathrm{g}$ and  $v^\mathrm{bb}=v_\mathrm{g}+v_\mathrm{p}$. To arrive at the expressions (\ref{eq:195}) and (\ref{eq:196}), we have set the masses of both dark matter and mesoscopic particles equal. We can do this without loss of generality since we are dealing with effective particles. We are now able to obtain quantitative results for the evolution of mixed dark and isothermal baryonic matter.

\subsection{Results at late times}
\label{sec:10.3}
Their density fluctuation power spectra are evolved from the epoch when matter and radiation decoupled at a redshift of $z=1100$ to a redshift of $z=100$. We choose a relatively high redshift for the final time since we currently perform the calculations only in the linear order of the resummed theory. Furthermore, the beginning of the reionization era makes it difficult to rigorously track the mean temperature evolution of the gas.

We use the same initial BBKS power spectra \cite{1986ApJ...304...15B} for both dark and baryonic matter. However, in Fig.\ \ref{fig:12}, we set the initial baryonic power spectrum to zero on wavenumbers $k \geq 1\,h\,\mathrm{Mpc}^{-1}$ as a crude approximation for the Silk damping, which suppressed baryonic structures at small scales during radiation decoupling \cite{1968ApJ...151..459S}.

\begin{figure}[ht]
  \includegraphics[width=\hsize]{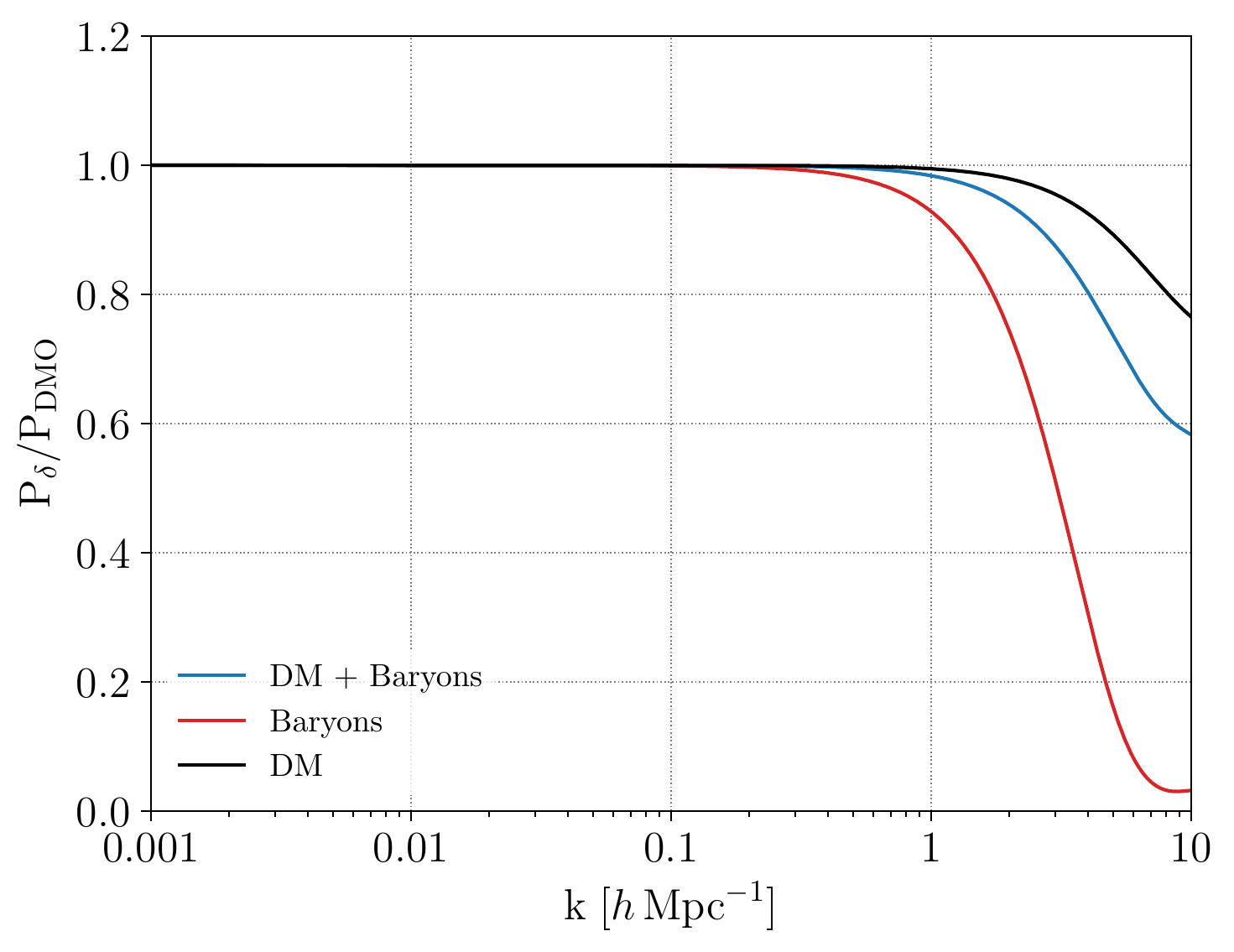}
\caption{Relative deviation of the baryon power spectrum (red), the dark matter power spectrum (black) and the total power spectrum (blue) of a mixed system as compared to a system consisting of dark matter only, evolved from the time of radiation decoupling until $z=100$, using slightly different cosmological parameters $\Omega_\mathrm{m} = 0.315$, $\Omega_{\Lambda} = 0.685$, $\Omega_\mathrm{b} = 0.049$ and $h = 0.673$. On small scales, the growth of baryonic as well as dark matter structures is suppressed by the baryon pressure.}
\label{fig:11}
\end{figure}

\begin{figure}[ht]
  \includegraphics[width=\hsize]{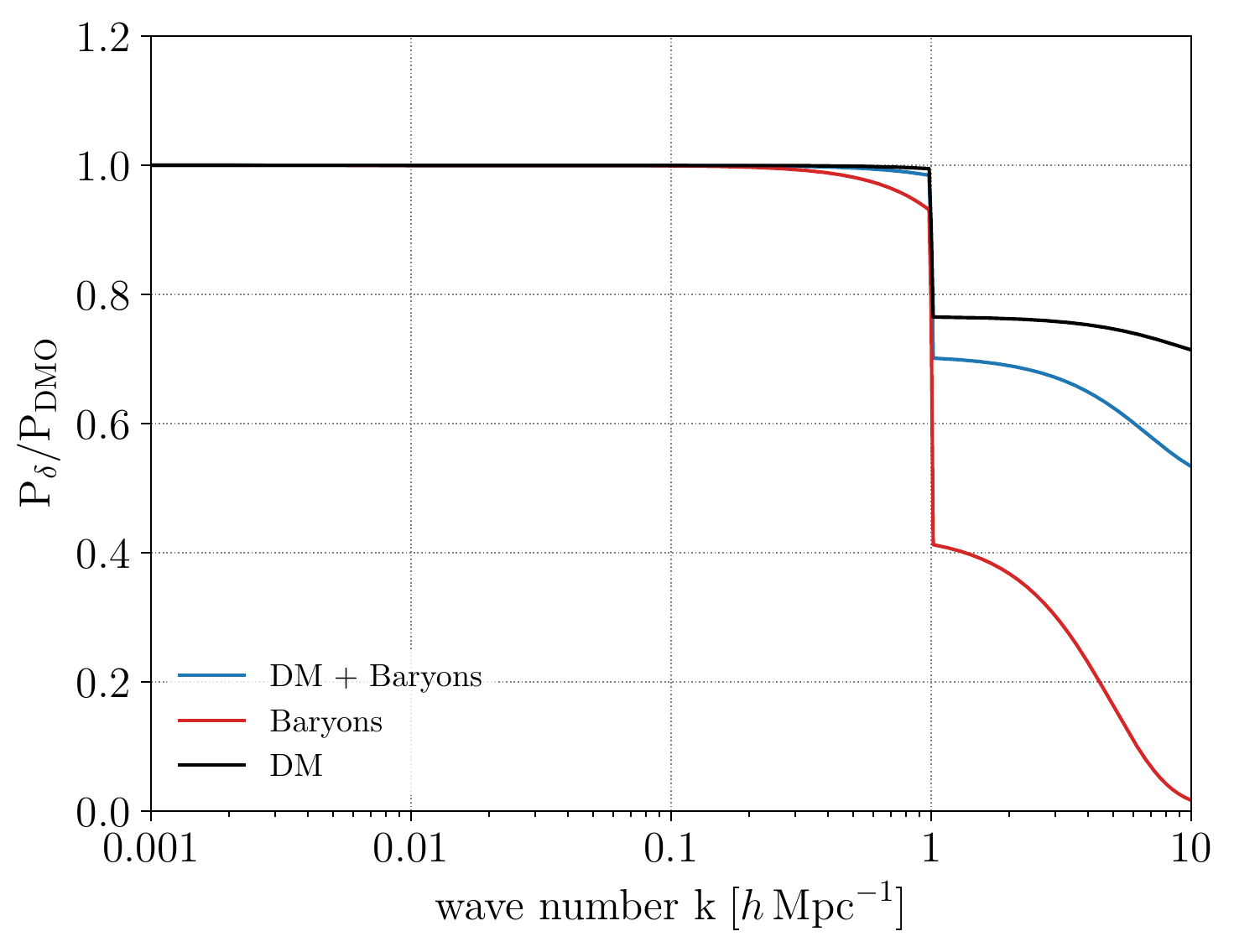}
\caption{Relative deviation of the baryon power spectrum (red), the dark matter power spectrum (black) and the total power spectrum (blue) of a mixed system as compared to a system consisting of dark matter only, evolved from the time of radiation decoupling until $z=100$, using the same cosmological parameters as in Fig.~\ref{fig:11}. At wavenumbers $k \geq 1\,h\,\mathrm{Mpc}^{-1}$ we set the initial baryonic power spectrum to zero as a crude model for the effect of Silk damping. Due to the gravitational interaction with the dark matter, baryonic structures are nevertheless build up on these scales.}
\label{fig:12}
\end{figure}

We plot the results in Figs.\ \ref{fig:11} and~\ref{fig:12}, where we have divided the obtained power spectra by the power spectrum for pure dark matter to show the relative deviations caused by the incorporation of baryonic matter.

The results show that the repulsive pressure interactions suppress structure growth at small scales. The fact that the dark-matter power spectrum is also suppressed shows that RKFT is able to capture the interplay between the baryonic pressure interactions and the gravitational interactions between the two particle species. This also allows to describe the build-up of baryonic structures on scales where the initial baryonic power spectrum was set to zero, as can be seen in Fig.\ \ref{fig:12}. Even the subsequent suppression of the dark-matter spectrum caused by the pressure of these new baryonic structures is captured.

The next steps towards gaining further insight into baryonic effects on structure formation after radiation decoupling are to drop the isothermal approximation and to develop effective potentials capturing additional physical properties of baryonic matter, e.g.\ radiative cooling effects.

\subsection{Mesoscopic particles with photons}
\label{sec:10.4}

\begin{figure}[ht]
  \includegraphics[width=\hsize]{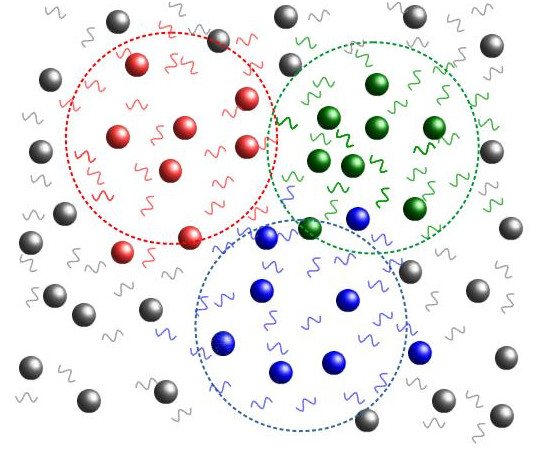}
\caption{Illustration of mesoscopic particles containing both baryons and photons. The color of the baryons and photons indicates to which mesoscopic particle they belong.}
\label{fig:13}
\end{figure}

To investigate structure formation before radiation decoupling at $z \approx 1100$, it is necessary to also take into account that the pressure forces acting on the baryons are mainly due to photons. Therefore, we modify the set-up of the mesoscopic particle by constructing it in such a way that it contains both baryons and photons (cf.\ Fig.\ \ref{fig:13}).

When investigating structure growth between matter-radiation equality and photon decoupling, we assume that the mass of the mesoscopic particle and hence its gravitational potential are dominated by its baryonic content. Hence, the gravitational potential of the particle is the same as in (\ref{eq:195}). However, the enthalpy of the particle is dominated by the photons since they are much more abundant -- the ratio in the number densities being given by $\eta \approx 6 \cdot 10^{-10}$ -- as well as due to their ultrarelativistic nature.

The resulting pressure potential in a baryon-photon fluid is given by
\begin{equation}
  \tilde{v}_{p,\gamma}(k,t) = \frac{\pi^4}{180 \zeta(3)}
  \frac{k_\mathrm{B}T_\mathrm{CMB}}{\bar{\rho}^\mathrm{b}
  \eta m_\mathrm{P} H_0^2 \Omega_{\mathrm{m},0}} \frac{a(t)}{g(t)}
\label{eq:197}
\end{equation}
with $k_\mathrm{B}$ being the Boltzmann constant, $m_\mathrm{P}$ the proton mass, $T_\mathrm{CMB}$ the current temperature of the CMB, $H_0$ the current value of the Hubble function and $\Omega_{\mathrm{m},0}$ the current dimensionless matter density parameter.

\begin{figure}[ht]
  \includegraphics[width = \hsize]{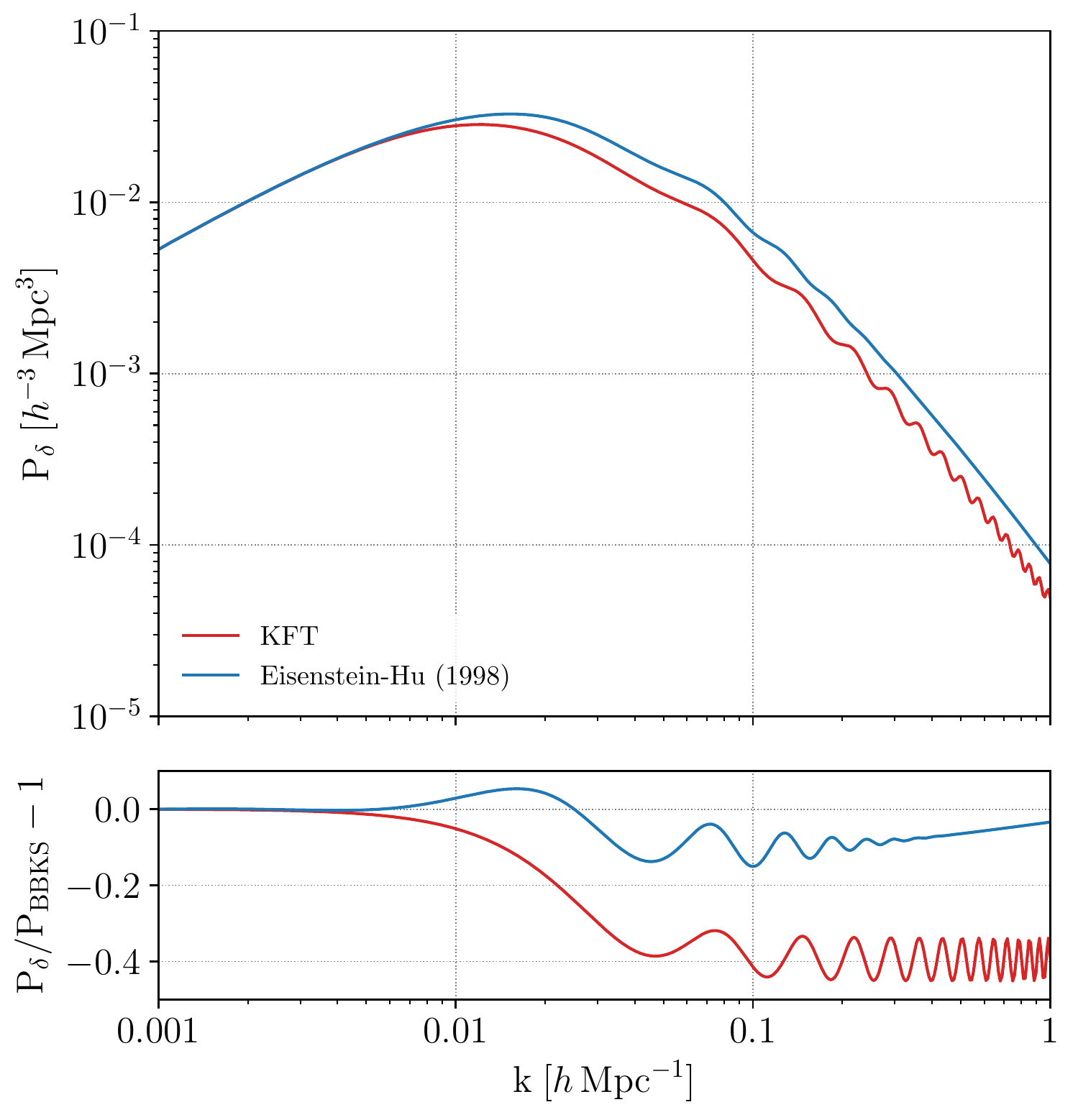}
\caption{(Upper panel) Comparison between the power spectrum at recombination obtained in our model (red) and the power spectrum obtained using the Eisenstein-Hu transfer function (blue), using the same cosmological parameters as in Fig.~\ref{fig:11}. (Lower panel) The relative deviation of both spectra from the linearly evolved initial BBKS spectrum \cite{1986ApJ...304...15B} is shown. The KFT spectrum reproduces oscillatory features close to the positions of the BAOs in the Eisenstein-Hu spectrum, but shows a stronger pressure-induced suppression of structure growth on intermediate and small scales. Furthermore, the effect of Silk damping is not included in our model.}
\label{fig:14}
\end{figure}

Assuming an initial BBKS power spectrum, the resulting matter power spectrum can be seen in Fig.\ \ref{fig:14}. It is compared to the power spectrum obtained using the Eisenstein-Hu transfer function \cite{1998ApJ...496..605E}.

We clearly see that our model qualitatively reproduces features of BAOs close to their positions in the semi-analytic Eisenstein-Hu model. The main difference between the two models is a stronger suppression of structures above a wavenumber of $k\approx 10^{-2}\,h\,\mathrm{Mpc}^{-1}$.

This suppression is a consequence of assuming a simplified initial BBKS spectrum, which does not take the effect of photons on the transfer function into account. In the future, we will correct for this to improve the quantitative agreement between our results and Eisenstein-Hu. Beyond this, the mesoscopic particles need to be modified to also take into account the dissolution of structures at small scales due to Silk damping. This can be achieved by treating the enthalpy as an additional dynamical property of the particles.

\section{Structure formation with modified theories of gravity}
\label{sec:11}

General relativity successfully describes the physics of gravitational interactions on a vast range of scales under the assumption that the total matter and energy budget of the Universe is to be augmented by the presence of three unknown ingredients, namely dark matter, dark energy in form of a cosmological constant, and the inflaton field. It is understood that the theory can only be used as an effective field theory valid up to the Planck scale, which is not renormalizable. Even though it is observationally in good agreement with cosmological data, general relativity suffers from severe and persistent theoretical obstacles, one of them being the cosmological constant itself. These have motivated the construction of alternative theories of gravity, both in the large- and in the small-scale regimes.

In the same spirit as general relativity, one can construct other effective field theories for the underlying gravitational interactions. Keeping the fundamental properties of general relativity intact (for instance unitarity, locality and invariance under Lorentz symmetry), any modification of gravity will inevitably introduce new propagating degrees of freedom. They usually come in form of additional scalar, vector, or tensor fields. Therefore, the majority of alternative theories of gravity can be classified by means of three important classes (see \cite{2018arXiv180701725H} for a recent review on alternative theories of gravity)
\begin{itemize}
  \item scalar-tensor theories: quintessence, Brans-Dicke, $k$-essence, Galileon, Horndeski, beyond-Horndeski, and DHOST-type theories;
  \item vector-tensor theories: vector-Galileon, generalized Proca theories, beyond-generalized Proca, or multi-Proca theories;
  \item tensor-tensor theories: massive gravity, bigravity, or multi-gravity theories;
\end{itemize}
If present, these additional degrees of freedom weaken gravity on cosmological scales. They either form a condensate whose energy density sources self-acceleration of the cosmic expansion, or they contribute as a condensate whose energy density compensates the cosmological constant. The latter is used to solve the cosmological-constant problem, while the former is useful for applications to dark energy and inflation. The presence of some of these degrees of freedom, for instance cosmic vector fields, could even give rise to a violation of the cosmological principle. Quite generically, these alternative theories of gravity will modify
\begin{itemize}
  \item the background evolution by altering the Hubble function $H$;
  \item perturbations by enforcing altered structure formation due to: an altered time sequence of gravitational clustering; the evolution of peculiar velocities and the number density of collapsed objects; modifications in the gauge-invariant matter density contrast and its relation to the gravitational potential; changes in the gravitational slip parameter, the effective gravitational potential, and the growth rate;
\end{itemize}
These alternative theories come with free parameters or even with free functions. Restricting their allowed parameter space is an indispensable task in order to test them against general relativity. At the background level, geometrical probes measuring the angular-diameter distance as a function of redshift (CMB and BAO) and the distance-redshift relation of supernovae can be used to constrain the allowed field space. Concerning linear perturbations, the CMB temperature anisotropies provide a supplementary tool to probe the evolution of the inhomogeneous perturbations. However, these probes are not sufficiently powerful to disentangle certain degeneracies between different alternative gravity theories. Hence, testing alternative theories of gravity on cosmological scales will require a thorough understanding of non-linear cosmic structure formation. Nevertheless, high-resolution simulations of cosmic structures in the non-linear regimes for the many existing alternative theories of gravity would not be affordable or almost impossible. Therefore, an essential task for future inference from cosmological data will be to develop analytical approaches to non-linear cosmic structure formation.

KFT offers a promising framework for this purpose. Modifications of gravity theories do not alter the KFT formalism itself. Many alternative theories of gravity can thus easily be incorporated into the fundamental KFT equations, and e.g.\ non-linear power spectra obtained from KFT can be directly compared to observations. The few necessary adaptations concern
\begin{itemize}
  \item the background evolution, encoded in the Hubble function $H$;
  \item the Hamiltonian of the system, including additional degrees of freedom;
  \item a possible evolution of the effective gravitational coupling constant $G_\mathrm{eff}$;
  \item the linear growth factor $D_+$; and
  \item changes to the gravitational potential.
\end{itemize}
\begin{figure}[t]
  \includegraphics[width=\hsize]{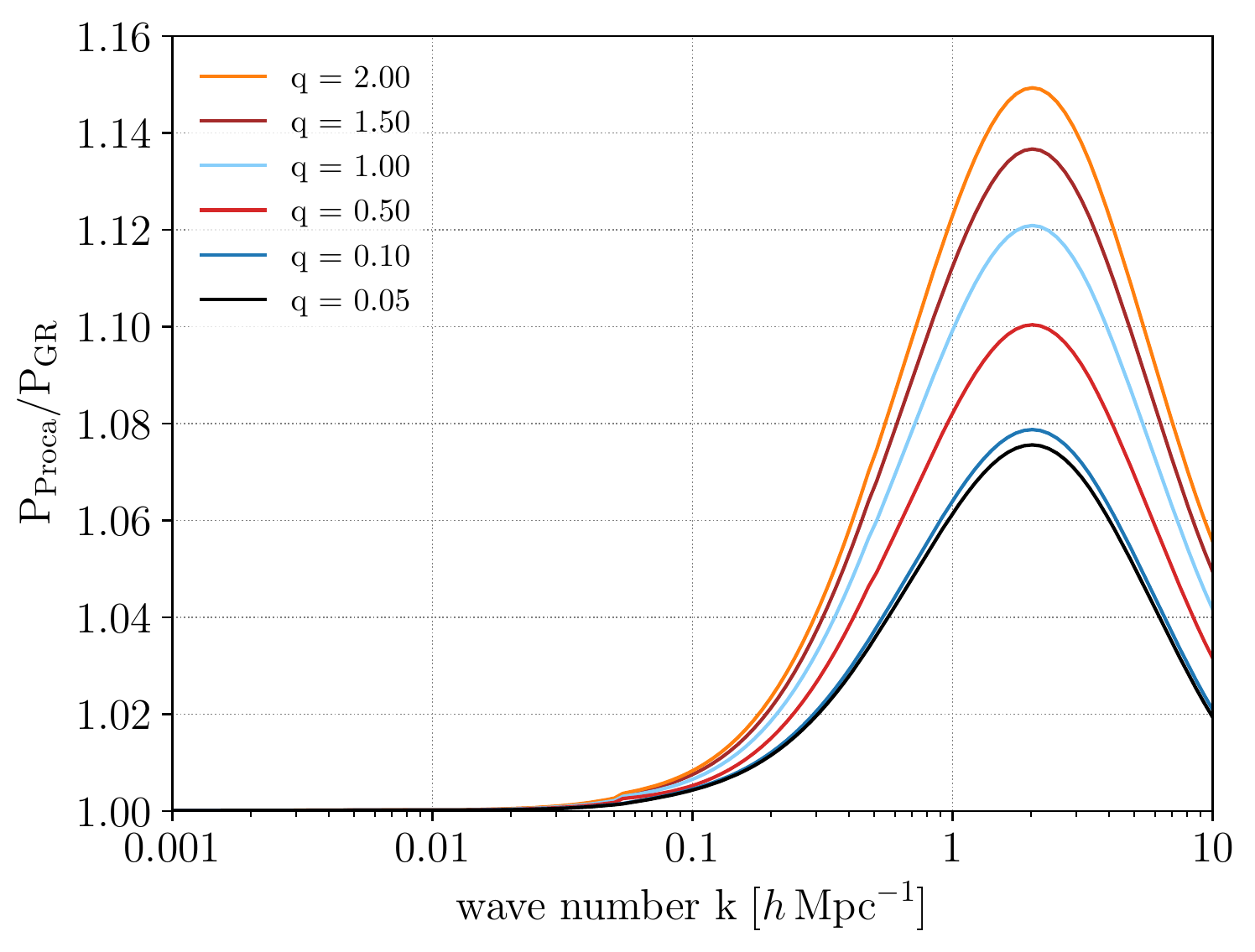}
\caption{The non-linear power spectrum of generalized Proca theories compared to the standard spectrum in general relativity as a function of the wave number $k$. The different models within this class of gravity theories are represented by $q_v$. }
\label{fig:15}
\end{figure}
As a proof of concept, we show here results obtained within a specific dark-energy model taken from the class of generalized Proca theories \cite{2014JCAP...05..015H, 2014JHEP...04..067T, 2016PhLB..757..405B, 2016JCAP...02..004A}. Concrete self-accelerating solutions have successfully been constructed \cite{2016PhRvD..94d4024D, 2016JCAP...06..048D, 2016PhRvD..93j4016D} and compared with background probes \cite{2017PhRvD..95l3540D}. The non-linear power spectrum obtained for this specific modification of gravity is shown in Fig.\ \ref{fig:15}. One immediate observation is the increase of power on scales $k\sim 2\,h\,\mathrm{Mpc}^{-1}$, which would give rise to enhanced structure formation on these scales. Direct comparisons of such results with $N$-body simulations would be highly interesting.

\section{Beyond cosmological applications}
\label{sec:12}

As stated at the beginning of this article, Kinetic Field Theory can be applied to any classical ensemble of particles in or out of equilibrium. To illustrate this, an interesting exercise is to consider an ensemble of Rydberg atoms out of equilibrium.

In the past decade Rydberg atoms -- atoms excited into a very high principal quantum number $n$ -- have enjoyed increasing popularity in theoretical as well as experimental physics. Due to the large separation of the outer electron and the nucleus, Rydberg atoms have a strong dipole moment. Thus, the interaction of atoms in Rydberg states with each other even when they are separated by a microscopic distance is still strong, whereas the van der Waals forces between two ground-state atoms separated by a macroscopic distance would be negligible. In addition, Rydberg atoms have a long lifetime of $\sim 100\, \mu\mathrm{s}$. This makes them excellent candidates for the study of interactions in many-body systems \cite{2009NatPh...5...91W, 2016PrMPh..68..178B}.

\begin{figure}[ht]
  \includegraphics[width=\hsize]{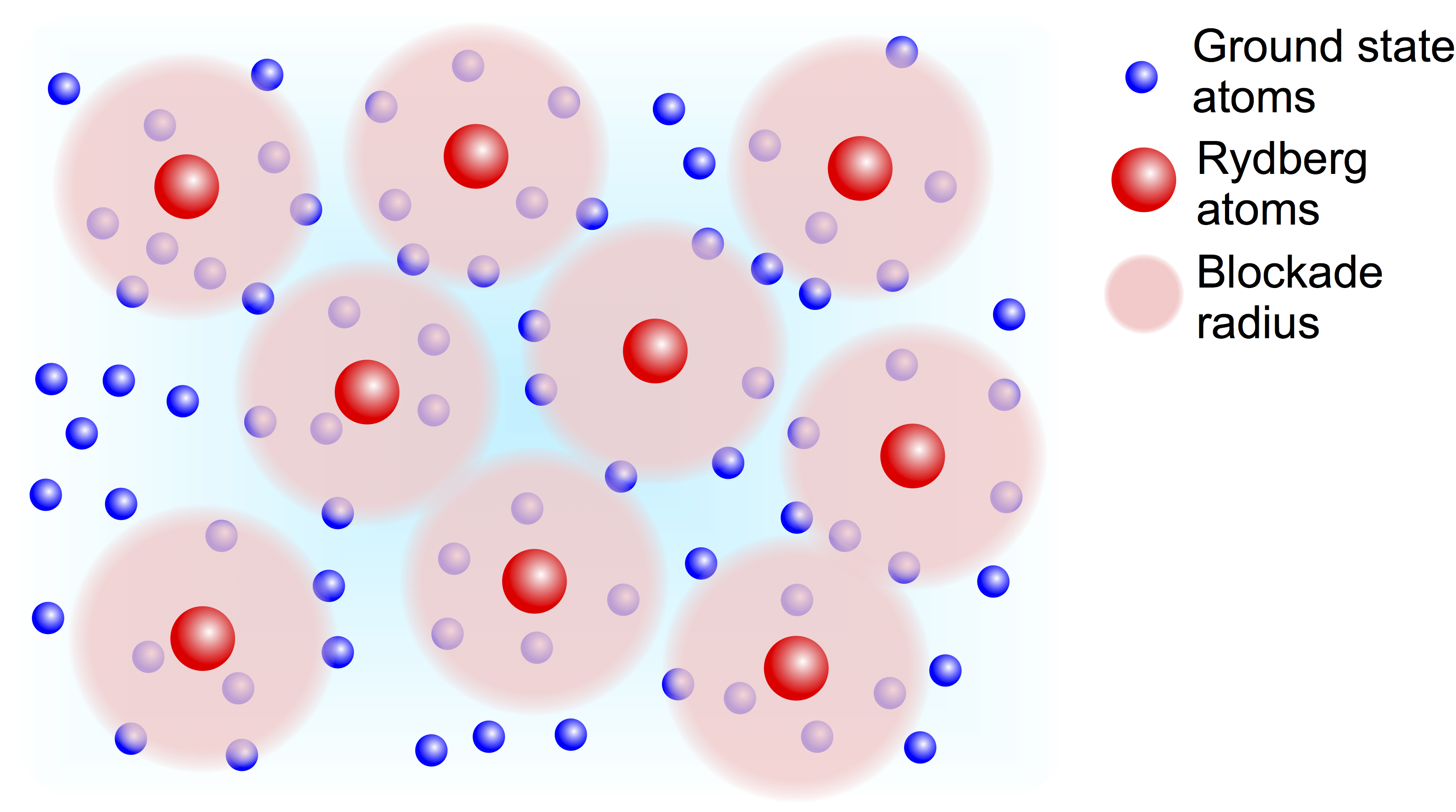}
\caption{Illustration of the Rydberg blockade in a gas. The ground-state atoms (blue) are exited into Rydberg states (red). However, due to the Rydberg blockade, there is a radius $R_b$ around each Rydberg atom where no other ground state atom can be excited.}
\label{fig:16}
\end{figure}

\begin{figure}[ht]
  \includegraphics[width=\hsize]{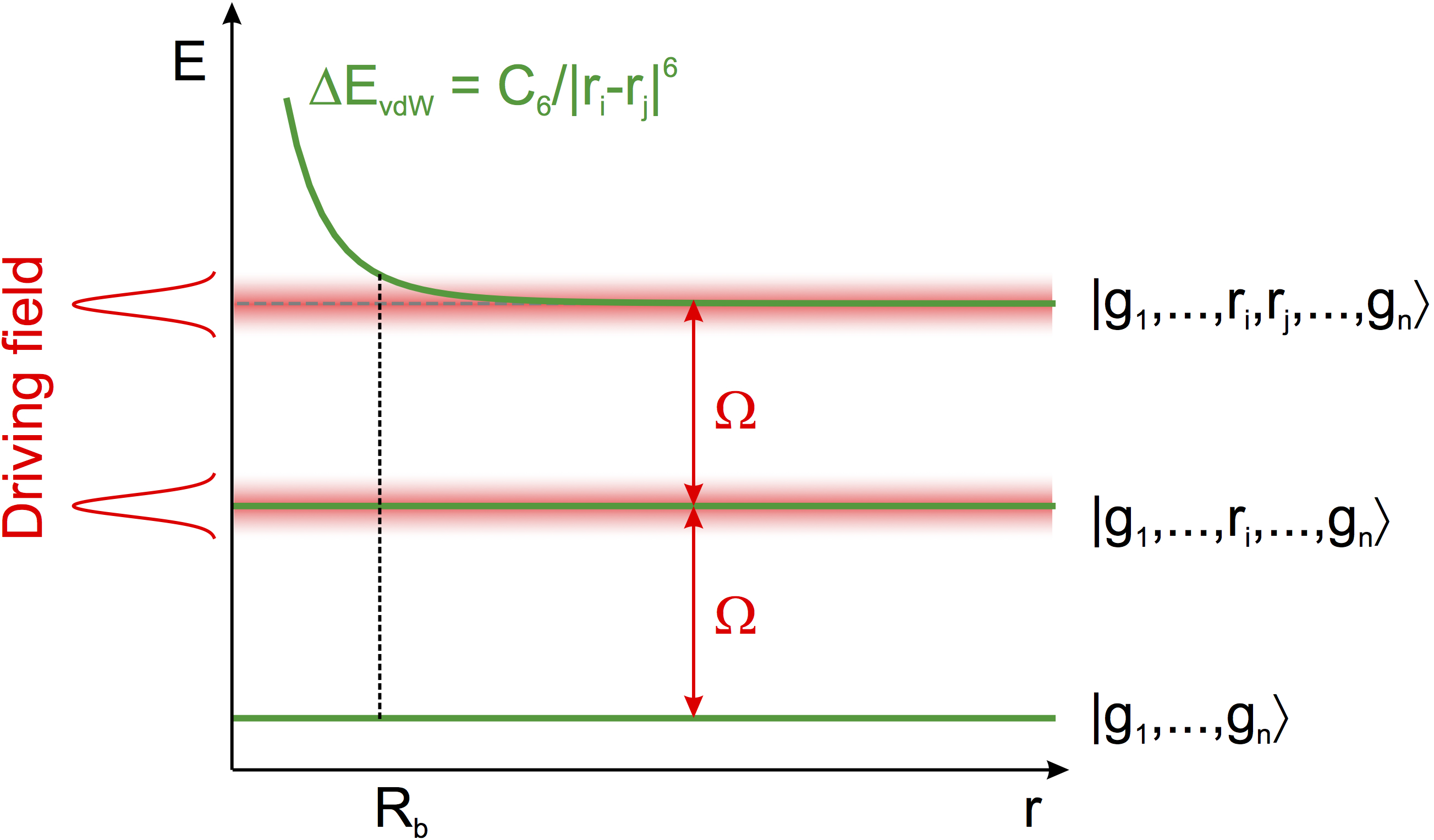}
\caption{Illustration of the energy shift due to the Rydberg blockade. A laser tuned to resonance with the excitation of one atom is not resonant with the excitation of the second atom inside the Rydberg blockade radius, given that the line width of the excitation is smaller than $\Delta E_\mathrm{vdW}$. When in the Rydberg-blockade regime, the system undergoes Rabi oscillations between the collective states $\vert gg ... g \rangle$ and $\vert gg ... r_j ... g \rangle$ at a frequency $\sqrt{N}\Omega$ with the single-atom Rabi frequency $\Omega$ characterizing the coupling between the ground and Rydberg state of a single atom.}
\label{fig:17}
\end{figure}

The two properties of Rydberg atoms that make them such interesting systems for KFT are their strong interactions and the Rydberg blockade \cite{2005rdat.book.....G, 2004PhRvL..93p3001S} illustrated in Fig.\ \ref{fig:16} and Fig.\ \ref{fig:17}. The Rydberg blockade provides a natural (anti-)correlation function for these systems, as no two Rydberg atoms can be closer than two times the Rydberg blockade radius $R_b$ initially.

Strictly speaking, a system of Rydberg atoms has to be treated in the scope of quantum mechanics. However, if treated as a classical system with given anti-correlation function due to the Rydberg blockade radius, they can be useful to distinguish between quantum and classical correlations in strongly-correlated many-body systems. Finding the barrier at which quantum effects can no longer be ignored is a central task in the field of quantum mechanics and may help to understand the nature of correlations due to quantum-mechanical effects.

\subsection{Rydberg systems and KFT}
\label{sec:12.1}

The initial correlation function is set up with the following picture in mind: $N$ ground-state atoms with a high packing fraction are simultaneously excited into the Rydberg state. Due to the Rydberg blockade, no two Rydberg atoms are closer than $2\, R_b$. Thus, the Rydberg atoms can initially well be approximated as hard spheres. The excitation is assumed to be a Gaussian random process, even though, in reality, Gaussianity is broken precisely because of the Rydberg blockade effect due to a strong interaction between the atoms. It stands to reason that, although Gaussianity is broken on very small scales around each Rydberg atom, it is again restored on scales outside the Rydberg blockade, where interactions between ground-state atoms are negligible and atoms are excited at random. Since the initial conditions are formulated at the scales of the Rydberg-blockade radius, and the volume of the Rydberg blockade is very small compared to the relevant scales, Gaussianity can be safely assumed. Thus, the initial distribution of Rydberg atoms will be described by a multivariate Gaussian with a correlation matrix determined by anti-correlations due to the Rydberg blockade. The initial correlations can thus be formulated just as in Sect.\ \ref{sec:3.3}.

\begin{figure}[ht]
  \includegraphics[width=\hsize]{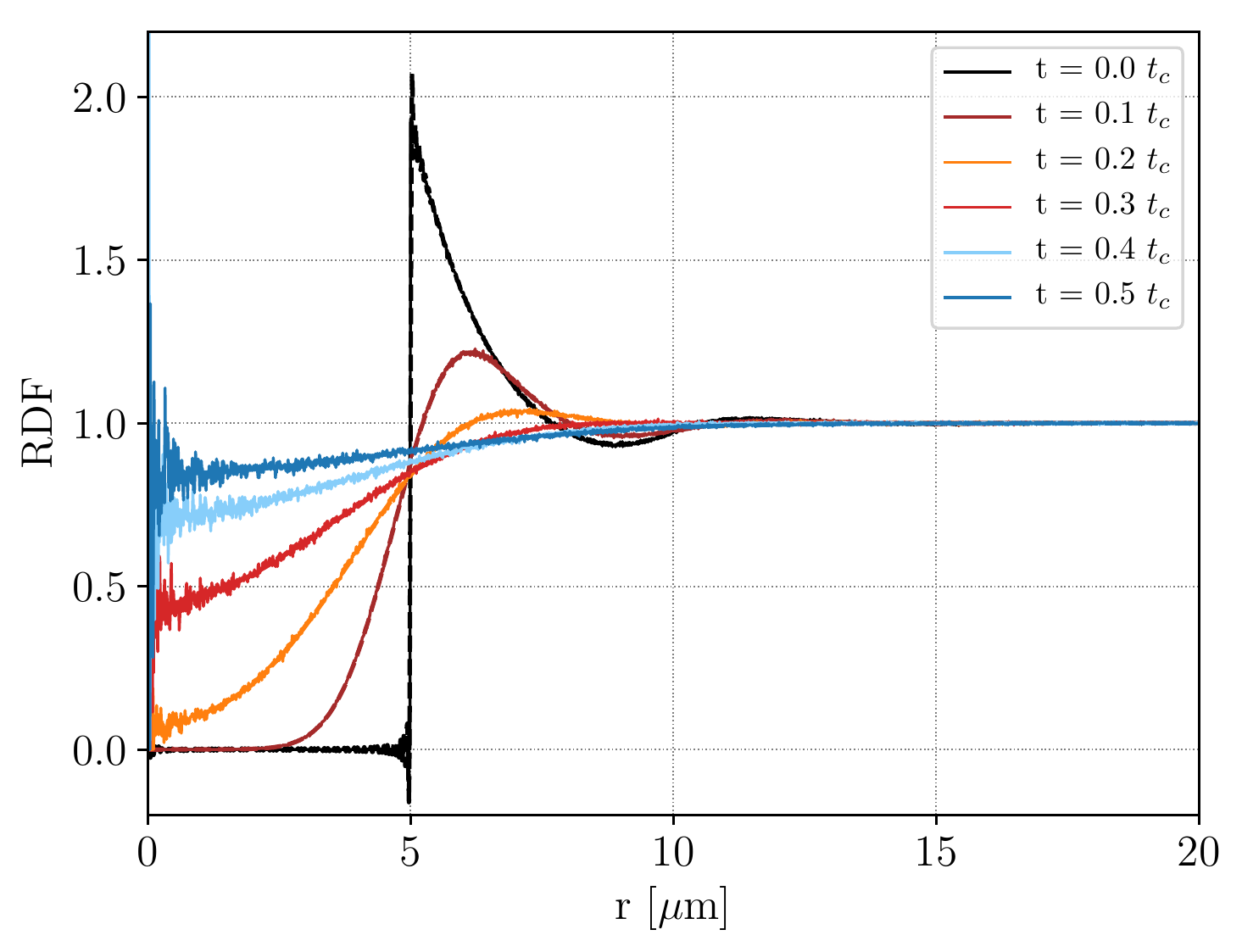}
\caption{Comparison between predictions from KFT (\emph{dashed}) and MD simulations (\emph{solid}) for the free evolution, i.e.\ no interaction potential between particles, of a system of initially anti-correlated Rydberg atoms. The dashed lines are barely visible due to the good agreement with the, slightly noisy, solid lines from the simulation. Note that the radial distribution function $\mathrm{RDF} = \xi(r) + 1$ is shown here instead of the two-point correlation function $\xi(r)$. The results shown here are for $n = 32000$ particles of mass $m = 1.44\cdot 10^{-25}\,\mathrm{kg}$ in a box of $V = 8\cdot 10^6\, \mu \mathrm{m}^3$ and a Rydberg radius of $R_b = 5\, \mu \mathrm{m}$. The results are given in terms of the collision time-scale $t_c = \bar d/v_\mathrm{th} \approx 64\, \mu\mathrm{s}$ with mean particle distance $\bar d$ and thermal velocity $v_\mathrm{th}$.}
\label{fig:18}
\end{figure}

The system of $N$ Rydberg atoms is assumed to be homogeneous and isotropic, with the particle positions initially being correlated due to the Rydberg blockade, but momentum correlations being absent except for a momentum dispersion due to a physical temperature that will be set externally. In Fig.\ \ref{fig:18} the initial correlation function which is directly sampled from a molecular-dynamics (MD) simulation is shown. When treating Rydberg atoms as classical particles, their trajectories will be subject to Hamilton's equations of motion.

The result for the free evolution, i.e.\ no interaction potential between particles, of the system is shown in Fig.\ \ref{fig:18}. The predictions from KFT for the free evolution of particles agree with those from the MD simulations on all scales. The initial structures due to the Rydberg blockade are gradually washed out by thermal motion. The inclusion of an interaction potential between particles is now straightforward in the scope of the KFT formalism. For this system the resummation scheme introduced in Sect.\ \ref{sec:9} is the most efficient approach to compute the density-fluctuation power spectrum with linear effects. With the tree-level results from Sect.\ \ref{sec:9}, it is already possible to capture all linear effects in the initial power spectrum to infinite order in the interaction potential.

\section{Summary and conclusions}
\label{sec:13}

Kinetic field theory applies the principles and concepts of a statistical field theory to ensembles of classical particles in or out of equilibrium. The particle ensemble is characterized by the probability distribution of its phase-space coordinates and by the equations of motion. An exact and complete generating functional can be defined such that the statistical properties of the particle ensemble at any later time can then be derived by applying suitable functional derivatives to this generating functional. Up to this point, the theory is applicable to wide classes of classical particle ensembles, or, more generally, to ensembles of classical degrees of freedom.

The focus of this review has been the application of KFT to cosmology. In contrast to other analytic approaches to cosmic structure formation, KFT has the decisive advantage of avoiding the notorious shell-crossing problem by construction. Compared to the structurally similar statistical quantum field theories, the Hamiltonian equations of motion for classical particles are deterministic and, by their symplectic nature, have unit functional determinant. The initial state for the particles in cosmology is empirically well-defined as a statistically homogeneous and isotropic, Gaussian random field. Based on these conditions, applying KFT to cosmology is quite straightforward.

Particle interactions have to be approximated, though. We have described three ways for doing so; first, by expanding the interaction operator into a Taylor series, leading to the equivalent of Feynman rules; second, by applying Born's approximation relative to suitably chosen inertial particle trajectories; and third, by a resummation scheme that emerges from a reformulation of KFT purely in terms of macroscopic fields. We have shown how non-linear power spectra for the cosmic density and velocity fields can be derived from KFT and how KFT can be used to explain the internal structure of gravitationally-bound, dark-matter structures.

We have further clarified the relation between KFT and the BBGKY hierarchy of more conventional kinetic theory. The application of KFT to fluids is possible by introducing mesoscopic pseudo-particles which, in addition to their microscopic interaction potential, acquire a repulsive potential for mimicking pressure forces. This approach then allows to describe mixtures between dark matter and gas, and we have shown how effects like the baryonic acoustic oscillations can be described in this way. As a further extension of KFT within cosmology which is quite straightforwardly possible, we have given a first example for its application to generalizations of general relativity \cite{2018arXiv180701725H}. Finally, we have shown an application of KFT to a completely different physical system, composed of cold Rydberg atoms, to illustrate the flexibility of the KFT approach.

KFT itself, but in particular its applications to cosmology, are in an early stage of development, and much work remains to be done. The main purpose of this review is three-fold: (1) It should demonstrate how natural the foundations of KFT are and how straightforward its generating functional can be derived. (2) It should sketch formal extensions of the theory, in particular in view of different ways to include particle-particle interactions and resummation schemes. (3) It should describe possibilities for manifold applications of KFT to cosmological problems, including the joint evolution of gas and dark matter, and give an outlook on applications of KFT to systems entirely unrelated to cosmology. Fundamental clarifications of statistical properties of self-gravitating systems are also in reach and have been initiated in terms of fluctuation-dissipation relations \cite{2018JPhCo...2b5020D}. Apart from this, calculating higher-order spectra of cosmological quantities and quantifying the accuracy of KFT are among the next important steps to be undertaken.

\begin{acknowledgements}
  First and foremost, we wish to acknowledge the numerous and most valuable contributions of many students to the development of KFT and its applications to cosmology and Rydberg gases, in particular Arne Bahr, Daniel Berg, Moritz Beutel, Sebastian Bieringer, Marc Boucsein, Igor Br\"ockel, Robin B\"uhler, Frederick del Pozo, Johannes Dombrowski, Leander Fischer, Marlene G\"otz, Lukas Heizmann, Patrick Jentsch, Felix Kuhn, Laila Linke, Julius Mildenberger, Robert Ott, Martin Pauly, Christophe Pixius, Thimo Preis, Philipp Saake, Andr\'e Salzinger, Jan Schneider, Alexander Schuckert, Johannes Schwinn, Christian Sorgenfrei, Marie Teich, and Stefan Zentarra. We are grateful to Manfred Salmhofer for his particularly precious help and advice, to J\"urgen Berges, Jan Pawlowski, Matthias Weidem\"uller, Christof Wetterich, Stefan Hofmann, Adi Nusser, Andreas Wipf, Ruth Durrer, Norbert Straumann, Luca Amendola, and Bj\"orn M.\ Sch\"afer for clarifying discussions and support. Martin Feix and Sven Meyer contributed creative ideas. The Heidelberg Graduate School for Fundamental Physics, the Centre for Quantum Dynamics of Heidelberg University, and the Max Planck Institute for Astronomy provided substantial financial support, partly from funds of the Excellence Initiative of the German Federal and State Governments. This work was supported in part by the Transregional Collaborative Research Centre ``The Dark Universe'' of the German Science Foundation.
\end{acknowledgements}

\onecolumn

\section*{Appendix}
\label{sec:appendix}

\subsection*{Appendix A: Guide to calculating the power spectrum in the perturbative approach}
\label{subsec:appA}

This appendix is intended as a comprehensive guide to provide the reader with a concise picture of the steps needed to arrive at the power spectrum presented in Fig. \ref{fig:3} of this review. \footnote{Note that all vectors in this appendix will be indicated by an arrow to help avoid possible confusion during implementation.}

We begin with (\ref{eq:40}) and apply 
\begin{enumerate}[(1)]
  \item a Taylor expansion of the exponential factor in terms of the interaction operator to first order; and
  \item a Taylor expansion of the exponential factor in the initial conditions (\ref{eq:67}) in terms of the initial correlations up to second order.
\end{enumerate}
Step 1 results in the following expression for the power spectrum
\begin{equation}
  P_\delta(k) \approx \left[1+ \mathrm{i} \hat{S}_{\mathrm{I}}\right] Z_0[\vc{L}, 0]\;.
\end{equation}
To prepare for step 2 we split the momentum correlations (\ref{eq:68}) into a damping term 
\begin{equation}
    Q_D := \frac{\sigma_1^2}{3}\left\langle\vc L_p,\vc L_p\right\rangle
\end{equation}
and a term contaning momentum correlations between any two distinct particles,
\begin{equation}
  Q := \vc L_p^\top\left[
    C_{p_jp_k}\otimes E_{jk}
  \right]\vc L_p\;.
\end{equation}
The Taylor expansion up to second order of step 2 then results in the following expressions,
\begin{align}
  Z_0^{(1)}[\vc L,0] &:= V^{-N}\E^{-Q_D/2}
  \int\D\vc q\,\left(
    1-\frac{Q}{2}+\mathrm{i}\sum_jC_{\delta_jp_k}\vec L_{p_k}+
    \frac{1}{2}\sum_{j\ne k}C_{\delta_j\delta_k}
  \right)
  \E^{\mathrm{i}\left\langle\vc L_q,\vc q\right\rangle}\;, \nonumber\\
  Z_0^{(2)}[\vc L,0] &:= \frac{V^{-N}}{8}\E^{-Q_D/2}
  \int\D\vc q\,Q^2\,
  \E^{\mathrm{i}\left\langle\vc L_q,\vc q\right\rangle}\;.
\label{eq:01-96}
\end{align}
$Z_0^{(1)}[\vc{L}, 0]$ contains all terms up to first order in initial correlations and $Z_0^{(2)}[\vc{L}, 0]$ only the second order terms. We can then write the free generating functional as
\begin{equation}
  Z_0[\vc{L}, 0] \approx Z_0^{(1)}[\vc{L}, 0] + Z_0^{(2)}[\vc{L}, 0]\;. 
\end{equation}

Note that we only consider higher-order momentum correlations here and ignore cross terms of the form $QC_{\delta_j\delta_k}$ and $QC_{\delta_jp_k}\vec L_{p_k}$ since terms containing momentum correlations dominate at late times due to the time dependence of the momentum propagator $g_{qp}(\tau) := g_{qp}(\tau, 0)$. 
The initial momentum correlations and, of course, the interaction potential between different particles is responsible for the growth of structures while the damping term is responsible for dissolving those structures. If we take initial momentum correlations as well as interactions into account order by order, we must make sure that the damping term does not enter exponentially into the low-order contributions since this would lead to an overestimation of the damping effect. We shall therefore later approximate the damping term $\exp(-Q_D/2)$ consistently at one order less than the term $\exp(-Q/2)$. This implies that damping will only be included in terms of at least second-order in the initial momentum correlations.

To first and second order in the initial correlations, with the improved Zel'dovich propagator $ g_{qp}(\tau)$ defined in (\ref{eq:59}) and suitably approximated damping terms, we thus find the following expressions for the power spectrum in our \emph{free} theory
\begin{align}
  P_\delta^{(1)}(k, \tau) &= P_\delta^\mathrm{(i)}\left(k\right)
  \left(1+ g_{qp}(\tau)\right)^2 \;,\nonumber\\
  P_\delta^{(2)}(k, \tau) &= \frac{ g^4_{qp}(\tau)}{2\Bigl(
    1+\frac{\sigma_1^2}{3} g^2_{qp}(\tau)k^2
  \Bigr)}
  \int_{k'}
  P_\delta^\mathrm{(i)}\left(k'\right)
  P_\delta^\mathrm{(i)}\left(\vec{k}-\vec{k}'\right)
  \left(\frac{\vec{k}\cdot\vec{k}'}{k'^2}\right)^2
  \left(
    \frac{\vec{k}\cdot(\vec{k}-\vec{k}')}{(\vec{k}-\vec{k}')^2}
  \right)^2\;,
\label{eq:01-182}
\end{align}
where we have specified for clarity that the power spectra on the right-hand sides are the power spectra $P_\delta^\mathrm{(i)}$ characterising the initial particle distribution.

In our perturbative approach the interaction potential is given by
\begin{equation}
  \tilde v(k, \tau) = -\frac{A(\tau)}{\bar\rho}
  \left(\frac{1}{k^2}+\frac{\bar\nu}{k}\right)\;,
\label{eq:01-193}
\end{equation}
where we have introduced an additional term as compared to (\ref{eq:82}) which corresponds to an adhesion term following the idea of the adhesion approximation \cite{1989RvMP...61..185S, 1989MNRAS.236..385G, 2012PhyU...55..223G}. This approximation was introduced in order to compensate the effect of free streaming in the Zel'dovich approximation once particle trajectories have crossed. The necessity for such a term in the perturbative scheme arises due to the fact, that to first order in the gravitational potential the overshooting of particle trajectories due to the improved Zel'dovich propagator is not adequately compensated, causing a loss of power on small scales, where structures are wiped out.
This overshooting is compensated by going to second-order perturbation theory, but has not yet been implemented in our code. In our more advanced schemes to include particle interactions which we present in Sect. \ref{sec:4.2} and \ref{sec:4.3} the adhesion term becomes obsolete, since the choice for the free propagator describing inertial motion and the form of the force term are consistently linked.\\

In (\ref{eq:01-193}), $\bar\nu$ is an amplitude with the dimension of a length scale. As a suitable length scale, we choose the velocity dispersion $\sigma_v$, propagated to the time $\tau$ by the improved Zel'dovich propagator $ g_{qp}(\tau)$,
\begin{equation}
  \bar\nu =  g_{qp}(\tau)\,\sigma_v\;.
\label{eq:01-194}
\end{equation}
We approximate $\bar\nu$ by its late-time behaviour, setting $ g_{qp}(\tau)\,\sigma_v\approx2$.\\

With the interaction potential (\ref{eq:01-193}), the first-order perturbation contributions to the non-linear power spectrum are
\begin{align}
\label{eq:01-190}
  \delta^{(1)}P_\delta^{(1)}(k, \tau) &=
  2\left(1+ g_{qp}(\tau)\right)P_\delta^\mathrm{(i)}(k)
  \int_{0}^{\tau}\D\tau'
  A(\tau') g_{qp}(\tau,\tau')\left(1+ g_{qp}(\tau')\right)\;,\\
  \delta^{(1)}P_\delta^{(2)}(k, \tau) &= \frac{2}{\Bigl(
    1+\frac{\sigma_1^2}{3} \bigl(T^2 + T'^{\,2} - \vec T \cdot \vec T'\bigr)
  \Bigr)} \int_0^\tau\D\tau'
  A(\tau') g_{qp}(\tau,\tau')
  \int_{k'} (Z_\mathrm{A} + Z_\mathrm{B} + Z_\mathrm{C} + Z_\mathrm{D})\;,\nonumber
\end{align}
where we introduced the notations $T := g_{qp}(\tau) \vec k$ and $T' := g_{qp}(\tau') \vec k'$ for brevity. The terms $Z_\mathrm{A}$, $Z_\mathrm{B}$, $Z_\mathrm{C}$, and $Z_\mathrm{D}$ have been derived in detail in \cite{2016NJPh...18d3020B} so that we merely state the results here,
\begin{align}
  Z_\mathrm{A} &= g^2_{qp}(\tau) \, g_{qp}(\tau') \, \frac{(\vec k\cdot\vec k')^2 \, \vec k\cdot(\vec k-\vec k') \,
    (\vec T - \vec T')\cdot(\vec k-\vec k')}{k'^{\,2} \, (\vec k-\vec k')^4}
  P_\delta\left(\vec k-\vec k'\right)P_\delta\left(\vec k'\right)\;,
\label{eq:01-168} \\
 Z_\mathrm{B} &= - g_{qp}(\tau) \, g^2_{qp}(\tau') \, \frac{(\vec k\cdot\vec k')^2 \, \vec k'\cdot(\vec k-\vec k') \,
    (\vec T - \vec T')\cdot(\vec k-\vec k')}{k^2 \, (\vec k-\vec k')^4}
  P_\delta\left(\vec k\right)P_\delta\left(\vec k-\vec k'\right)\;,
\label{eq:01-169} \\
  Z_\mathrm{C} &= - g_{qp}(\tau) \, g_{qp}(\tau') \, \frac{(\vec k\cdot\vec k') \, \vec k\cdot(\vec T - \vec T') \, \vec k'\cdot(\vec T - \vec T')}{k^2 \, k'^{\,2}}
  P_\delta\left(\vec k\right)P_\delta\left(\vec k'\right)\;,
\label{eq:01-170} \\
  Z_\mathrm{D} &=
  \frac{1}{2} \, g^2_{qp}(\tau) \, g^2_{qp}(\tau') \, k^2 \,
  \frac{(\vec k\cdot\vec k')^2 \, \bigl(\vec k\cdot(\vec k-\vec k')\bigr)^2}{k'^{\,4} \, (\vec k-\vec k')^4}
  P_\delta\left(\vec k-\vec k'\right)P_\delta\left(\vec k'\right)\;.
\label{eq:01-171}
\end{align}

The terms $P_\delta^{(1)}$ and $\delta^{(1)}P_\delta^{(1)}$ that are proportional to the initial power spectrum must reproduce the linear growth of the initial power spectrum that can be given as $D^2_+P_\delta^\mathrm{(i)}(k)$. We therefore replace the terms linear in the initial power spectrum in (\ref{eq:01-182}) and (\ref{eq:01-190}) by $D^2_+P_\delta^\mathrm{(i)}(k)$.
The final result for the non-linear density-fluctuation power spectrum to first order in the interaction potential and to second order in the initial correlations is then given by
\begin{equation}
  P_\delta (k, \tau) = D^2_+(\tau)P_\delta^\mathrm{(i)}(k) + P_\delta^{(2)}(k, \tau) + \delta^{(1)}P_\delta^{(2)}(k, \tau)\;.
  \label{eq:01-172}
\end{equation}
We still need to evaluate the integrals over $\tau'$ and $k'$ that are left in (\ref{eq:01-172}) which is done numerically.

The final ingredients still required are the cosmological model of our Universe and an initial density-fluctuation power spectrum. The cosmological parameters are needed for the computation of the expansion function of the Universe which, in turn, is needed for the growth factor $D_+(a)$ that serves as a time coordinate for us. We have, of course, now the complete freedom to take the cosmological model of our choosing; we can choose any initial power spectrum that we like, start our calculations at the appropriate scale factor $a$ at which our initial power spectrum is defined and can then compute the non-linear power spectrum (\ref{eq:01-172}) at a desired final scale factor. To arrive at Fig. \ref{fig:3} we made the following choices:
\begin{itemize}
  \item  $\Lambda$CDM Universe with $\Omega_\mathrm{m} = 0.3$, $\Omega_{\Lambda} = 0.7$, $\Omega_\mathrm{b} = 0.04$, $h = 0.7$, $\sigma_8 = 0.8$.
  \item The initial power spectrum was generated with CAMB \cite{2000ApJ...538..473L} at redshift $z=1100$.
  \item Our final time is chosen to be today (corresponding to $z=0$).
\end{itemize}
There are no other parameters in KFT.

\twocolumn

\bibliographystyle{andp2012}
\bibliography{main}

\end{document}